\documentclass[preprint2]{aastex}
\usepackage{float}

\newcommand{\sgr}{\mbox{SGR\,J0501+4516~}}
\newcommand{\sgrnos}{\mbox{SGR\,J0501+4516}}

\slugcomment{Not to appear in Nonlearned J., 45.}

\shorttitle{\textit{Fermi}/GBM Observation of \sgrnos}
\shortauthors{Lin et al.}

\begin{document}

\title{\textit{Fermi}/GBM Observations of \sgr Bursts}

\author{Lin Lin\altaffilmark{1, 2}, Chryssa Kouveliotou\altaffilmark{3}, Matthew G. Baring\altaffilmark{4}, Alexander J. van der Horst\altaffilmark{5}, Sylvain Guiriec\altaffilmark{2}, Peter M. Woods\altaffilmark{6}, Ersin G\"o\u{g}\"u\c{s}\altaffilmark{7}, Yuki Kaneko\altaffilmark{7}, Jeffrey Scargle\altaffilmark{8}, Jonathan Granot\altaffilmark{9}, Robert Preece\altaffilmark{2}, Andreas von Kienlin\altaffilmark{10}, Vandiver Chaplin\altaffilmark{2}, Anna L. Watts\altaffilmark{11}, Ralph A.M.J. Wijers\altaffilmark{11}, Shuang Nan Zhang\altaffilmark{1, 12}, Narayan Bhat\altaffilmark{2}, Mark H. Finger\altaffilmark{5}, Neil Gehrels\altaffilmark{13}, Alice Harding\altaffilmark{13}, Lex Kaper\altaffilmark{11}, Victoria Kaspi\altaffilmark{14}, Julie Mcenery\altaffilmark{13}, Charles A. Meegan\altaffilmark{5}, William S. Paciesas\altaffilmark{2}, Asaf Pe'er\altaffilmark{15}, Enrico Ramirez-Ruiz\altaffilmark{16}, Michiel van der Klis\altaffilmark{11}, Stefanie Wachter\altaffilmark{17}, Colleen Wilson-Hodge\altaffilmark{3}}
\email{lin.lin@uah.edu}

\altaffiltext{1}{National Astronomical Observatories, Chinese Academy of Sciences, Beijing 100012, China}
\altaffiltext{2}{CSPAR, University of Alabama in Huntsville, Huntsville, AL 35805, USA}
\altaffiltext{3}{Space Science Office, VP62, NASA/Marshall Space Flight Center, Huntsville, AL 35812, USA}
\altaffiltext{4}{Department of Physics and Astronomy, Rice University, MS-108, P.O. Box 1892, Houston, TX 77251, USA}
\altaffiltext{5}{Universities Space Research Association, NSSTC, Huntsville, AL 35805, USA}
\altaffiltext{6}{Corvid Technologies, 689 Discovery Drive, Huntsville, AL 35806, USA}
\altaffiltext{7}{Sabanci University, Faculty of Engineering and Natural Sciences, Orhanli- Tuzla, \.{I}stanbul 34956, Turkey}
\altaffiltext{8}{Space Science and Astrobiology Division, NASA/Ames Research Center, Moffett Field, CA 94035-1000, USA}
\altaffiltext{9}{Centre for Astrophysics Research, University of Hertfordshire, College Lane, Hatfield, Herts, AL10 9AB, UK}
\altaffiltext{10}{Max Planck Institute for extraterrestrial Physics, Giessenbachstrasse, Postfach 1312, 85748, Garching, Germany}
\altaffiltext{11}{Astronomical Institute "Anton Pannekoek," University of Amsterdam, Postbus 94249, 1090 GE Amsterdam, The Netherlands}
\altaffiltext{12}{Key Laboratory of Particle Astrophysics, Institute of High Energy Physics, Chinese Academy of Sciences, P.O. Box 918-3, Beijing 100049, China}
\altaffiltext{13}{NASA Goddard Space Flight Center, Greenbelt, MD 20771, USA}
\altaffiltext{14}{Department of Physics, Rutherford Physics Building, McGill University, Montreal, QC H3A 2T8, Canada}
\altaffiltext{15}{Harvard-Smithsonian Center for Astrophysics, Cambridge, MA 02138, USA}
\altaffiltext{16}{Department of Astronomy and Astrophysics, University of California, Santa Cruz, CA 95064, USA}
\altaffiltext{17}{Spitzer Science Center, California Institute of Technology, 1200 E. California Blvd., MC 220-6, Pasadena, CA 91125, USA}

\begin{abstract}
We present our temporal and spectral analyses of 29 bursts from \sgrnos, detected with the Gamma-ray Burst Monitor onboard the {\it Fermi}
Gamma-ray Space Telescope during the 13 days of the source activation in 2008 (August 22 to September 3). We find that the $T_{90}$ durations
of the bursts can be fit with a log-normal distribution with a mean value of $\sim 123$\,ms. We also estimate for the first time event
durations of Soft Gamma Repeater (SGR) bursts in photon space (i.e., using their deconvolved spectra) and find that these are very similar to
the $T_{90}$s estimated in count space (following a log-normal distribution with a mean value of $\sim 124$\,ms). We fit the time-integrated
spectra for each burst and the time-resolved spectra of the five brightest bursts with several models. We find that a single power law with an
exponential cutoff model fits all 29 bursts well, while 18 of the events can also be fit with two black body functions. We expand on the
physical interpretation of these two models and we compare their parameters and discuss their evolution. We show that the time-integrated and
time-resolved spectra reveal that $E_{\rm{peak}}$ decreases with energy flux (and fluence) to a minimum of $\sim30$\,keV at
$F=8.7\times10^{-6}$ erg cm$^{-2}$ s$^{-1}$, increasing steadily afterwards. Two more sources exhibit a similar trend: SGRs J$1550-5418$ and
$1806-20$. The isotropic luminosity, $L_{\rm iso}$, corresponding to these flux values is roughly similar for all sources
($0.4-1.5\times10^{40}$ erg s$^{-1}$).
\end{abstract}


\section{Introduction \label{intro}}

Magnetars are slowly rotating neutron stars associated with extreme magnetic fields ($B>10^{14}$ G). Several obscure neutron star
subpopulations haven been claimed as magnetar candidates, in particular Soft Gamma Repeaters (SGRs) and Anomalous X-ray Pulsars (AXPs); for
reviews see \citet{woods2006,mereghetti2008}. Most magnetars have been discovered either from their persistent X-ray emission properties or
when they enter into randomly occurring outbursts, during which they emit a multitude of short ($\sim100$ milliseconds), soft $\gamma-$/hard
X$-$ray bursts. Thus far, approximately twenty magnetar sources are known and most of them reside in the Galactic Plane with a higher
concentration close to the center; two are located in the Magellanic Clouds.

\sgr was discovered with \textit{Swift} on 2008 August 22, when it emitted a series of bright, soft, short bursts
\citep{holland2008,barthelmy2008}. The first burst also triggered the Gamma-ray Burst Monitor (GBM) onboard the {\it Fermi} Gamma-ray Space
Telescope ({\it FGST}; hereafter {\it Fermi}). Soon after, our Target of Opportunity (ToO) observations with the {\it Rossi X-ray Timing
Explorer} ({\it RXTE}) established a period of $\sim$ 5.76 s in the persistent X-ray emission of the source \citep{gogus2008}. Further observations with \textit{RXTE} and the \textit{Swift}/X-ray Telescope (XRT) revealed a spin-down rate of $1.5(5)\times10^{-11}$ s s$^{-1}$, indicating a dipole
magnetic field of $2.0\times10^{14}$ G \citep{woods2008, rea2009, gogus2010}. Our subsequent $\textit{Chandra}$ ToO observations established an
accurate location of the source at $R.A. (J2000) = 05^{\rm h} 01^{\rm m} 06\fs76$, $Dec (J2000) = +45\arcdeg 16\arcmin 33\farcs92$, with an
1$\sigma$ uncertainty of $0\farcs11$ \citep{gogus2010}. This is the first magnetar location at roughly the Galactic anticenter direction, placing \sgr most likely at the Perseus arm of our Galaxy at $\sim2$ kpc \citep{xu2006}.

The \sgr outburst lasted approximately 2 weeks, during which several bright bursts were detected with \textit{Swift}, GBM, {\it RXTE}, {\it
KONUS}-Wind, and {\it Suzaku} (Enoto et al. 2009; Aptekar et al. 2009; Kumar et al. 2010; see also Table \ref{obs}). After the burst activity ceased, the source flux decreased exponentially with an e-folding time of $27.9\pm2.5$ days \citep{gogus2010}. During the entire outburst interval, GBM triggered on 26 bursts; in addition, an untriggered event search in the daily data sets revealed seven more bursts. We present here our
analyses of the 29 GBM bursts from \sgr for which we have high-resolution data; the properties of the X-ray persistent source emission have
been published by \citet{rea2009} and \citet{gogus2010}. In Section \ref{data}, we describe the instrument, observations and data selection. In
Sections \ref{tempo} and \ref{spec}, we present our temporal and spectral results, respectively. We discuss the interpretation of our results
in Section \ref{discussion}.

\section{Instrumentation and Data \label{data}}

The \textit{Fermi}/GBM has a wide field of view (8 sr; un-occulted) and a continuous broad-band energy coverage (8 keV -- 40 MeV). It consists
of 12 NaI detectors (8 -- 1000 keV) arranged in 4 clusters of three each and 2 BGO detectors (0.2 -- 40 MeV) placed at opposite sides of the
spacecraft \citep{meegan2009}. In trigger mode GBM provides three types of science data: CSPEC with continuous high spectral resolution (1024
ms and 128 energy channels), CTIME for continuous high time resolution (64 ms and 8 energy channels), and Time-Tagged photon Event (TTE) data
(2 $\mu$s and 128 energy channels). For a detailed description of the instrument and data types, see \citet{meegan2009}. When a GBM trigger
occurs, TTE data are provided from $\sim 30$ s pre-trigger to $\sim 300$ s post-trigger. With its very high temporal resolution the TTE
datatype is most suitable for the detailed temporal and spectral analyses of very short events like SGR bursts and is, therefore,  the only
datatype used throughout this paper.

During the burst active period of the source (2008 August 22 -- 2008 September 3), GBM triggered on 26 bursts. We further
implemented the algorithm described in \citet{kaneko2010} to search for any untriggered events in the daily data sets of 2008 August 21 through
2008 September 14, and found seven individual short bursts located in the same direction as \sgrnos. Among these only three had TTE data and
were, therefore, included in our analyses. Two of the 29 bursts were very bright, causing saturation of the high speed science data bus; the
saturated parts of these bursts were excluded from any spectral analysis. For all spectra and durations we selected the NaI detectors with an
angle to the source smaller than $50^{\circ}$ to avoid attenuation effects. We also excluded any detectors blocked by the {\it Fermi}/Large Area Telescope (LAT), or by the spacecraft radiators or solar panels. The BGO detectors were not used as there was no obvious emission in the NaIs above 200 keV. In columns two through four of Table \ref{obs} we list for each of the 29 GBM events their trigger numbers, times, and the selected NaI detectors used for the following analyses.

\section{Temporal Analysis \label{tempo}}

\begin{figure*}[p3t]
\includegraphics*[bb=0 0 490 340, scale=0.4]{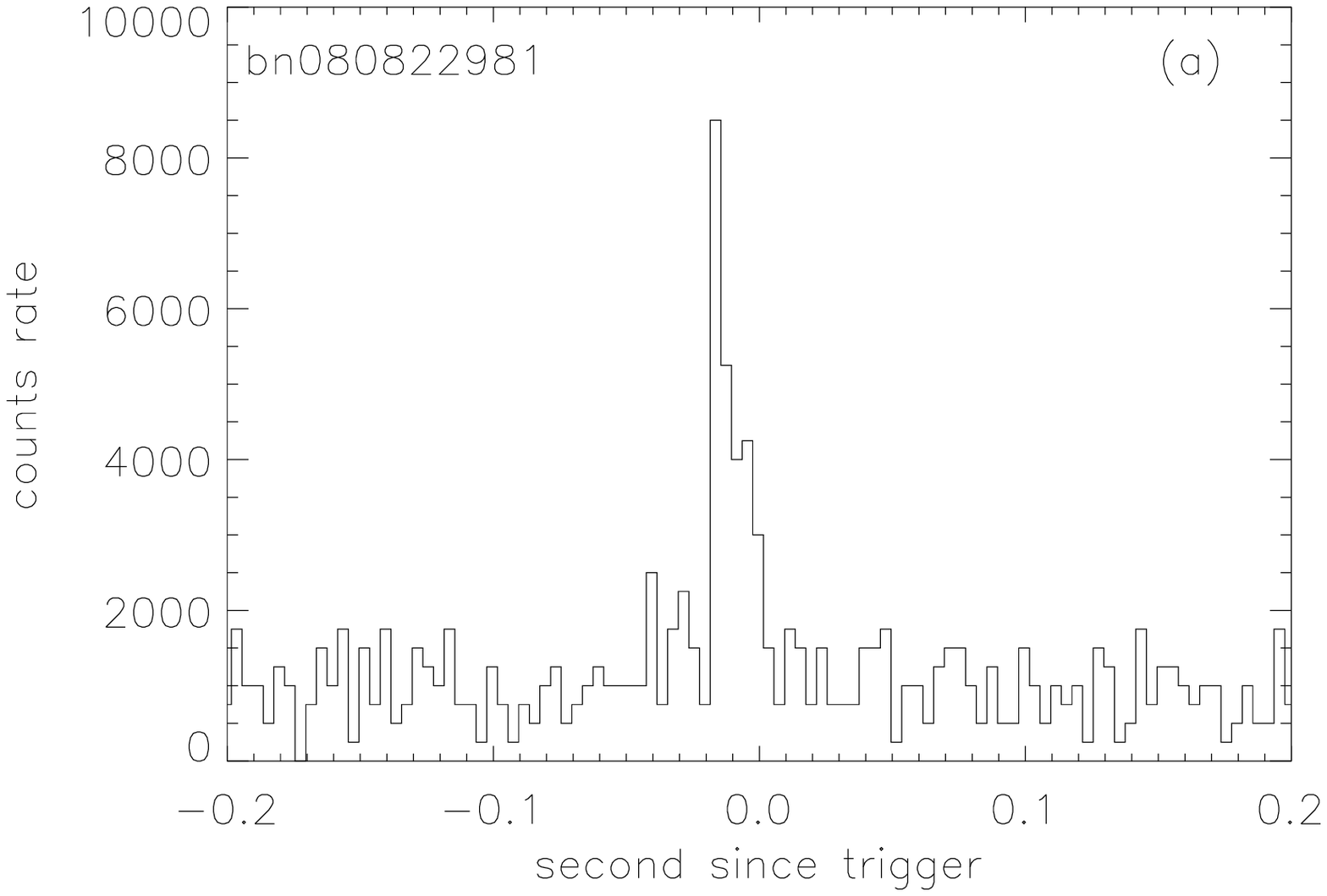}
\includegraphics*[bb=0 0 490 340, scale=0.4]{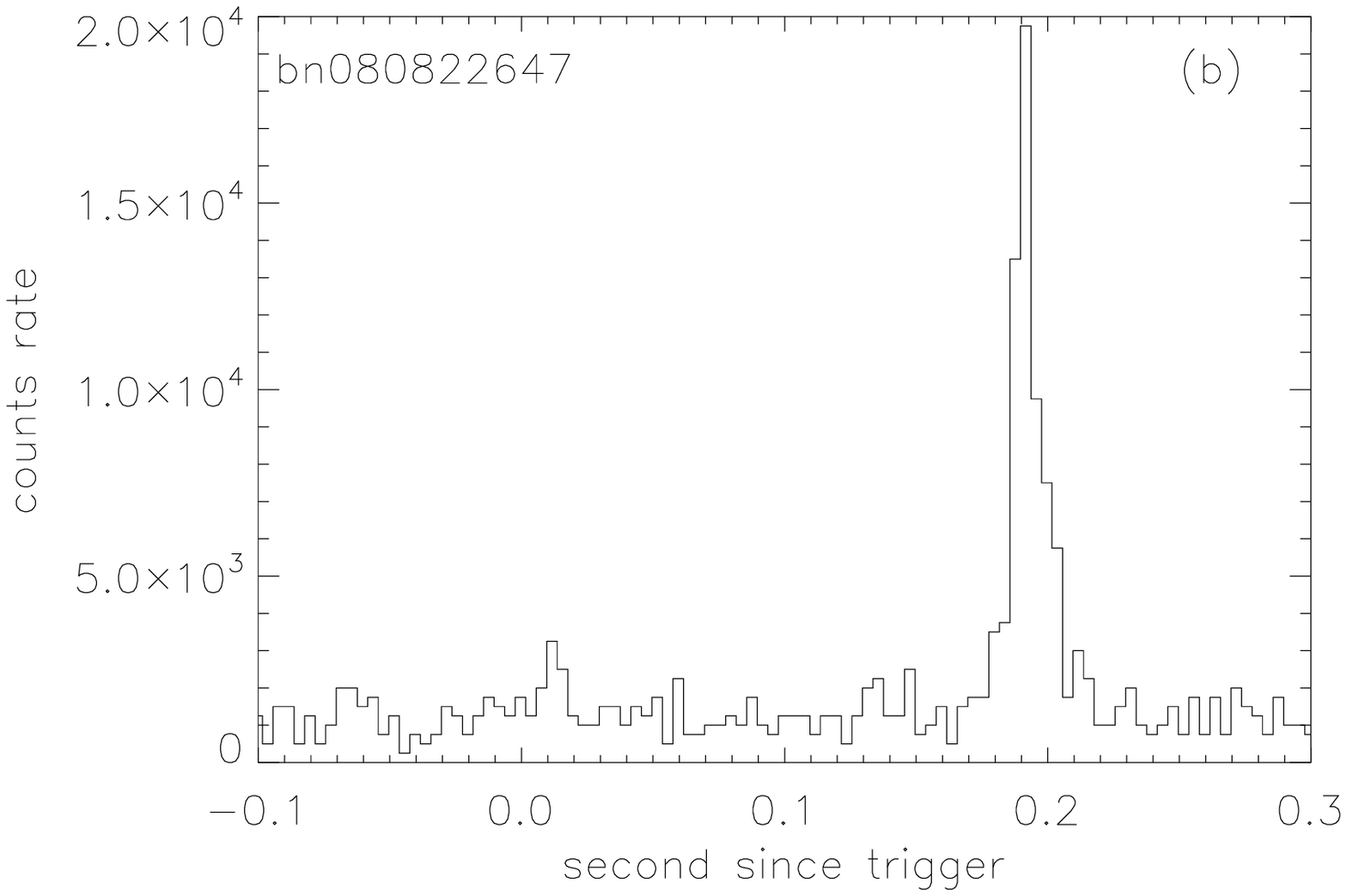}\\
\includegraphics*[bb=0 0 490 340, scale=0.4]{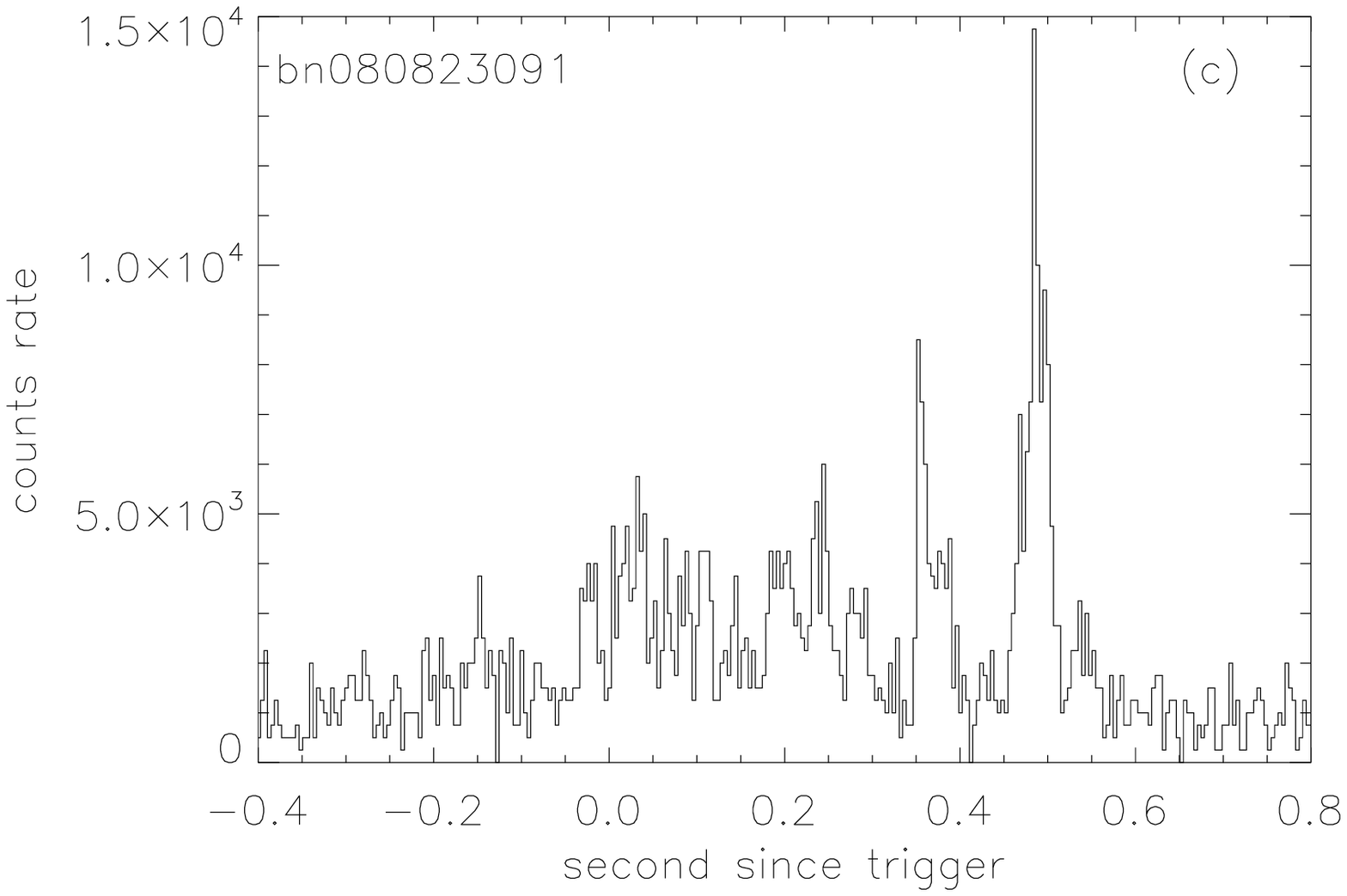}
\includegraphics*[bb=0 0 490 340, scale=0.4]{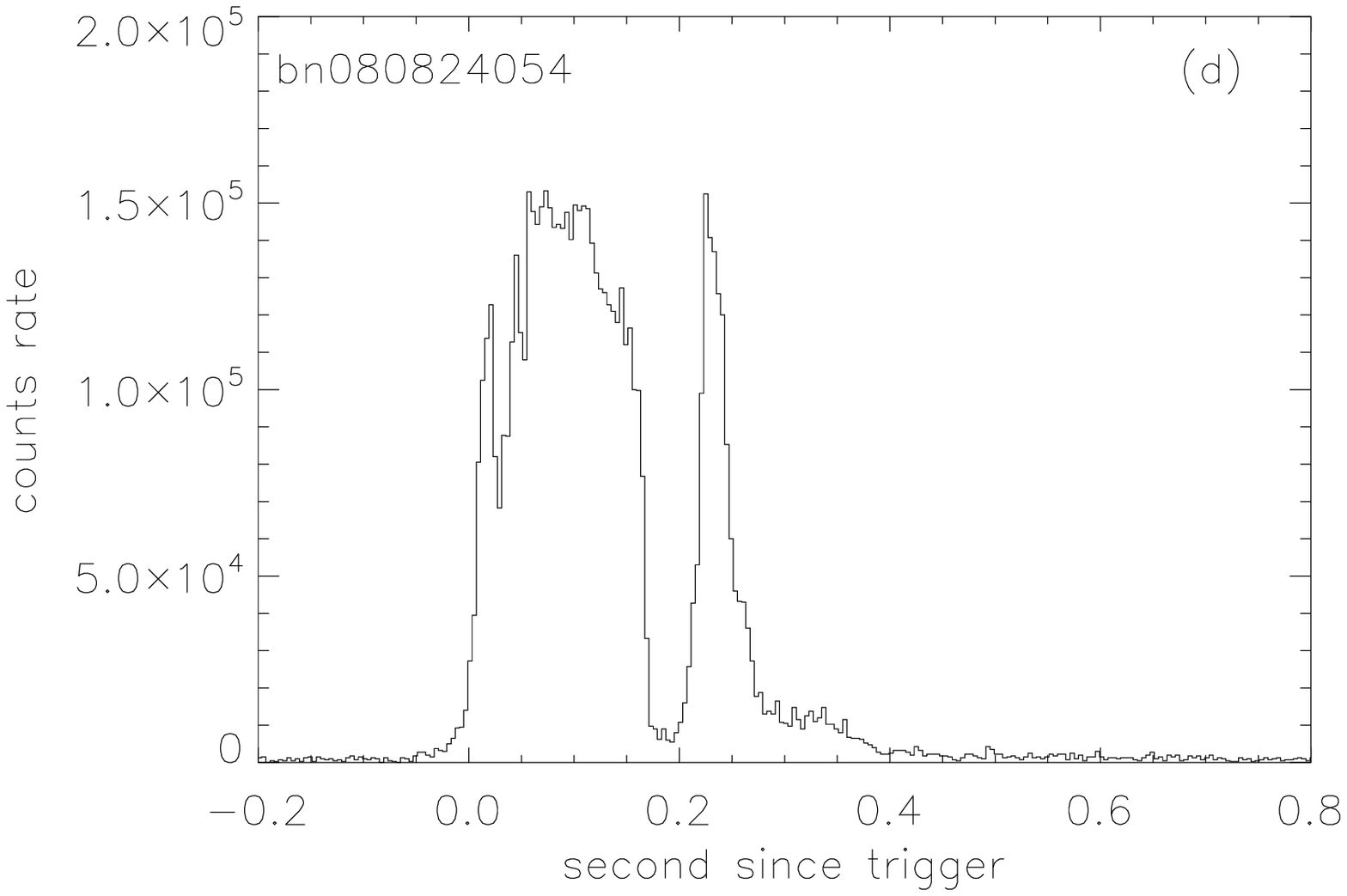}
\caption{Light curves of four bursts from \sgr integrated with 4\,ms bin size over $8-200$\,keV.  \label{lc}}
\end{figure*}

Figure \ref{lc} exhibits the light curves of four representative \sgr bursts. Their profiles vary from a single short pulse (Figure \ref{lc}a)
to a multi-pulse event (Figure \ref{lc}c). We note that the flat top in the event of Figure \ref{lc}d is caused by saturation. We have used the
GBM TTE data to estimate the event durations in both count and photon space. Although the former process has been used exclusively in the past
to estimate SGR and AXP burst durations \citep{gogus2001,gavriil2004}, the latter is used for magnetar bursts for the first time here. We
describe the methods and the results below; Table \ref{durations} contains the mean values and widths of the distributions of all temporal
parameters.

\subsection{$T_{90}$ and $T_{50}$ in count space}

For the $T_{90}$\footnote{$T_{90}$ ($T_{50}$) is the duration during which the background-subtracted cumulative counts increase from $5\%$
($25\%$) to $95\%$ ($75\%$) of the total counts.} estimate we used the algorithm originally developed by \citet{ck1993} for Gamma Ray Bursts (GRBs)
and later adapted for SGR bursts \citep{gogus2001, gavriil2004}, modified slightly to accommodate the GBM TTE data. Each duration was calculated
between $8-100$\,keV with 2\,ms bins as follows. First we fit the burst background using two time intervals before and after each burst ([-2 s,
-0.5 s] and [0.5 s, 2 s]) with a first order polynomial; these intervals were kept mostly the same for all bursts unless a precursor and/or a tail
were detected. Then we fit the background-subtracted, cumulative burst counts, with a linear plus a step function simultaneously; the linear
part was user-selected (before and after the burst), while the step part was determined from the fit linear trend. The height of the step
function was then used to represent the net total counts ($\textit{N}$) of the burst, subsequently used for the $T_{90}$ determination.

\begin{figure*}[p5t]
    \includegraphics*[bb=60 10 500 340, scale=0.5]{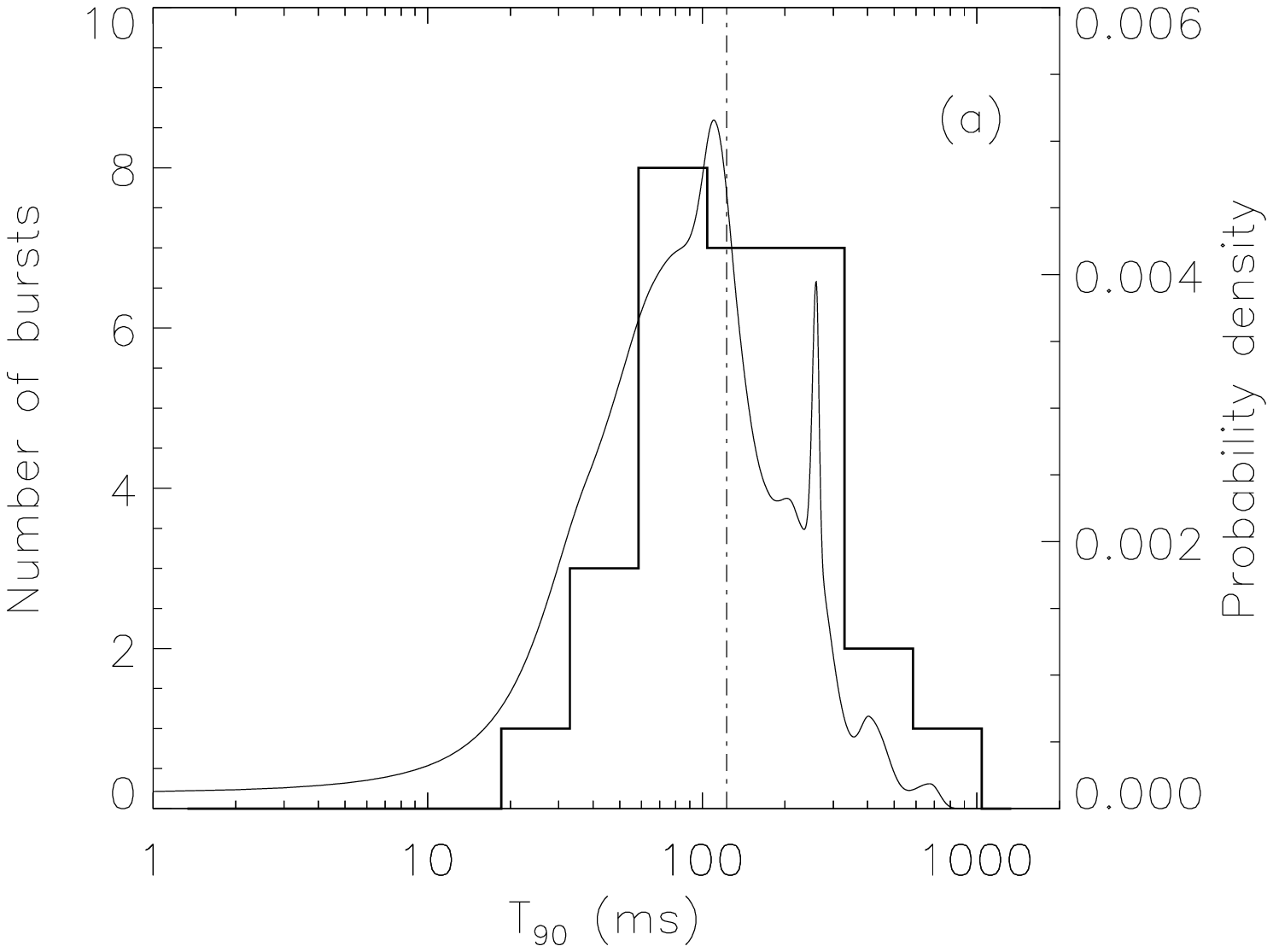}
    \includegraphics*[bb=60 10 500 340, scale=0.5]{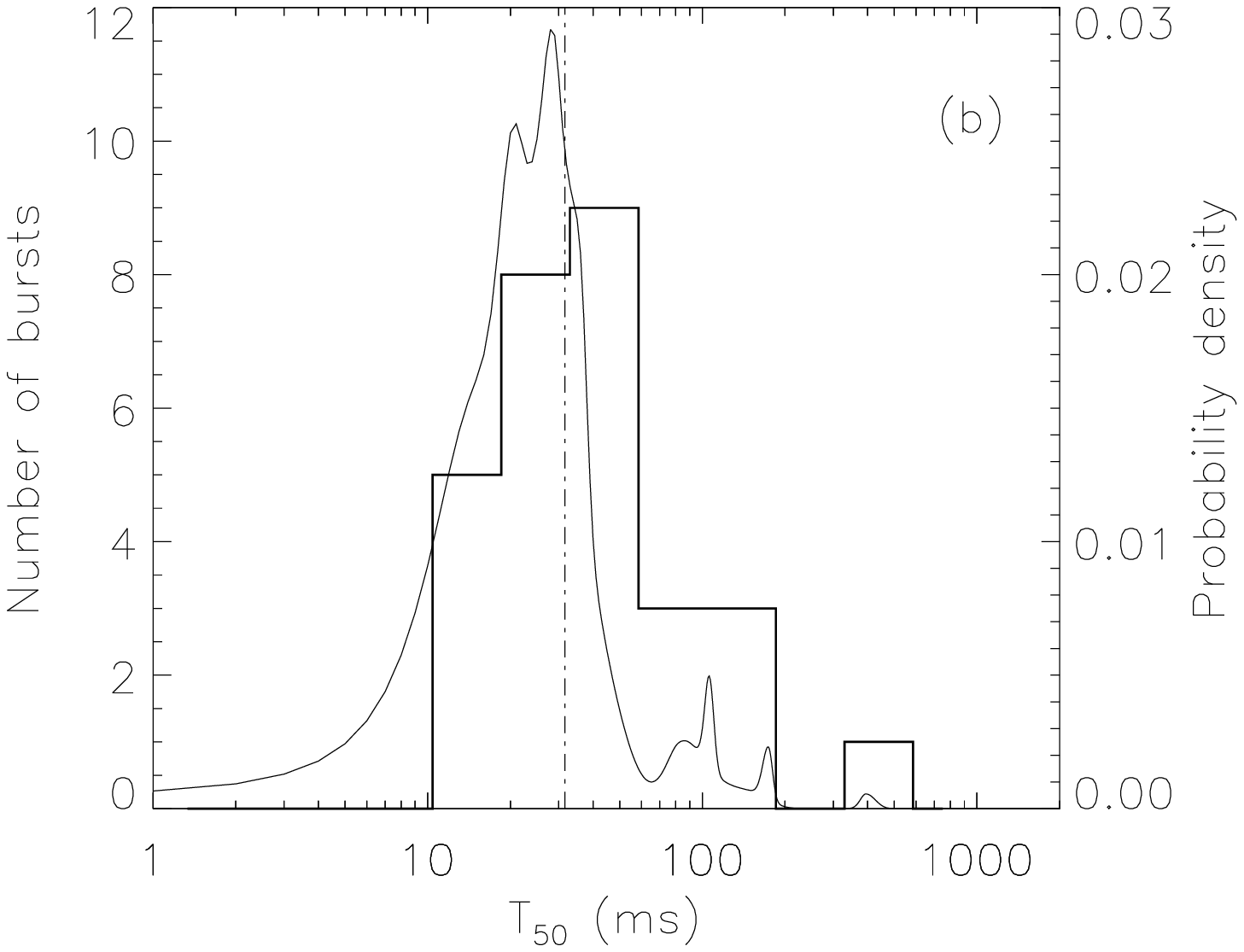} \\
    \includegraphics*[bb=60 10 500 340, scale=0.5]{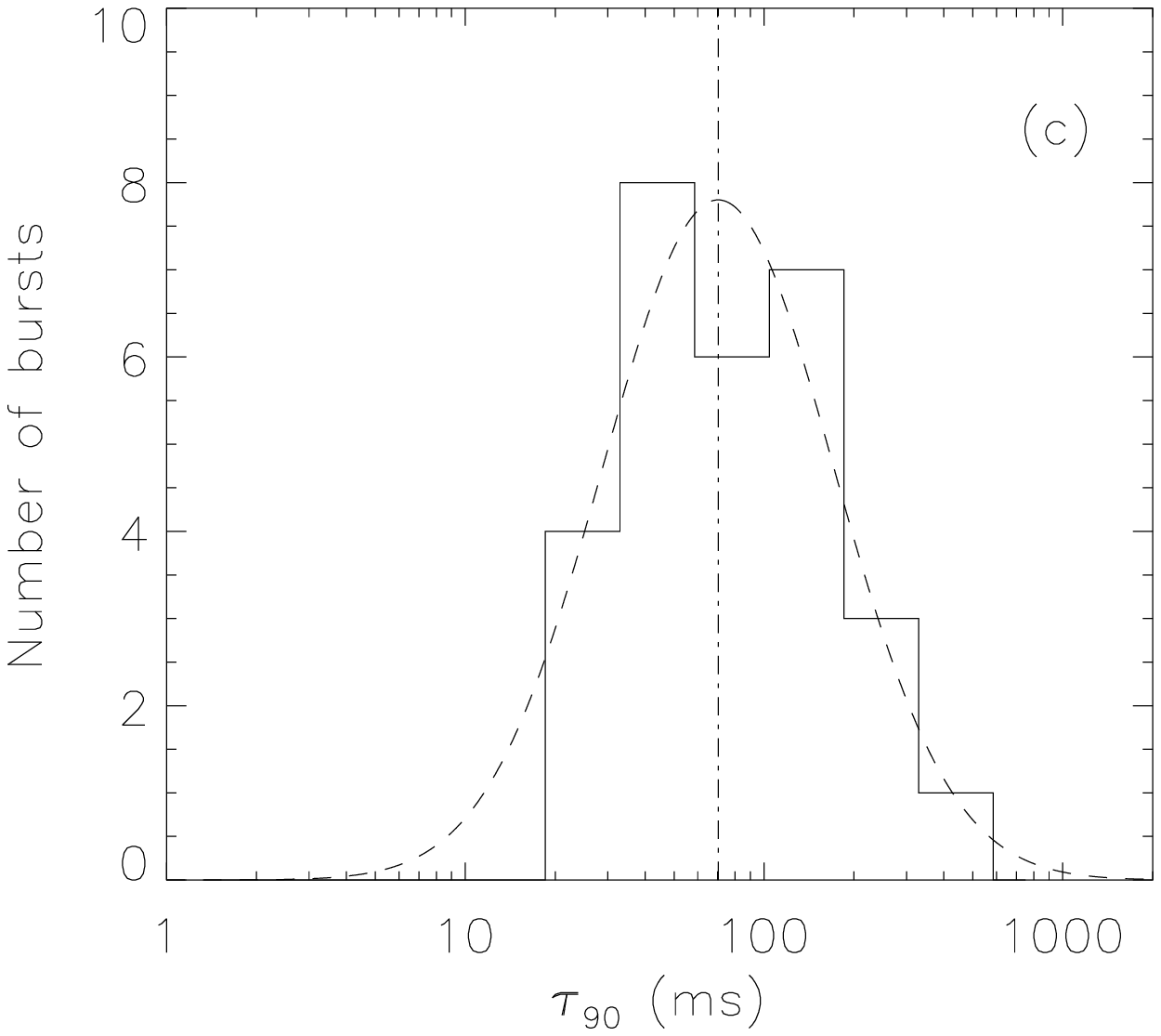}
    \includegraphics*[bb=60 10 500 340, scale=0.5]{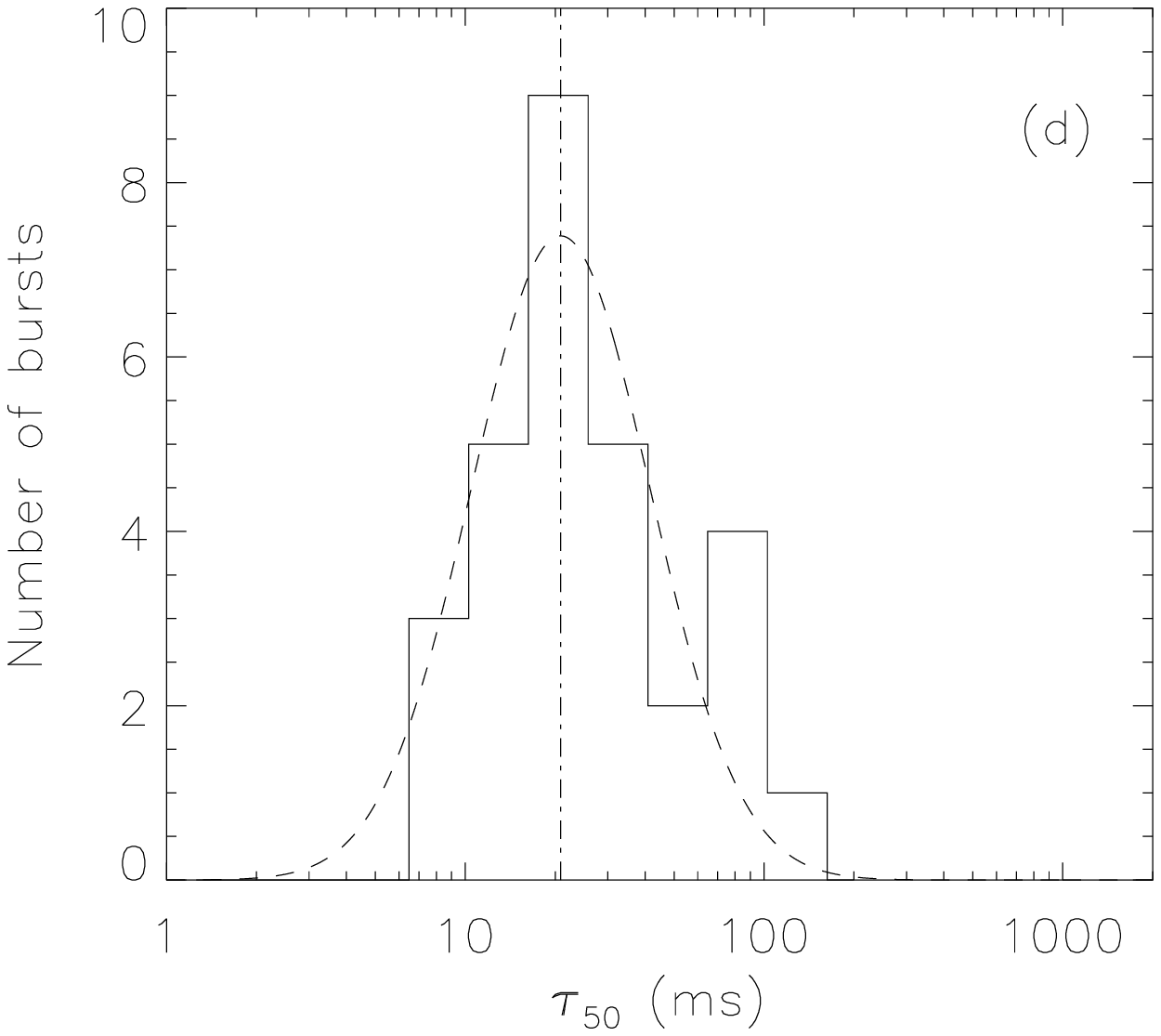} \\
    \includegraphics*[bb=60 10 500 340, scale=0.5]{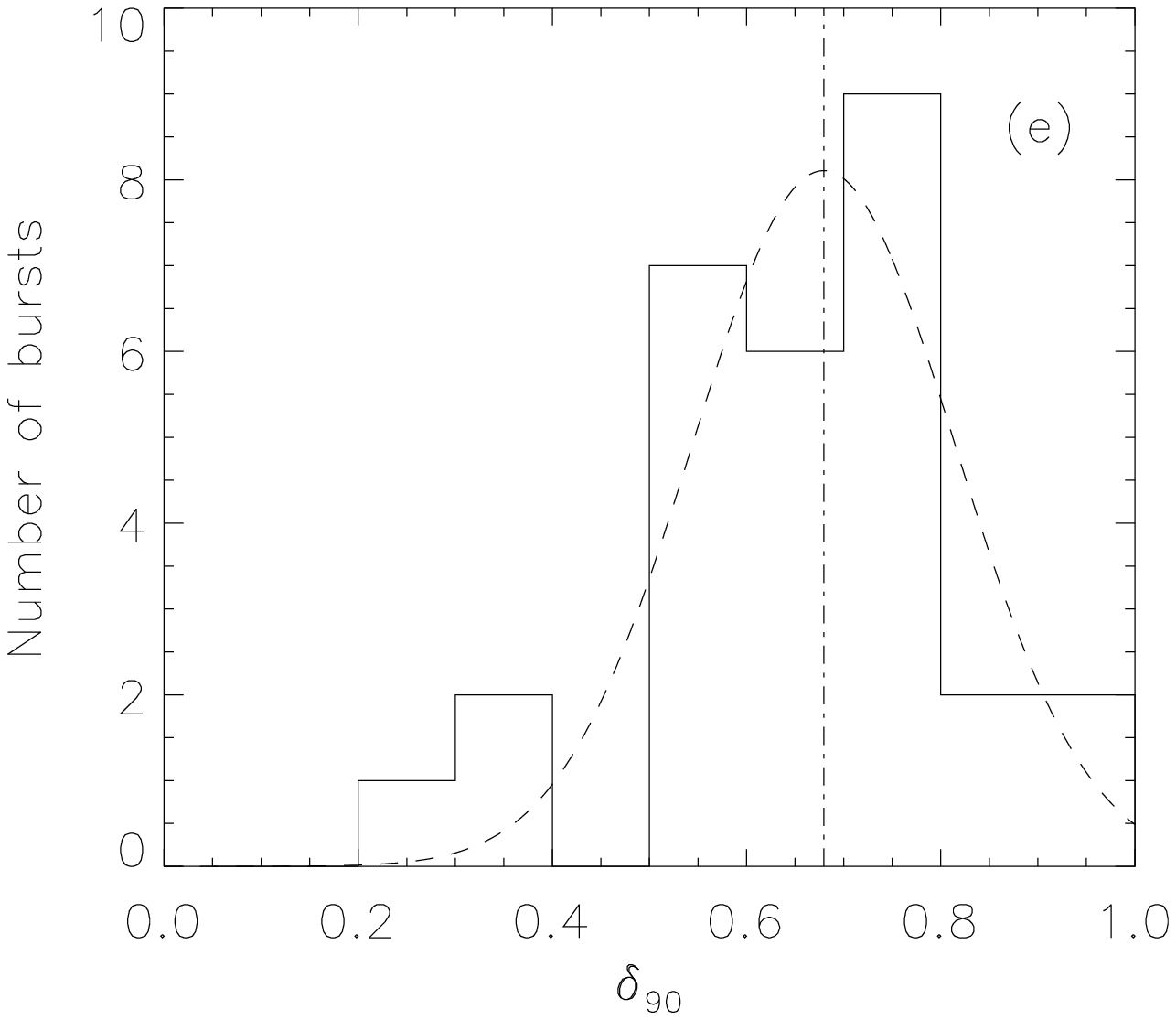}
    \includegraphics*[bb=60 10 500 340, scale=0.5]{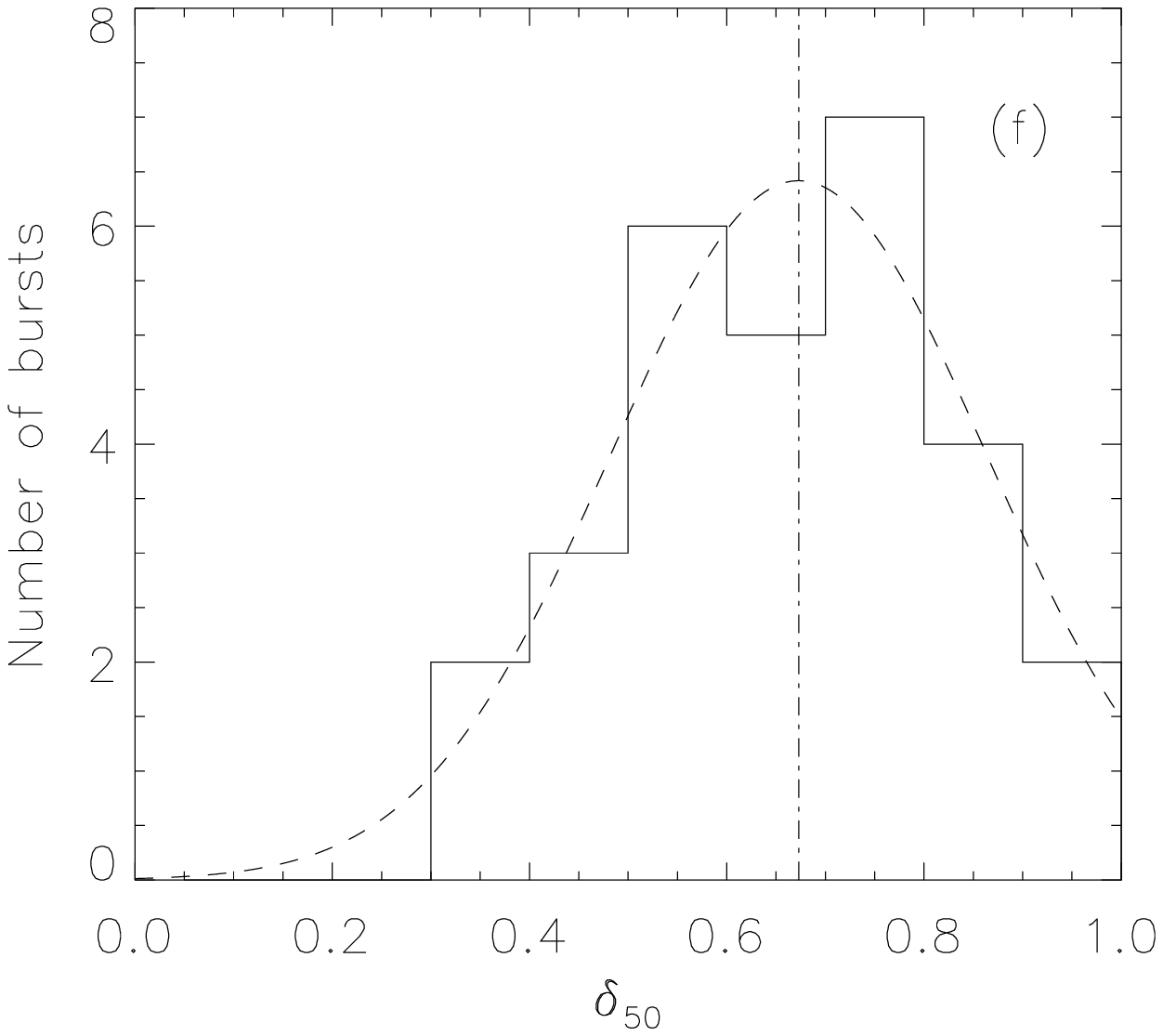}
    \caption{Distributions of $T_{90}$ \emph{(a)}, $T_{50}$
    \emph{(b)}, $\tau_{90}$ \emph{(c)}, $\tau_{50}$ \emph{(d)}, $\delta_{90}$
    \emph{(e)}, $\delta_{50}$ \emph{(f)}. The solid lines in panels (a) and (b) show the probability distribution functions of $T_{90}$ and $T_{50}$, respectively. The dashed curves show the best fits with log-normal (c, d) or normal (e, f) distributions. The vertical dot-dashed lines indicate the mean values of the fits for each histogram distribution.  \label{duration_dis}}
\end{figure*}

Panels a and b in Figure \ref{duration_dis} show the distributions of $T_{90}$ and $T_{50}$ for all 29 bursts; all individual $T_{90}$ values can be
found in Table \ref{obs} (column 4). For comparison reasons with other magnetar duration distributions (see section 3.4) we fit each distribution with log-normal functions and obtained $\langle T_{90} \rangle = 122.6_{-7.5}^{+7.9}$ ms ($\sigma = 0.35 \pm 0.03$, where $\sigma$ is the width of the distribution in the log-frame) and $\langle T_{50} \rangle = 31.6_{-2.3}^{+2.5}$ ms ($\sigma = 0.30 \pm 0.03$). The average values of the raw data weighted by their errors are $\langle T_{90}^{\rm w} \rangle = 138.3_{-20.5}^{+1.07}$ ms, and $\langle T_{50}^{\rm w} \rangle = 32.4_{-0.8}^{+0.9}$ ms.

The distributions above did not account for the uncertainties in the durations. We estimate the (max, min) errors ($\Delta{T_{\rm 90MAX}}$,
$\Delta{T_{\rm 90MIN}}$) for each $T_{90}$ (similarly $T_{50}$) as:
\begin{eqnarray}
T_{90\rm{MAX}} = t_{0.95(N+\sqrt{N})} - t_{0.05(N-\sqrt{N})} \nonumber \\
T_{90\rm{MIN}} = t_{0.95(N-\sqrt{N})} - t_{0.05(N+\sqrt{N})} \nonumber \\
\Delta{T_{90\rm{MAX}}}={T_{90\rm{MAX}}}-{T_{90}} \\
\Delta{T_{90\rm{MIN}}}={T_{90}}-{T_{90\rm{MIN}}} \nonumber
\end{eqnarray}
where $t_{n}$ is the time when the cumulative light curve reaches \textit{n} counts. We then created probability distribution functions (PDFs)
for each $T_{90}$ and $T_{50}$, as described by \citet{starling2008} and \citet{evans2009}. Each PDF was constructed using a two-sided normal
distribution, where the width of each half ($\sigma$) was set to the uncertainty of the duration ($\Delta{T_{\rm
90MAX}}$, $\Delta{T_{\rm 90MIN}}$, $\Delta{T_{\rm 50MAX}}$, and $\Delta{T_{\rm 50MIN}}$):
\begin{eqnarray}
P(x|\bar{x}, \sigma_1, \sigma_2) = \frac{\sqrt{2}}{\sqrt{\pi}(\sigma_1 + \sigma_2)} \{
\begin{array}{cc}
e^A & (x \leq \bar{x}) \\
e^B & (x > \bar{x})\\
\end{array}\cr
{\rm with} \{
\begin{array}{cc}
A={-(x - \bar{x})^2 / 2\sigma_{1}^{2}} \\
B={-(x - \bar{x})^2 / 2\sigma_{2}^{2}}
\end{array}
\label{pdf}
\end{eqnarray}
Each PDF describes the likelihood to obtain \textit{x} (i.e., $T_{90}$ or $T_{50}$) given its measured value $\bar{x}$. Finally, we averaged all sample PDFs to create the total PDF of $T_{90}$ ($T_{50}$) shown in Figure \ref{duration_dis} panels (a) and (b). Note that the shorter (fainter) events have larger errors, resulting in ``pulling'' the PDF towards shorter durations.

\subsection{$T_{90}^{\rm{ph}}$ and $T_{50}^{\rm{ph}}$ in photon space}

The photon-based durations, $T_{90}^{\rm{ph}}$, are estimated with an algorithm similar to the one used above over each burst cumulative
fluence in erg cm$^{-2}$. We used the same time resolution (2\,ms) and energy range ($8-100$\,keV) as in the count durations. The essential
difference here, is that these measurements utilize the intrinsic (deconvolved) burst spectra instead of the detector recorded counts to define
the burst intrinsic durations independent of different instruments. To perform these estimates, we used the GBM public software tool RMFIT
v3.3\footnote{http://fermi.gsfc.nasa.gov/ssc/data/analysis/user/} \citep[for a description of this tool see also][]{kaneko2006} and the new
datatype CTTE specially created to facilitate analyses of short events. This datatype simply bins the 128 TTE energy channels into the same 8
bins as the CTIME data. The errors in the duration estimates are taken from \citet{koshut1996a} and \citet{koshut1996b}. 

A detailed description of the photon-based durations can be found in the First Two Years GRB Catalog of the {\it Fermi}/GBM \citep{paciesas2011}. In short, an adequate background interval is selected before and after each burst and fit with the lowest acceptable order of a polynomial to determine the background model parameters. Next the entire burst interval is fit to determine the default set of photon model parameters. The model used in these fits is a power law with an exponential cutoff (COMPT; described in detail in Section 4). When all background and source model selections are determined for each 2\,ms time bin, we subtract the background, fit its spectrum using the COMPT model, and calculate its photon flux. These values are then used as inputs for the $T_{90}^{\rm{ph}}$ ($T_{50}^{\rm{ph}}$) estimates, performed with the same algorithm described above.

\begin{figure}[h]
    \includegraphics*[bb=30 10 500 350, scale=0.5]{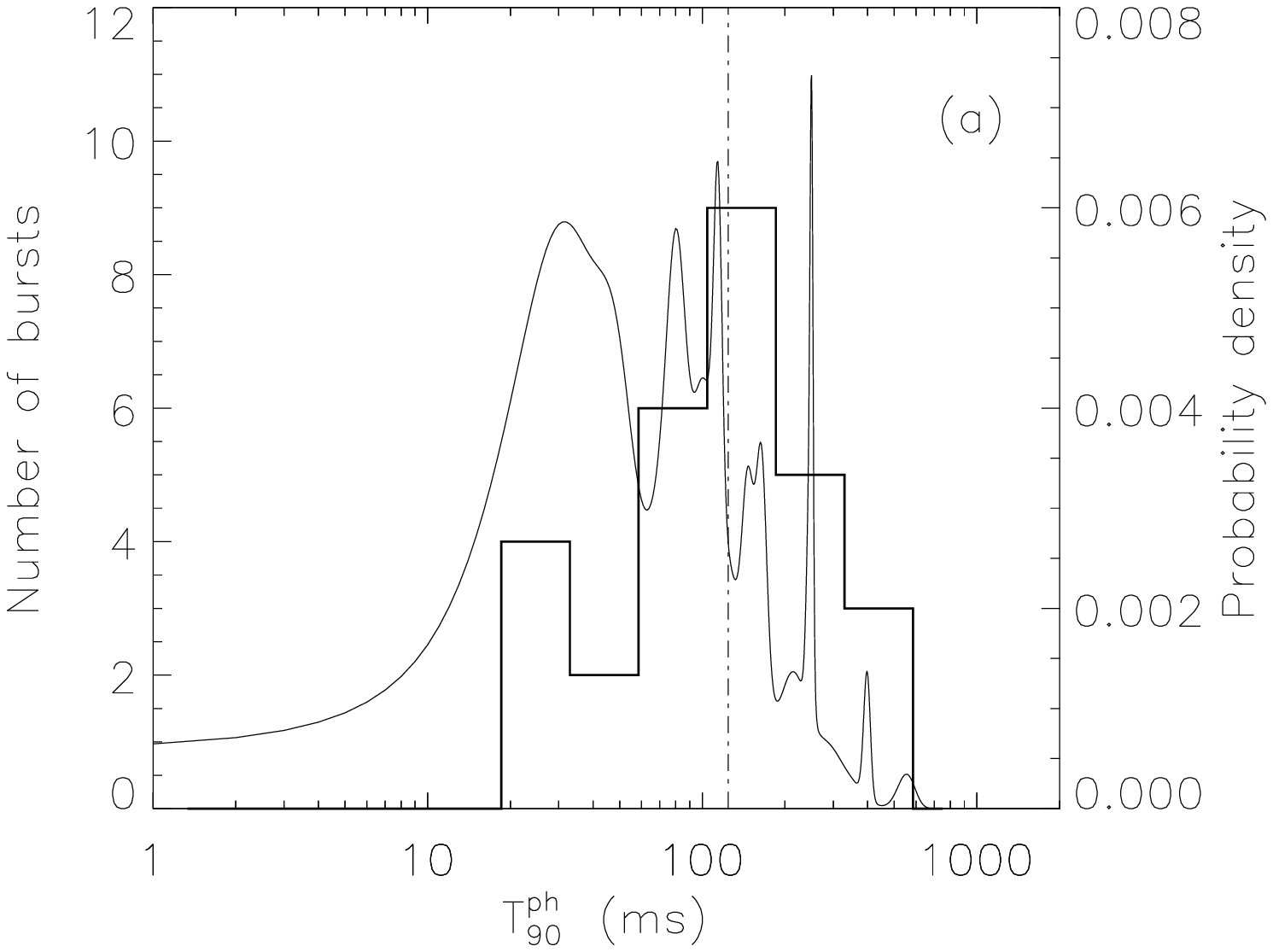}
    \includegraphics*[bb=30 10 500 350, scale=0.5]{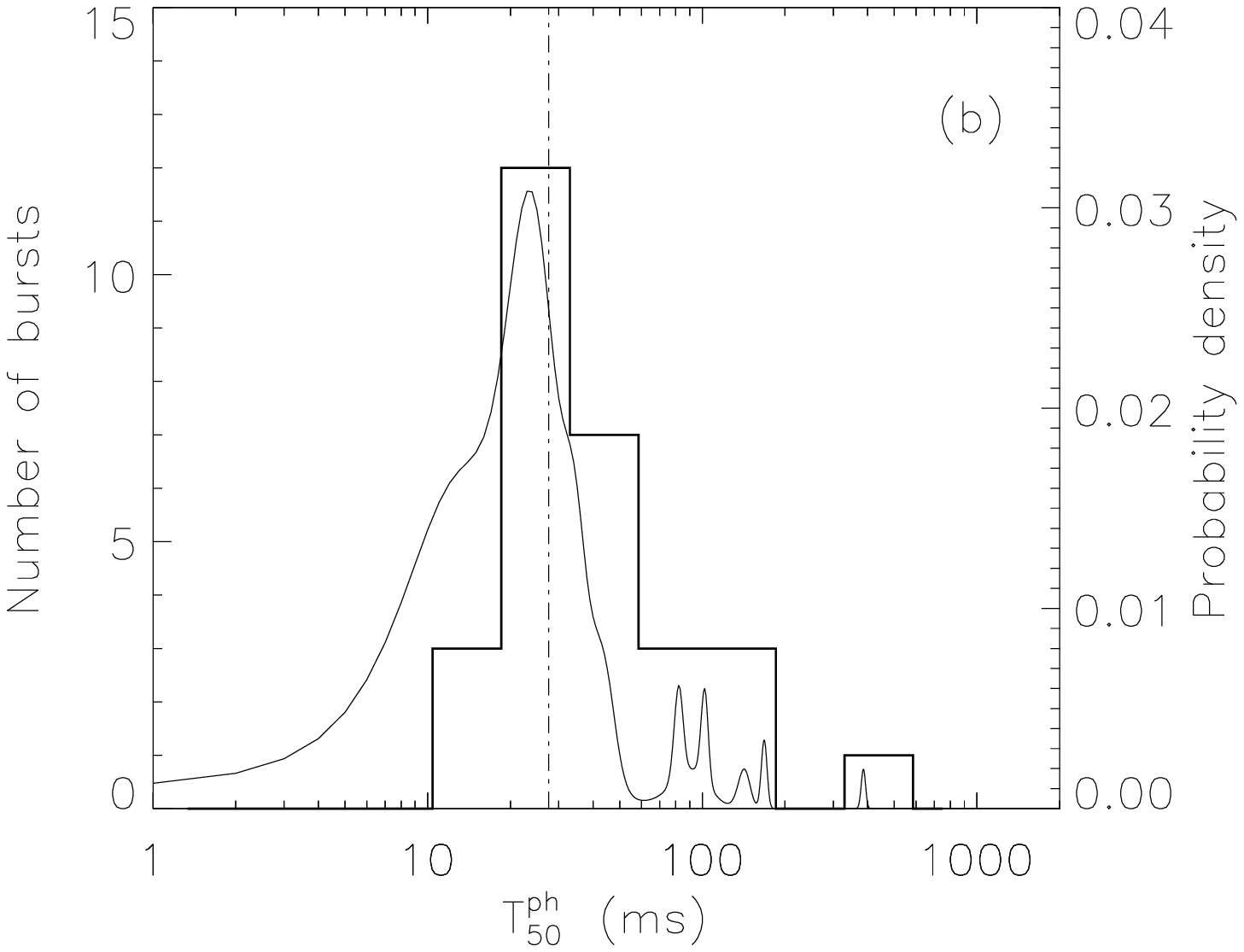}
\caption{Distributions and probability distribution functions of $T_{90}^{\rm{ph}}$ (a) and $T_{50}^{\rm{ph}}$ (b). The histograms show the raw
data and the solid curves show the PDFs. The vertical dot-dashed lines indicate the mean value of the log-normal fits of the histograms.
\label{duration_photon}}
\end{figure}

Figure \ref{duration_photon} shows the distributions of $T_{90}^{\rm{ph}}$ ($T_{50}^{\rm{ph}}$) fit with a log-normal function (panels a and b),
obtaining $\langle T_{90}^{\rm{ph}} \rangle = 124.2_{-15.2}^{+17.3}$ ms ($\sigma = 0.38 \pm 0.06$, where $\sigma$ is the width of the
distribution in the log-frame) and $\langle T_{50}^{\rm{ph}} \rangle = 27.6_{-1.7}^{+1.8}$ ms ($\sigma = 0.21 \pm 0.03$). The average values of
the raw data weighted by their errors are $\langle T_{90}^{\rm{ph_w}} \rangle = 161.2_{-1.6}^{+1.6}$ ms, and $\langle T_{50}^{\rm{ph_w}}
\rangle = 49.2_{-0.8}^{+0.8}$ ms. The individual $T_{90}^{\rm{ph}}$ values can be found in Table \ref{obs} (column 5).

\begin{figure}[h]
    \includegraphics*[bb=50 200 570 620, scale=0.45]{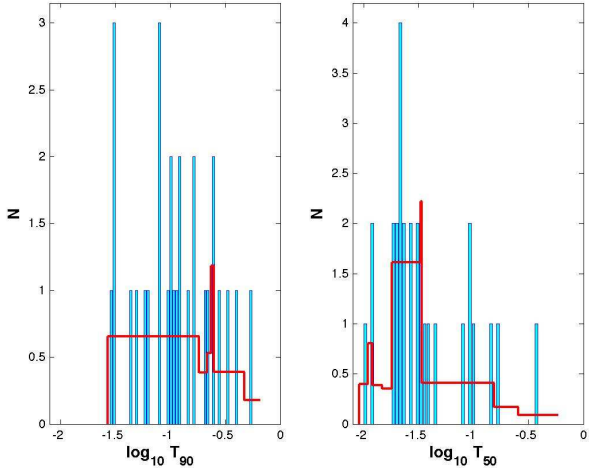}
\caption{Bayesian block representation of the $T_{90}^{\rm{ph}}$ ({\it left}), and $T_{50}^{\rm{ph}}$ ({\it right}), taking into account
measurement errors. The histograms show the raw data and the solid curves show the Bayesian blocks.  \label{duration_BB}}
\end{figure}

The solid curves in Figure \ref{duration_photon} exhibit the PDFs for $T_{90}^{\rm{ph}}$ and $T_{50}^{\rm{ph}}$ after taking into account the symmetrized errors in the values. This plot is an adaptive kernel density estimation: it was made by adding up a set of normalized Gaussian functions, one at each of the data points and with a 1$\sigma$ width given by the corresponding error estimate. The multiple narrow spikes in this figure, mostly at the long-duration end of the plot, are largely due to the fact that the corresponding errors (in contrast to those of shorter and fainter durations) are relatively small, yielding tall and narrow Gaussian components.  This is a rather high-variance estimate of the distribution function.  As an alternative we show in Figure \ref{duration_BB} a completely different density estimate based on an adaptation of the Bayesian block algorithm for histogramming of the $T_{90}^{\rm{ph}}$ and $T_{50}^{\rm{ph}}$ data, taking into account measurement errors in the independent variable \citep{scargle1998,scargle2011}. This algorithm finds the optimal piece-wise constant model to represent the data; the optimization corresponds to the maximum likelihood for a constant-rate Poisson model for the data in each of the bins.  The sizes and locations of the bins are all determined by this optimization, not pre-defined as in ordinary histograms.  The solid curve is the Bayesian block representation superimposed on such an ordinary histogram with bins chosen so that the value plotted is more than one only if there are duplicate values. This analysis suggests that the spiky structure of the PDFs in Figure \ref{duration_photon} is mostly due to noise fluctuations (the spikes corresponding to a small number of points with small formal measurement errors) and does not support a quantized or a multimodal duration distribution for the bursts from \sgrnos.

We compare the two duration distributions (in count and photon space) in section \ref{comp_durations}.

\subsection{$\tau_{90}$ ($\tau_{50}$) and $\delta_{90}$ ($\delta_{50}$) in count space}

The emission time $\tau_{90}$ ($\tau_{50}$) for each burst was determined by adding the time bins (2\,ms each) of high fluence in decreasing fluence rank until 90\% (50\%) of the fluence was reached. The emission time interval, thus, characterizes the duration of high-fluence emission. This parameter was also first introduced for GRBs as complementary to their $T_{90}$ ($T_{50}$) duration measures \citep{mitrofanov1999}. Panels c and d in Figure \ref{duration_dis} show the distributions of $\tau_{90}$ and $\tau_{50}$. These are also fit with log-normal functions obtaining
$\langle \tau_{90} \rangle = 70.3_{-6.5}^{+7.2}$ ms ($\sigma = 0.39 \pm 0.04$) and $\langle \tau_{50} \rangle = 20.9_{-2.3}^{+2.5}$ ms
($\sigma = 0.30 \pm 0.05$).

The ratio $\delta_{90} = \tau_{90}/T_{90}$ ($\delta_{50}=\tau_{50}/T_{50}$) is defined as the duty cycle of each burst by
\citet{mitrofanov1999}. This value should not be over 1, because the emission time excludes low fluence intervals in the burst, which sometimes
are included in the duration. Panels e and f in Figure \ref{duration_dis} present the distributions of $\delta_{90}$ and $\delta_{50}$, which
were fit with normal distributions with $\langle \delta_{90} \rangle = 0.68 \pm 0.03$ ($\sigma = 0.14 \pm 0.03$) and $\langle \delta_{50}
\rangle = 0.68 \pm 0.02$ ($\sigma = 0.19 \pm 0.02$). Both duty cycle distributions have the same mean value of 0.68.

\subsection{Comparisons of durations}
\label{comp_durations}

\begin{figure}[h]
    \includegraphics*[bb=25 15 500 325, scale=0.4]{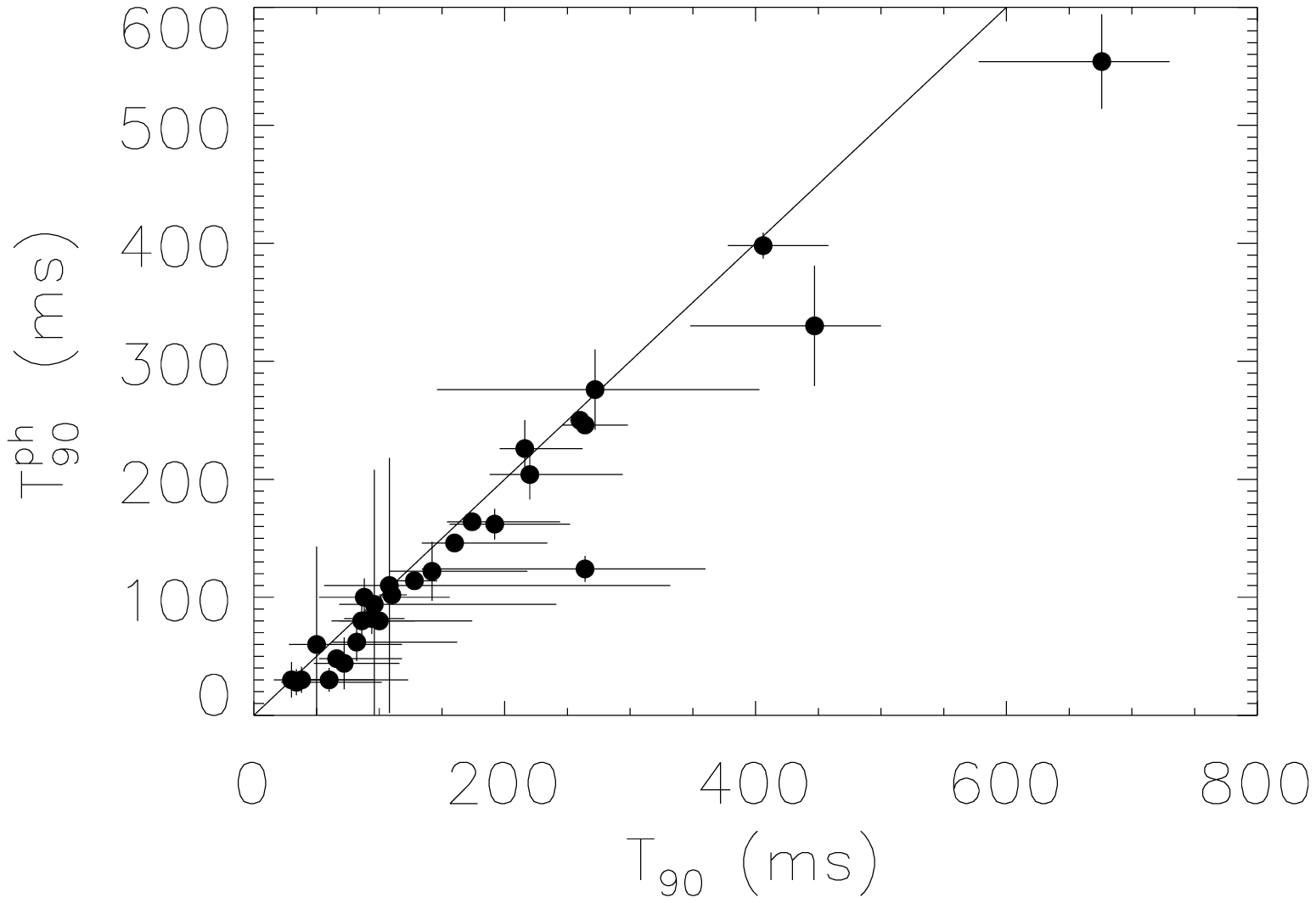}
    \includegraphics*[bb=25 15 500 325, scale=0.4]{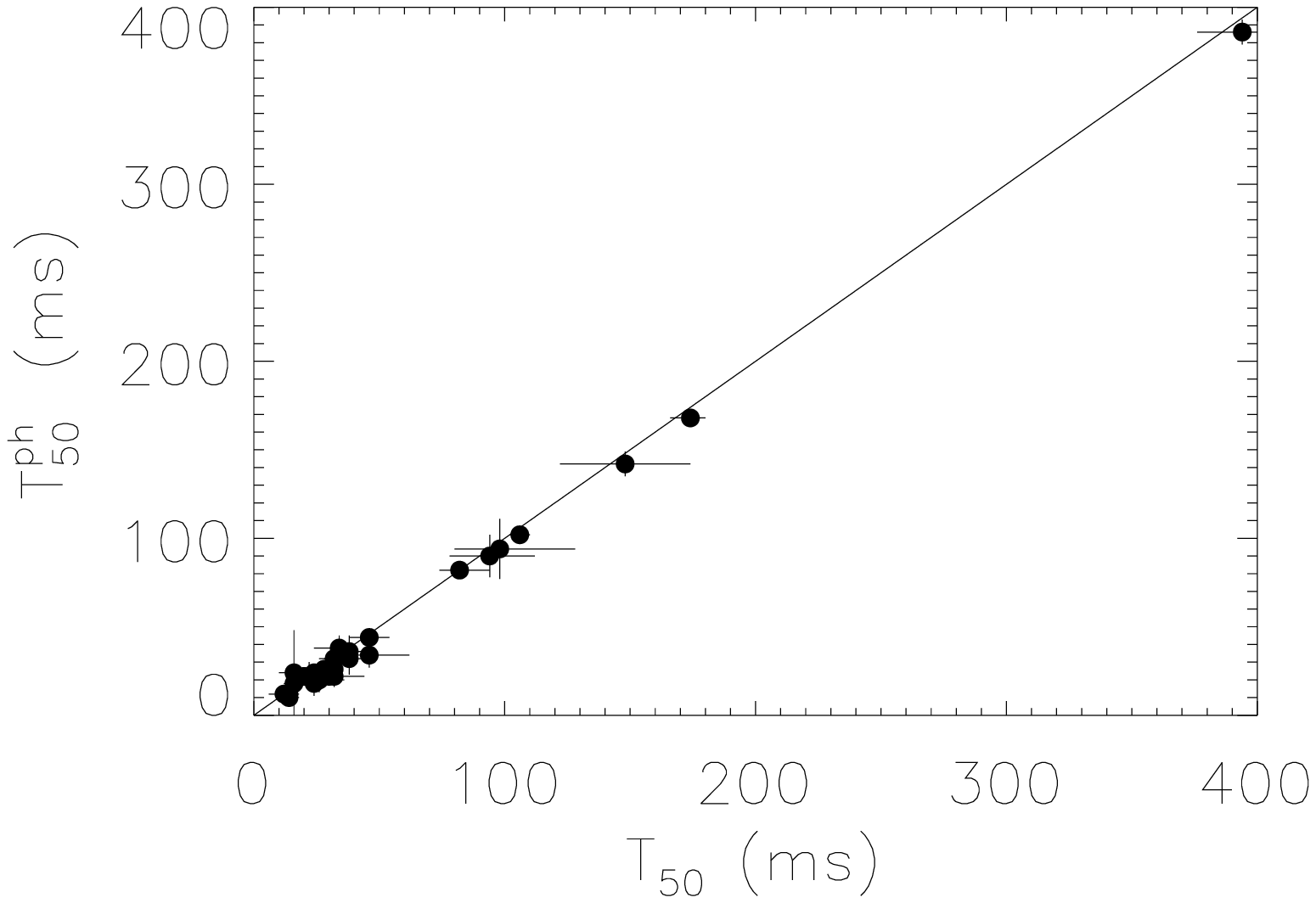}
\caption{{\it Top panel}: The comparison of the photon space $T_{90}^{\rm{ph}}$ and the count space $T_{90}$. {\it Bottom panel}: the similar plot for $T_{50}$. The solid lines are $x=y$. \label{cr_ph}}
\end{figure}

The durations in photon space are estimated using the deconvolved source spectra with the response of the GBM detectors. In principle, since the
spectral energies where SGR bursts emit most of their photons are relatively narrow ($\sim 8-100$\,keV) there should not be a large difference
between photon and count durations. Figure \ref{cr_ph} shows $T_{90}^{\rm{ph}} (T_{50}^{\rm{ph}})$ {\it versus} $T_{90} (T_{50})$. We notice
that in general the count space $T_{90}$ values tend to be larger by a very small amount, mainly due to the fact that they take
into account number of counts irrespective of their energy content. Since SGRs have mostly soft spectra, the counts corresponding to the lowest
energy photons often do not contribute much flux in the durations (i.e., less than 5\%). The $T_{50}$ estimates, however, are perfectly aligned
along the $x=y$ line, making these spectrally-independent measures for durations (as also noted for GRBs by Bissaldi et al. 2011). It is also important to note here that these photon durations validate earlier duration measurements, which have been done in count space for all magnetar candidates.

\begin{figure}[h]
\includegraphics*[bb=20 15 490 340, scale=0.5]{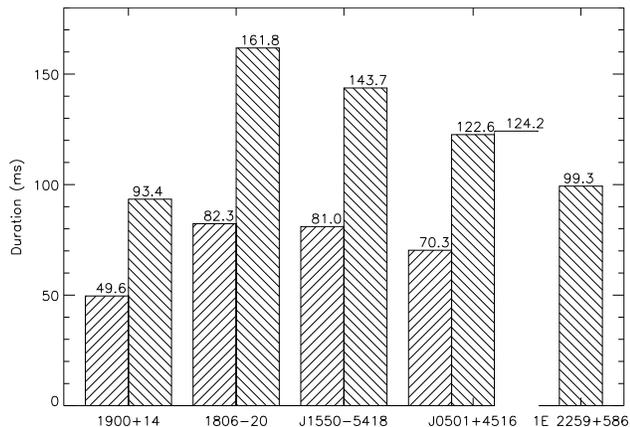}
\caption{Mean values of count space $T_{90}$ (right hatched bars) of four SGRs and one AXP and of $\tau_{90}$ (left hatched bars) of four SGRs. The photon space $T_{90}$ of \sgr is indicated with a blank bar. Values are marked in ms above each bar. \label{duration_com}}
\end{figure}

We now compare in Figure \ref{duration_com} the mean values of the count space $T_{90}$ and $\tau_{90}$ of \sgrnos, with those of four other
magnetar candidates. SGR J$1550-4518$ was also observed with GBM and the temporal parameters are estimated using exactly the same procedures as
here \citep{vdh2011}. SGRs $1806-20$ and $1900+14$ were estimated using {\it RXTE}/PCA observations in $2-60$\,keV by
\citet{gogus2001}. AXP 1E$2259+586$ durations are from \citet{gavriil2004} and are also estimated using {\it RXTE}/PCA observations in
$2-60$\,keV. It is obvious from the figure that all durations fall well within the same order of magnitude ($\sim 100-150$ ms), indicating a
similar origin for the bursts across the magnetar population.

\subsection{Relative timing between burst peaks and persistent emission pulse phase}

Using the pulse ephemeris reported in \citet{gogus2010}, we have aligned each burst peak to the pulse phase of the spin period of \sgrnos, to
search for possible correlations of the burst activity with rotational phase. In an effort to confirm the GBM barycentric correction for this
analysis, we compared the barycenter corrected light curves of burst bn080823.020 (Table 1) for the GBM and {\it RXTE}/PCA data. Note that the
pulse ephemeris is based primarily upon {\it RXTE}/PCA data. A cross-correlation of the GBM and {\it RXTE}/PCA burst time histories indicates
no significant shift with an upper limit of 4\,ms. Given the 5.76\,s pulse period of this SGR, even a 4 ms shift is negligible. As a final
check, the barycentering software was also tested by epoch folding TTE data to obtain a pulse profile of the Crab pulsar using the Jodrell Bank
ephemeris (http://www.jb.man.ac.uk/~pulsar/crab.html). We found that the phase of the first peak agreed within 200 $\mu$s of the {\it RXTE
result} shown by Rots et al. (2004).

\begin{figure}[h]
\includegraphics*[bb=20 15 490 340, scale=0.5]{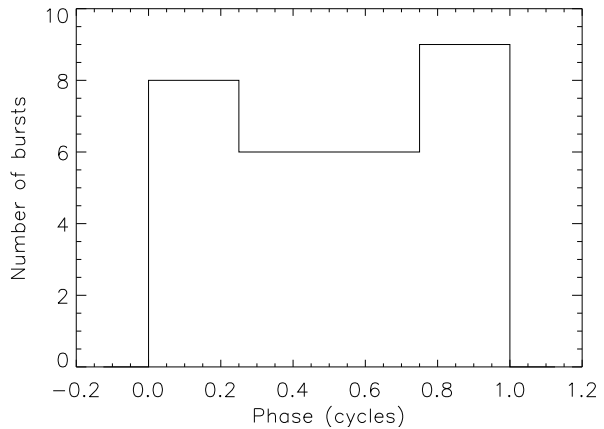}
\caption{The distribution of the phase offsets of all burst peaks relative to the pulsar phase. \label{phase_offset}}
\end{figure}

Burst peak phases were computed using the event times recorded in the TTE data in the energy range 8$-$60 keV. For each event time history, the
burst peak time was defined as the average event time for the six most closely spaced counts in a 4 sec interval surrounding the trigger time.
The phase of pulse maximum was defined by fitting an inverted parabola to the folded {\it RXTE}/PCA 2$-$10 keV light curve. Figure
\ref{phase_offset} shows the distribution of the phase offsets of all burst peaks relative to the pulse maximum. The average offset (average of
the absolute value) is 0.285 cycles.  For a random phase distribution (i.e., null hypothesis), one would expect an average offset of 0.25
cycles. A Monte Carlo simulation of 10$^5$ realizations for 29 draws from a random distribution shows that the probability of getting an offset
value of 0.285 for 29 samples is $\sim26$\% or roughly $1\sigma$. We conclude that there is no correlation of the \sgr burst peaks with pulse
phase.

\section{Spectral Analysis \label{spec}}

We performed all spectral analysis using RMFIT v3.3 and we generated response files for each burst using the GBM response generator {\it gbmrsp
v1.9}; Table \ref{obs} lists the detectors used for each event. Each event spectrum was restricted within $8-200$\,keV, as we did not detect any counts above 200\,keV. However, roughly one third of the bursts have significant emission in the $150-200$\,keV band: the average count rate in the brightest detector over the $T_{90}$ interval was larger than 1\,count/s. Most of the bursts were faint and they only had good statistics for time-integrated spectral analysis described below. Five events (including the two saturated bursts mentioned in Section \ref{data}) were very bright and we were able to perform time-resolved analysis as described in Section \ref{trs}. Finally, to account for the Iodine K-edge effects at 33.2\,keV, we excluded from our spectral fits the region between $30-40$\,keV.

\subsection{Time-integrated spectra \label{tis}}

We fitted the time-integrated spectra of all 29 bursts with several models: a single power law (SPL), an Optically Thin Thermal Bremsstrahlung
(OTTB), a single black body (BB), a power law with an exponential cut-off (COMPT)
\footnote{The analytic expression for this model is:
\begin{displaymath}
f = A \exp[-E(2+\lambda)/E_{\rm{peak}}] (E/E_{\rm{piv}})^\lambda,
\end{displaymath}
where $f$ is the photon number flux in photons s$^{-1}$ cm$^{-2}$ keV$^{-1}$, $A$ is the amplitude in the same units as $f$, $E_{\rm{peak}}$ is
the peak energy in keV, $\lambda$ is the photon index, and $E_{\rm{piv}}=20$keV is the pivot energy.}
, a
two black body spectrum (BB+BB), a single black body with a power law (BB+PL), a single black body with an OTTB (BB+OTTB), and finally, a
single black body with an exponential cut-off power law (BB+COMPT). The parameters of the last two models (BB+OTTB, BB+COMPT) could not be
constrained by most burst data, while a SPL was always a bad fit; all three models were, therefore, rejected from further spectral analysis. To
determine the goodness of fit for the remaining models we used the Castor modified Cash-statistic (C-stat). This is a modified maximum likelihood
estimator which asymptotes to $\chi^2$, used when there are small numbers of counts/bin (Poisson regime), which is the case for most of the SGR
events (especially in the higher energy bins).

\begin{figure}[p7]
    \includegraphics*[bb=20 30 540 730, scale=0.27, angle=90.0]{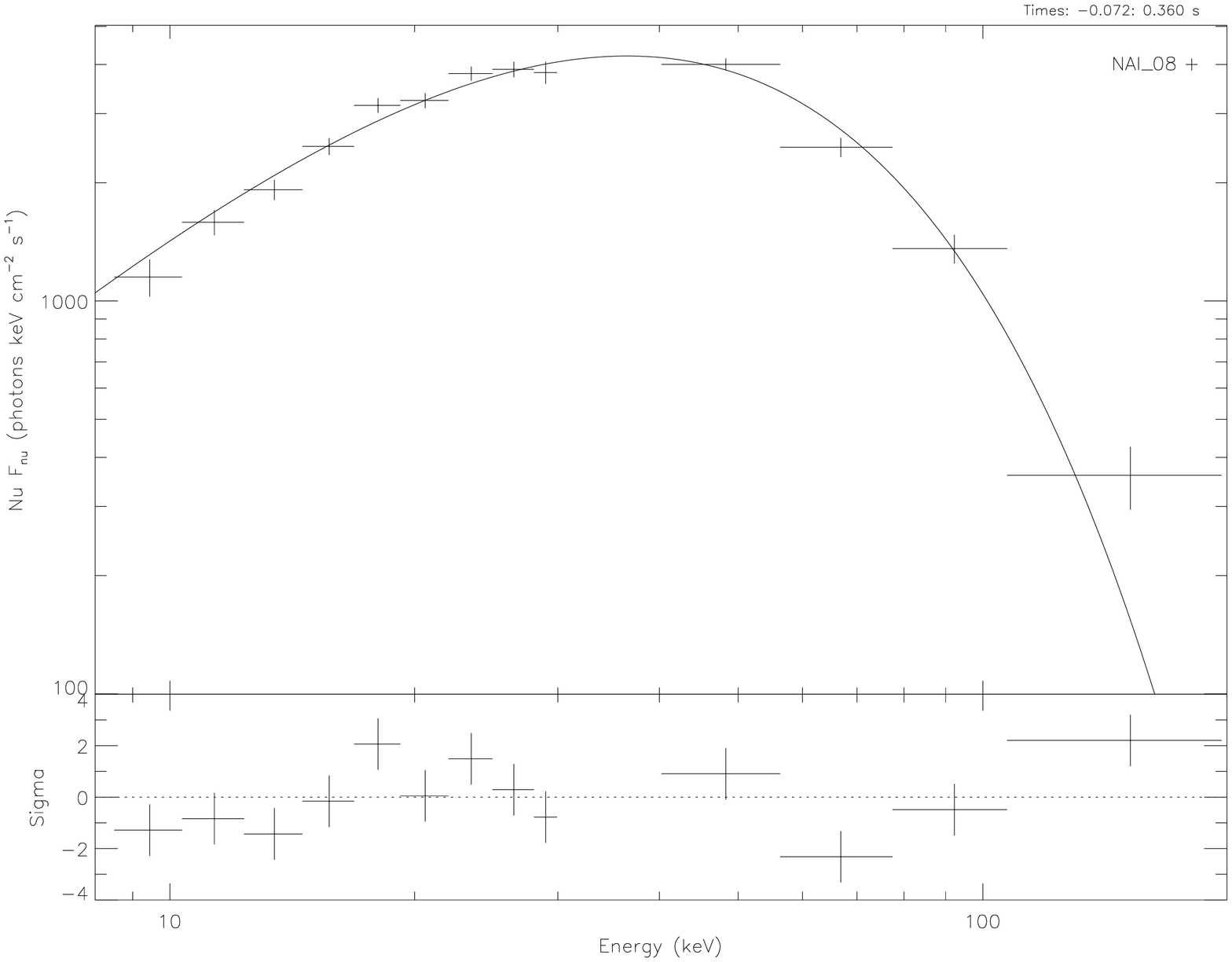}
    \includegraphics*[bb=20 30 540 730, scale=0.27, angle=90.0]{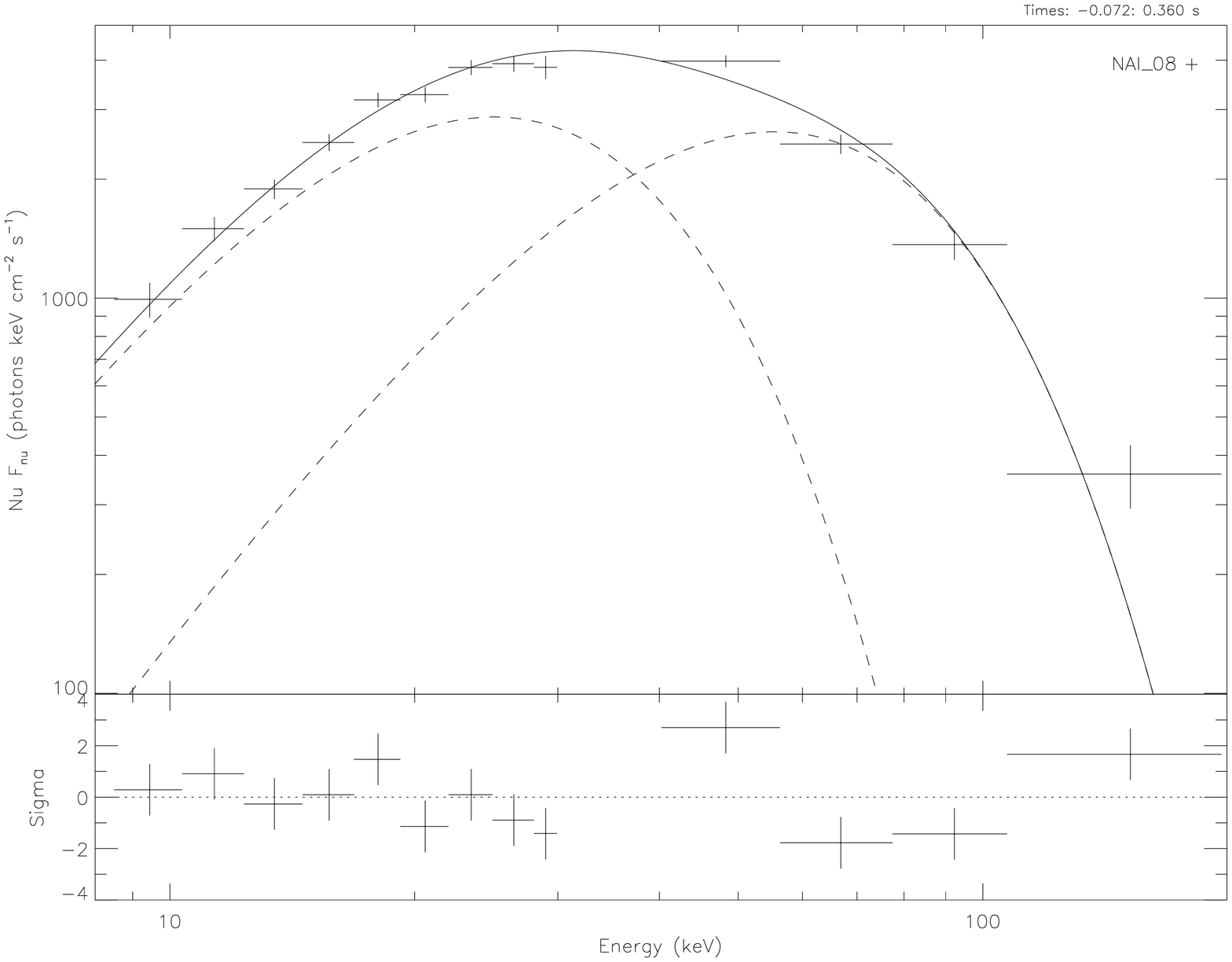}
    \includegraphics*[bb=20 30 540 730, scale=0.27, angle=90.0]{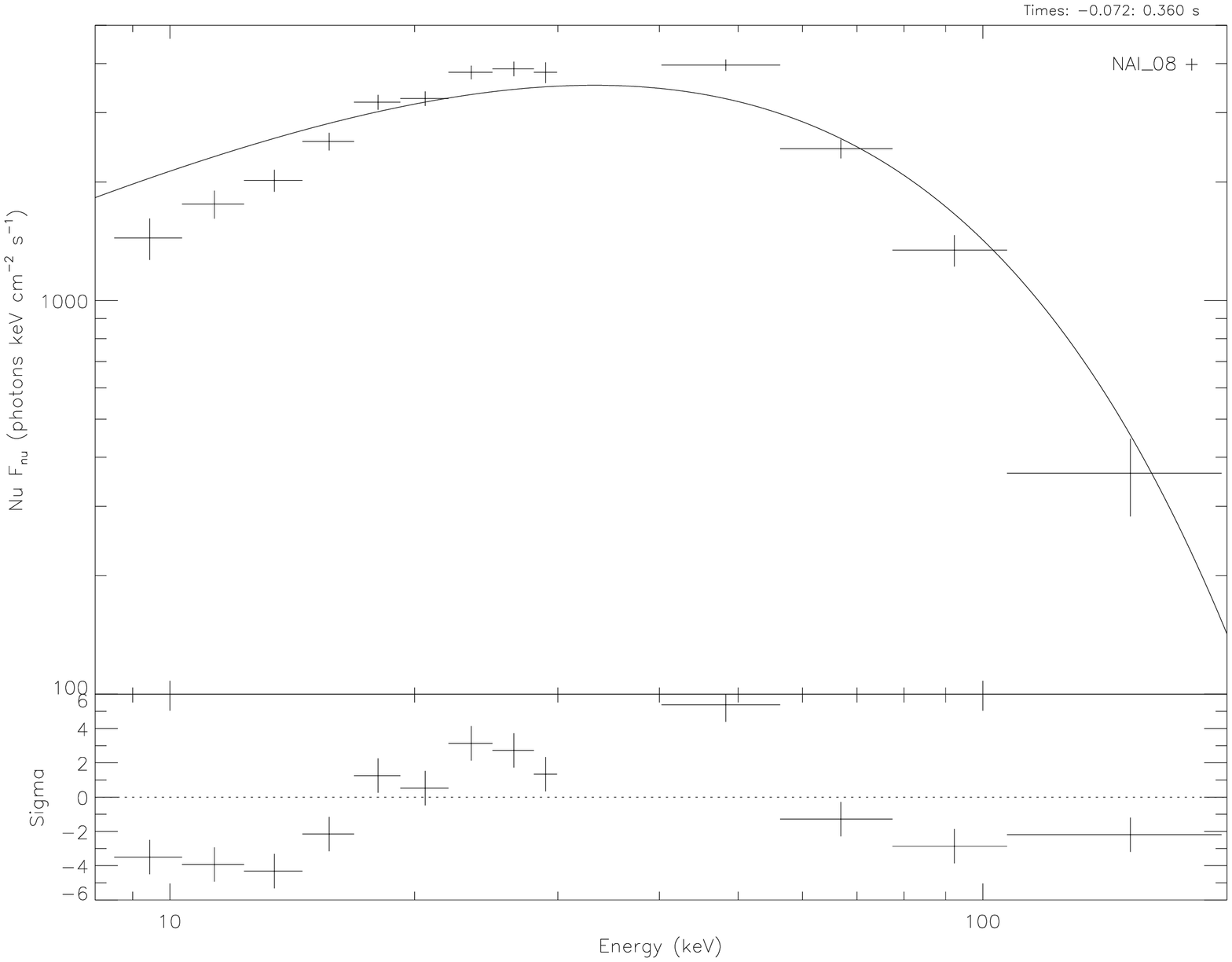}
    \includegraphics*[bb=20 40 540 740, scale=0.27, angle=90.0]{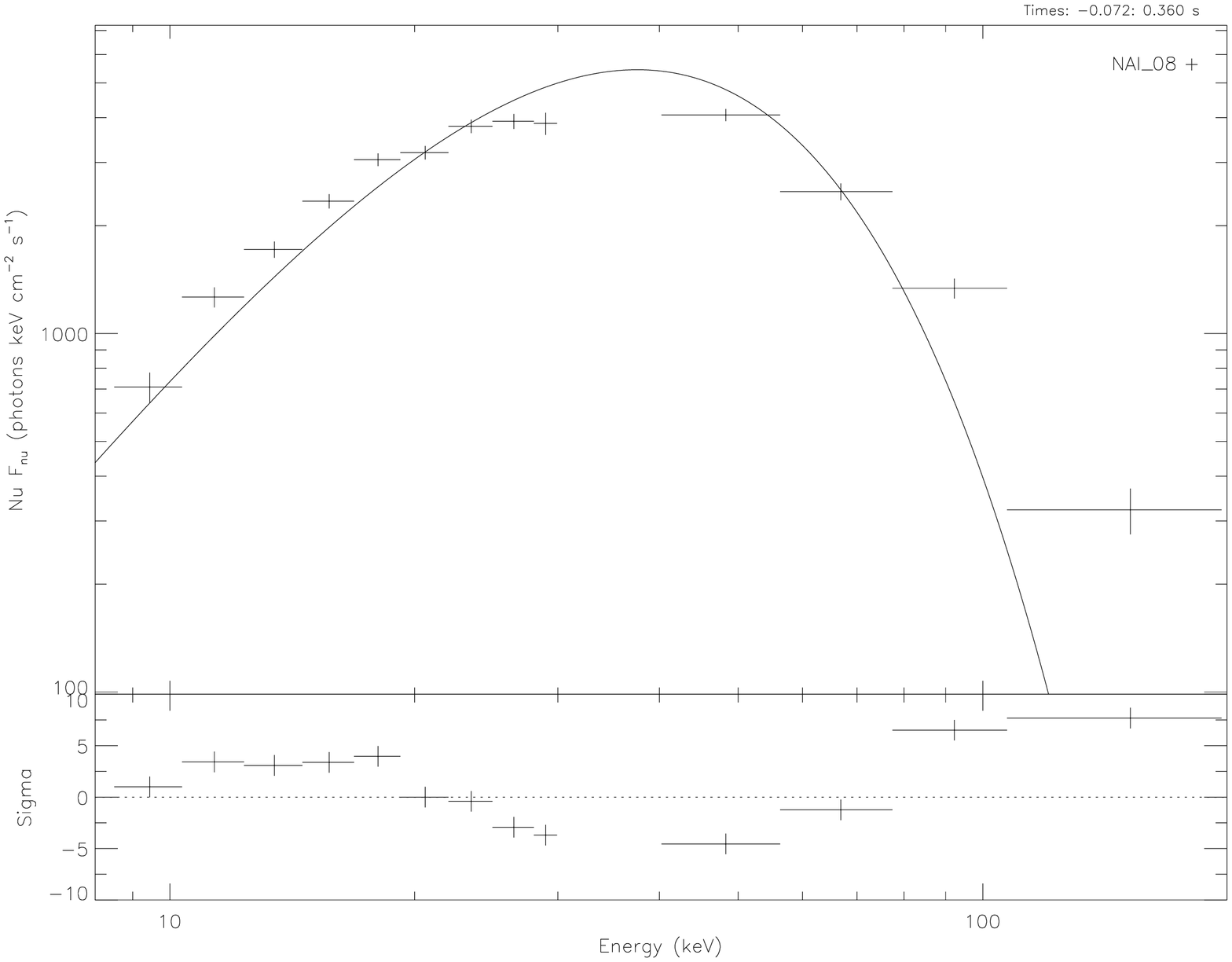}
    \caption{The spectrum of a bright \sgr burst (bn080826.136) fit with COMPT, BB+BB, OTTB, and BB models (from top to bottom). \label{spec_exa}}
\end{figure}

Using C-stat we were able to further reject single BB, OTTB and BB+PL models. The first two models fit only part of the weaker burst set; for
these the COMPT model also did not give a significantly smaller C-stat. It was not excluded, however, because this model, contrary to the first
two, fit all bursts. Moreover, a COMPT index of $1$ or $-1$ reproduces the BB or OTTB spectral shape. The BB+PL model had overall worse C-stat
values compared to the remaining two models (COMPT and BB+BB). The relative goodness of fit among these models is exhibited in Figure
\ref{spec_exa}, which shows spectral fits with COMPT, BB+BB, OTTB, and a single BB of one bright burst from \sgr (bn080826.136). From the
figure, we see that the COMPT and BB+BB models can fit the data equally well; the residuals, however, of the OTTB and the BB model fits are
unacceptably large.

The COMPT model fits all 29 bursts well, with the BB+BB model giving equally good fits in only 18 events, where we have enough statistics to constrain the model parameters. To determine whether the COMPT or the BB+BB model fit the data best, we simulated (using RMFIT) a large set of bursts with different intensity and spectral shape parameters, using COMPT and BB + BB as input models, and then fit them with both the COMPT and BB+BB models (see van der Horst et al. 2011 in preparation, for a detailed description of the simulations). We show that the C-stat improvement is not significant to conclude that BB+BB (with one more parameter) is better than COMPT. A set of simulations using Xspec gave similar results with RMFIT. Below we discuss our COMPT model fits for all 29 bursts. Since the BB+BB fits provide significant information on the source parameters \citep{olive2004,israel2008}, we also describe the results of these fits for 18 bursts. Section \ref{discussion} expands on the importance of each model.

\subsubsection{COMPT model fits}

\begin{figure}[h]
\includegraphics*[bb=60 15 480 325, scale=0.5]{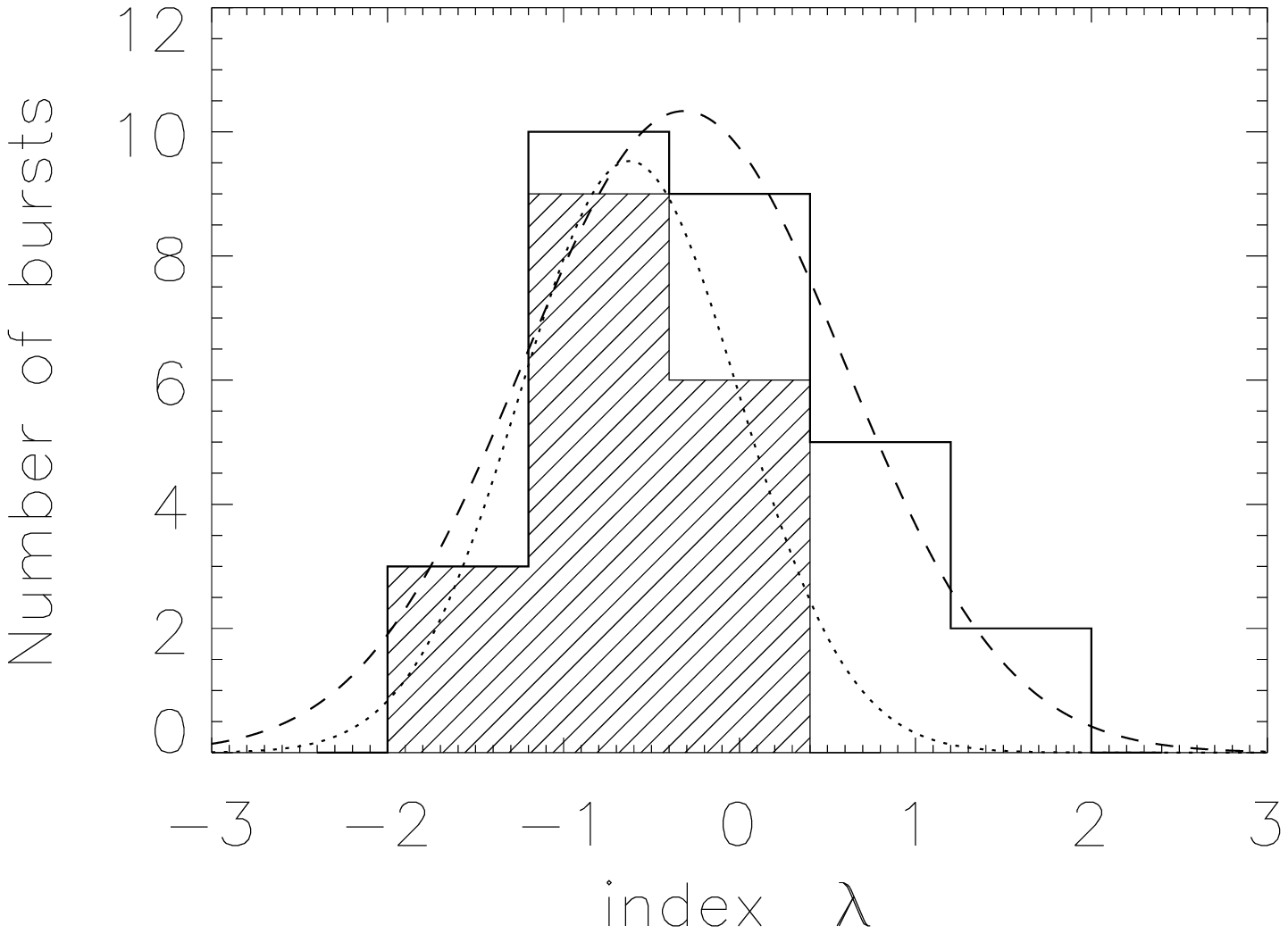}
\includegraphics*[bb=60 15 480 325, scale=0.5]{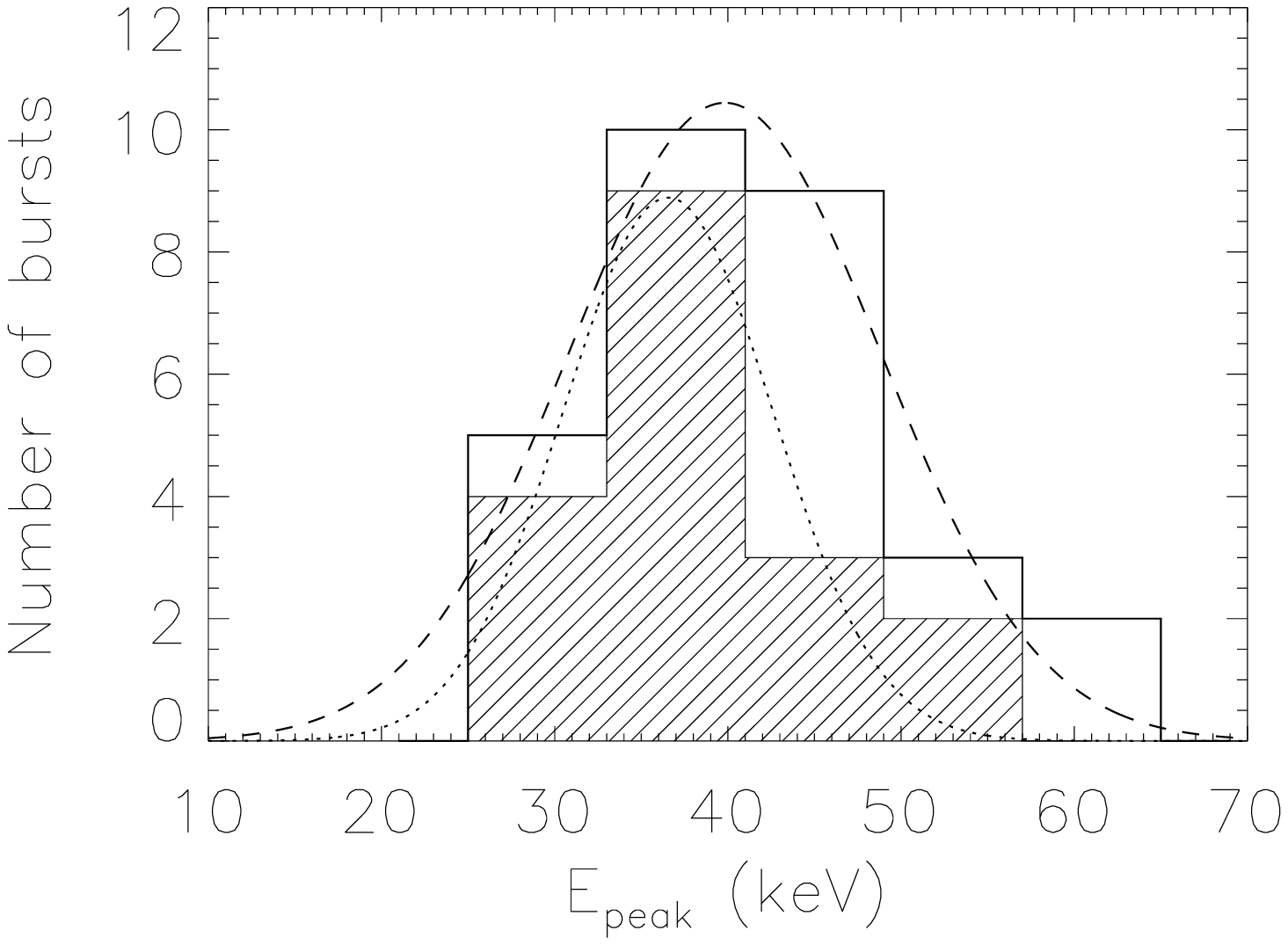}
\caption{Distributions of index (\textit{top panel}) and $E_{\rm{peak}}$ (\textit{bottom panel}) of the COMPT model fits for 29 bursts from \sgrnos. The right slashed bars represent the subset of the 18 bursts that can be fit with the BB+BB model as well. The dashed and dotted lines are the best fits with normal distributions of all bursts and of the subset of 18, respectively. \label{spec_distribution}}
\end{figure}

All \sgr bursts were well fit with the COMPT model. We list the model parameters and statistics in Table \ref{obs} (columns 6-10). The
distributions of index and $E_{\rm{peak}}$ are displayed in Figure \ref{spec_distribution}. The top panel shows the spectral index
distribution, which is centered around zero and is best fit with a normal function with an average of $-0.32\pm0.11$ ($\sigma = 0.9 \pm 0.1$).
This index distribution clearly excludes a pure OTTB or BB fit for all bursts, as such fits would require indices of $\sim-1$, $\sim+1$,
respectively. The bottom panel of Figure \ref{spec_distribution} exhibits the $E_{\rm{peak}}$ distribution, which also follows a normal
function with mean at 39.8$\pm$0.9 keV ($\sigma = 9.0\pm1.0$ keV). In both panels, the hatched areas highlight the distribution of 18 bursts
also fit with the BB+BB model. This subsample was fit with a normal distribution with $\langle E_{\rm{peak}} \rangle = 36.5 \pm 1.6$ keV,
$\sigma = 6.1 \pm 1.4$ keV and $\langle$index$\rangle = -0.63 \pm 0.04$, $\sigma = 0.62 \pm 0.04$. Most bursts here have negative photon index
and lower $E_{\rm{peak}}$, indicating a softer spectrum.

\begin{figure}[h]
\includegraphics*[bb=45 15 500 325, scale=0.5]{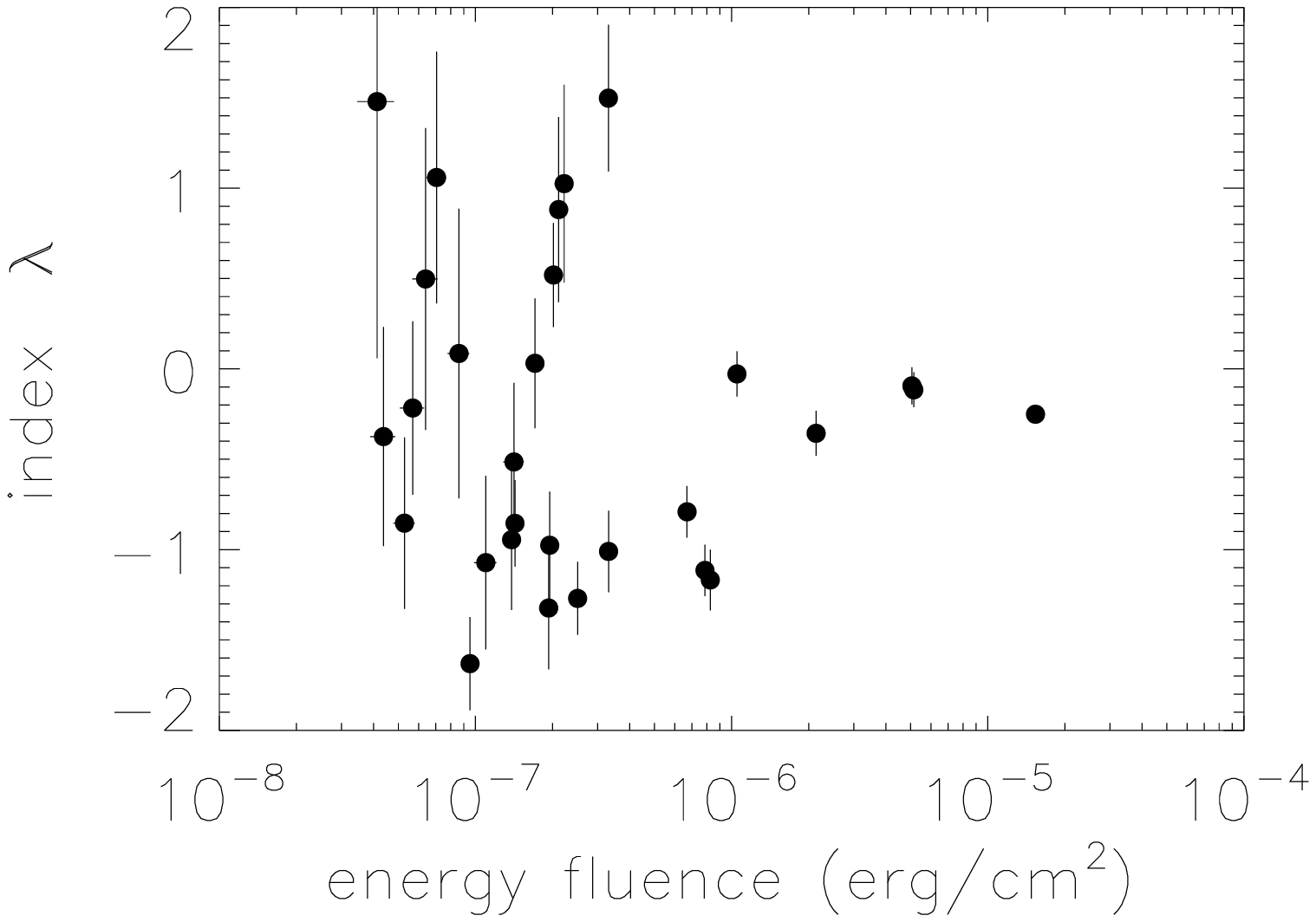}
\includegraphics*[bb=45 15 500 325, scale=0.5]{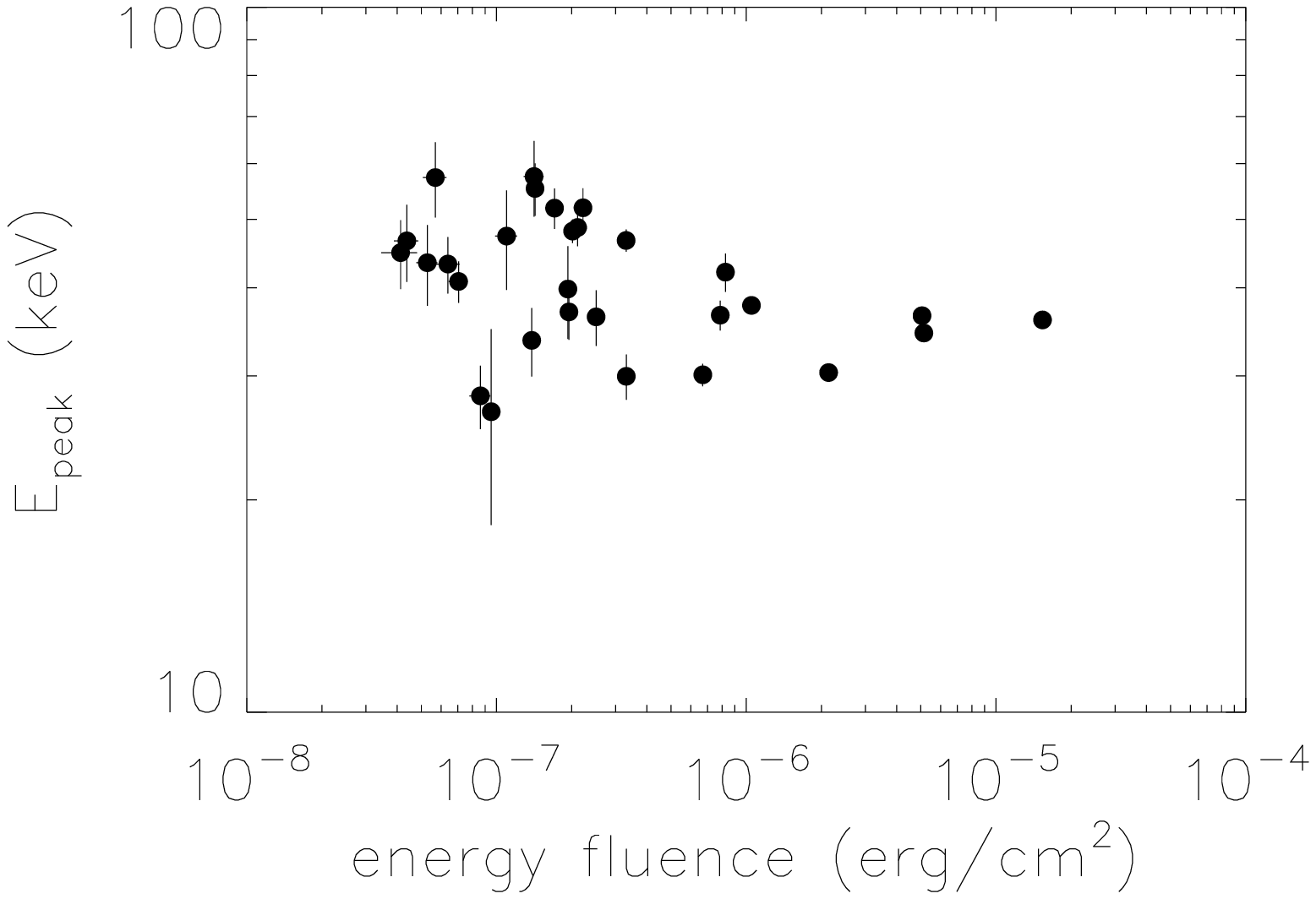}
\caption{Scatter plots of the COMPT index (\textit{top panel}) and $E_{\rm{peak}}$ (\textit{bottom panel}) {\it versus} the event energy fluence ($8-200$ keV) for 29 bursts from \sgrnos. \label{spec_correlation}}
\end{figure}

Using the COMPT spectral fits we estimated the event fluences ($8-200$ keV, also listed in Table \ref{obs}) and plotted in Figure
\ref{spec_correlation} their correlation with spectral indices (top panel) and with $E_{\rm{peak}}$ values (bottom panel). We note that the brightest
events have a constant index value of $\sim0$, which progresses to lower (softer) values for weaker ones and is widely scattered with larger
errors for the faintest events.  Similarly, a simple trend can not describe the relation between the hardness ($E_{\rm{peak}}$) and fluence.
For bursts with high energy fluence (e.g., $>5\times 10^{-7}$ erg cm$^{-2}$), $\textit{E}_{\rm{peak}}$ values are constant and cluster around 35
keV, while the values of weaker bursts cluster around 45\,keV with a large scatter range between 30 and 60 keV. We note here the large
span of fluence in our data ($3\times 10^{-5}$ to $3\times 10^{-8}$ erg cm$^{-2}$), which is one order of magnitude broader than earlier
results \citep{fenimore1994,gogus2001,gavriil2004}. We expand on these trends in Section \ref{discussion}.

\begin{figure}[h]
\includegraphics*[bb=10 15 500 340, scale=0.45]{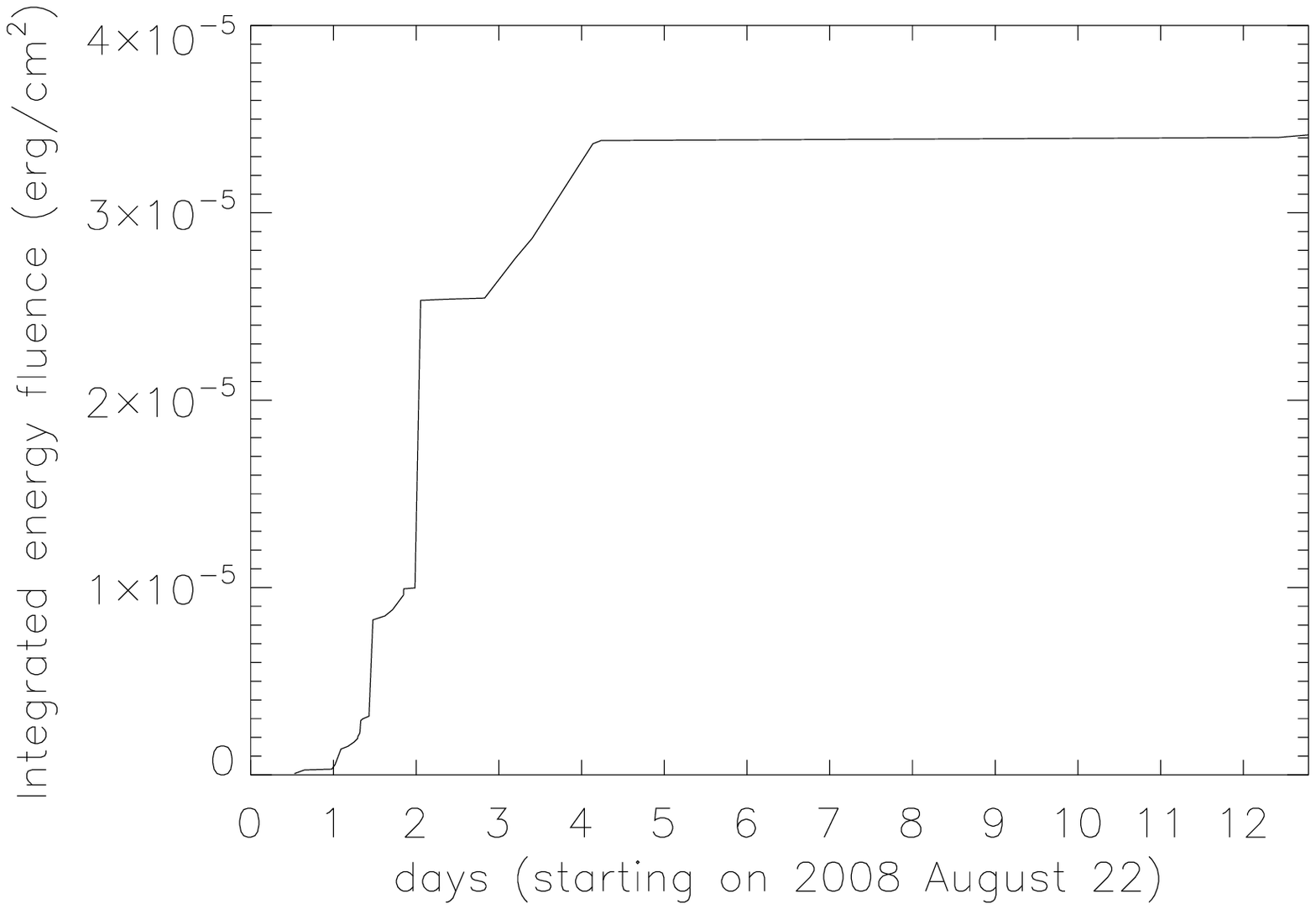}
\includegraphics*[bb=50 15 510 325, scale=0.5]{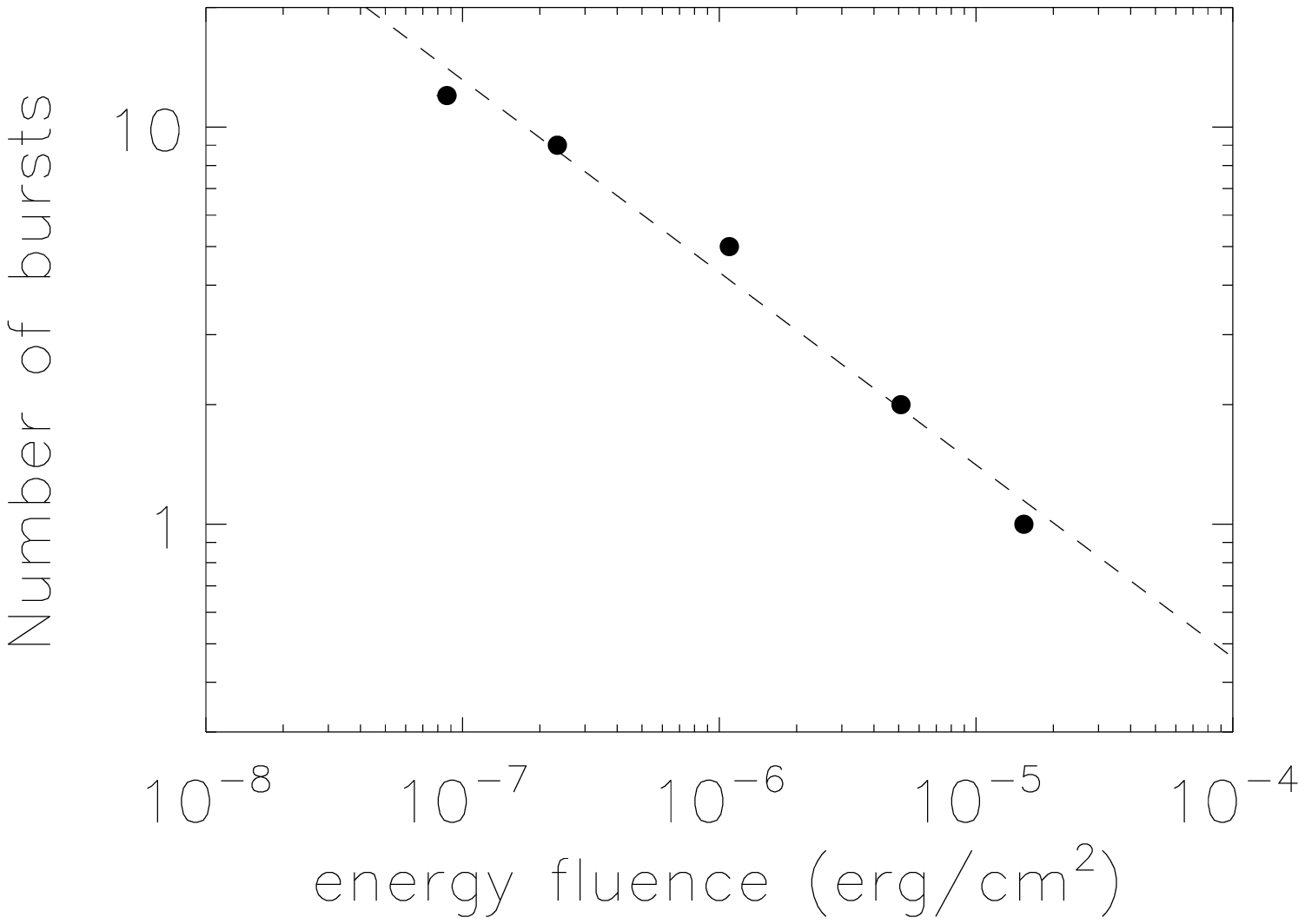}
\caption{\textit{top panel}: The evolution of the integrated
    energy fluence of \sgrnos, calculated with a COMPT spectral model. The cumulative plot starts at the first trigger on 2008 August 22.
    \textit{bottom panel}: The differential distribution of energy fluence estimated with a COMPT model. The dashed line is the best fit with a single power law of index $-0.48 \pm 0.02$. \label{energetic}}
\end{figure}

We now proceed to estimate the cumulative energy fluence during the active period of \sgrnos, shown at the top panel of Figure \ref{energetic}.
The plot levels off at $3.75 \times 10^{-5}$ erg $\rm{cm^{-2}}$, which should be considered a lower limit (as we have not taken into account
event saturation, untriggered events without TTE data, and additional bursts seen, e.g., with {\it Swift}). Assuming that the source is 2 kpc away from the Earth, then this fluence would correspond to a total energy of at least 1.8 $\times 10^{40}$ erg emitted from the magnetar in bursts during this 13-day active period. The bottom panel in Figure \ref{energetic} presents the differential distribution of energy fluence (log$N-$log$S$). The best fit with a power law function is overplotted in the figure. The index of the power law is $-0.48 \pm 0.02$, which corresponds to $dN/dF \propto F^{-1.48}$. This slope is similar to the one estimated with the {\it RXTE}/PCA for SGR\,$1806-20$ \citep{gogus2000}, but differs from all other SGR and AXP slope estimates, which are all very close to $-1.7$ \citep{woods2006}. It also differs from the slope estimate of the SGR\,$1806-20$ events detected with the International Cometary Explorer (ICE) and the Burst And Transient Source Experiment (BATSE) data, which is $-1.7$ as well \citep{gogus2000}. However, the GBM fluence range covers the higher end of fluence values, and the analyzed burst sample is most likely incomplete at the lower fluence end (which contributes most of the events in all other sources).

\subsubsection{BB+BB model fits}

Eighteen bursts were spectrally fit equally well with the BB+BB model. We chose to also present these results here, although this model is more
complicated than the COMPT model (with one more parameter), because it has been successfully used in the past for the spectral analysis of SGR
bursts \citep{olive2004,feroci2004,israel2008}; we will compare our spectral results in Section \ref{discussion}. Assuming that the two BB components arise
from two hot spots on the surface or photosphere of the source, each emission area, $R^2$, can be calculated from the temperature, $T$, of the corresponding black body spectrum as:
\begin{equation}
R^{2}=FD^{2}/\sigma T^{4} \label{r2}
\end{equation}
where \textit{F} is the average energy flux per event (total burst fluence divided by the spectral integration time), \textit{D} is the distance to the magnetar (assumed to be 2 kpc), and $\sigma$ is the Stefan Boltzmann constant.

\begin{figure}[h]
\includegraphics*[bb=5 15 500 350, scale=0.45]{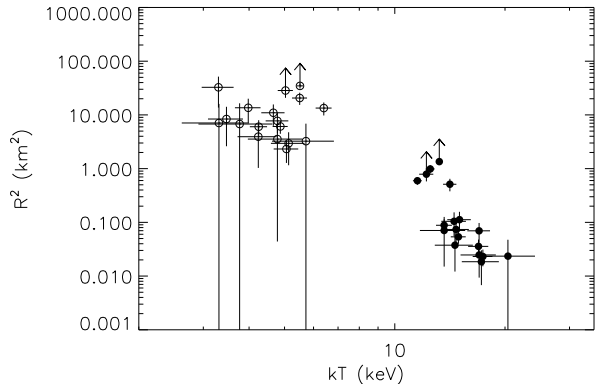}
\caption{Emission area as a function of black body temperature for time-integrated spectra. The dots mark the black body component with the higher temperature, while the circles represent the lower temperature black body. The upward arrows indicate the saturated bursts. \label{r2_kt}}
\end{figure}

\begin{figure}[h]
\includegraphics*[bb=10 10 500 340, scale=0.45]{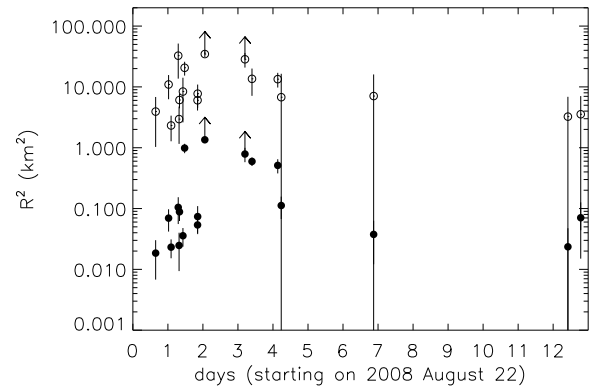}
\caption{Evolution of the emission areas of lower temperature black body (circles) and higher temperature black body (dots). The upward arrows indicate the saturated bursts. \label{r2_evo}}
\end{figure}

\begin{figure}[h]
\includegraphics*[bb=20 10 500 350, scale=0.5]{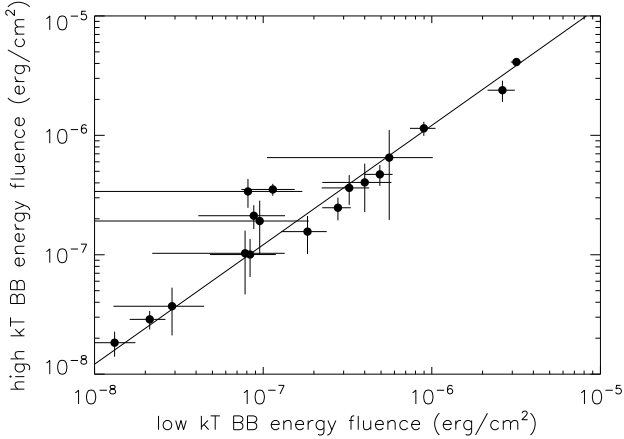}
\caption{The correlation between time-integrated fluence of cooler BB and hotter BB.  \label{lbb_hbb}}
\end{figure}

We plot the emission area of both BB components as a function of \textit{kT} in Figure \ref{r2_kt}. As shown earlier by \citet{israel2008} for SGR$1900+14$, we also note a clear separation of the
two temperatures and emission areas of the cooler and hotter black bodies. The cooler BB has a larger emission area, and the
temperature spreads between $3-7$ keV. The emission area of the hotter BB with temperatures in the $10-20$ keV range is much smaller. The
emission areas of both BB components have similar evolution through time during the active period of \sgr (Figure \ref{r2_evo}). The relative
increase in emission area, however, is higher (by a factor of ten) for the hotter BB component. Finally, we notice in Figure \ref{lbb_hbb} that the total fluence of the burst is divided equally between the two black body components (we find a correlation coefficient of 0.94 corresponding to a chance probability of $5.21 \times 10^{-9}$). A single power-law with an index of $1.00\pm0.05$ fits the data well.  

\subsection{Time-resolved spectra \label{trs}}

\begin{figure*}[p11t]
    \includegraphics*[bb=10 10 445 340, scale=0.35]{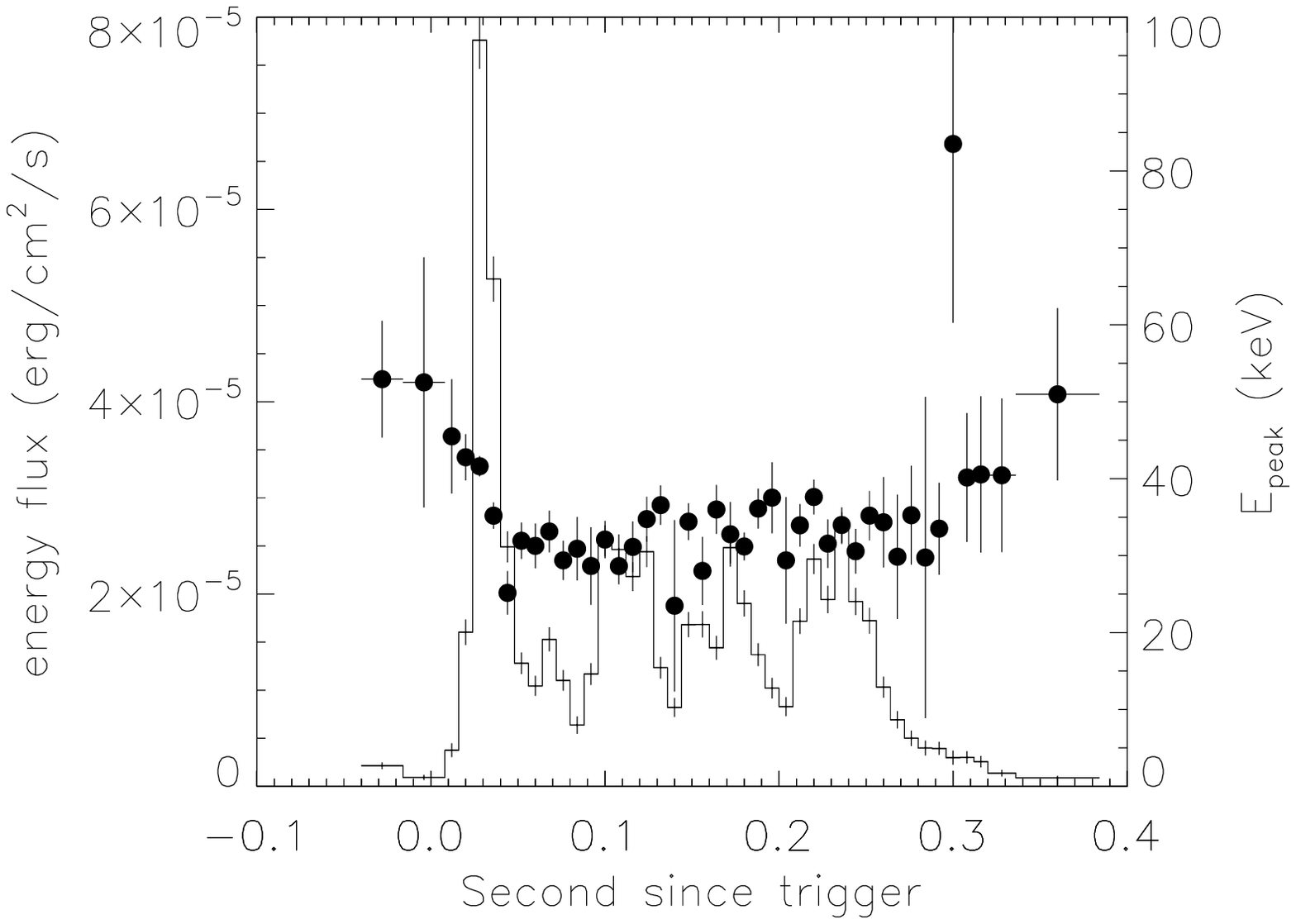}
    \includegraphics*[bb=50 10 450 330, scale=0.35]{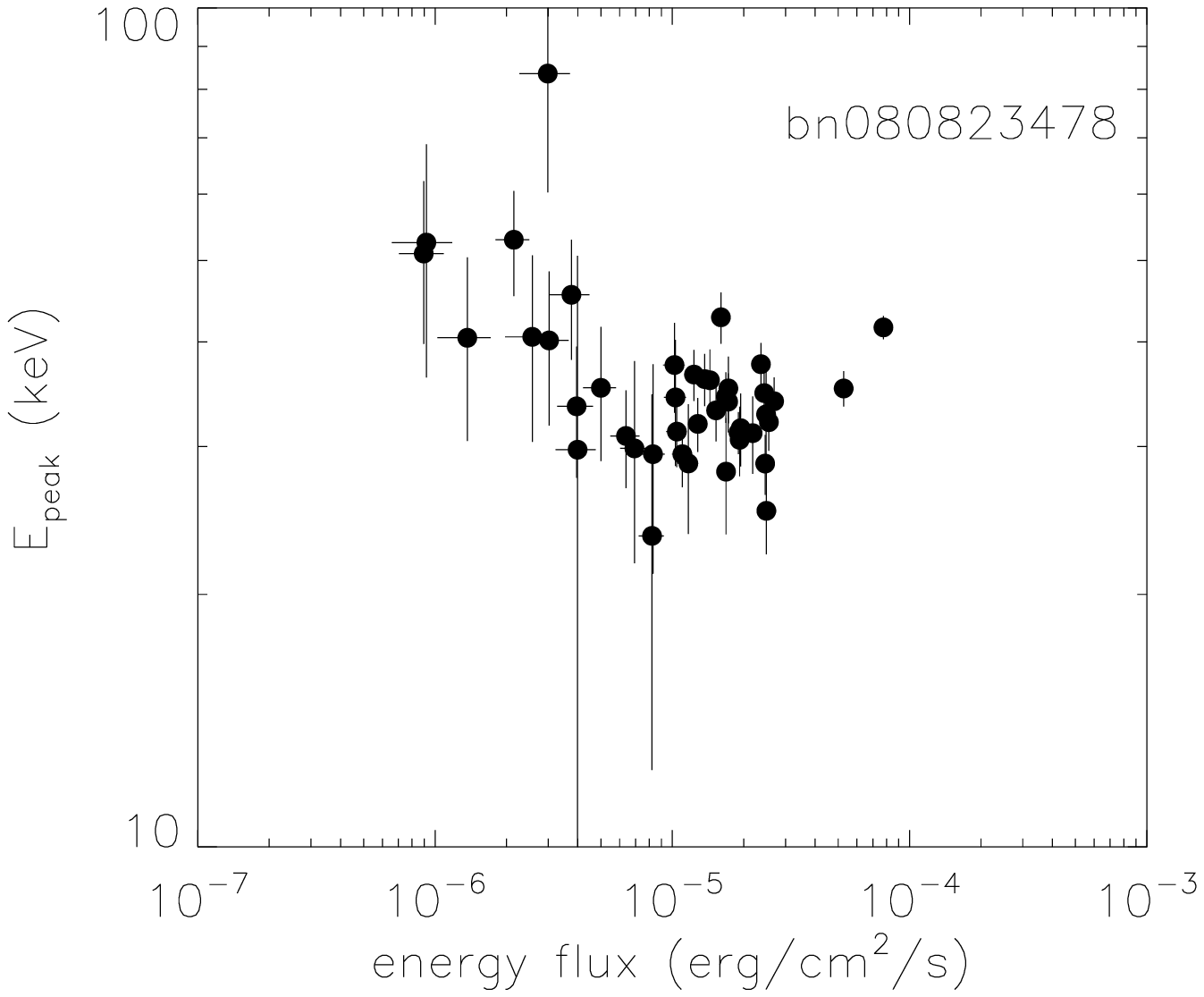}\\
    \includegraphics*[bb=10 10 445 340, scale=0.35]{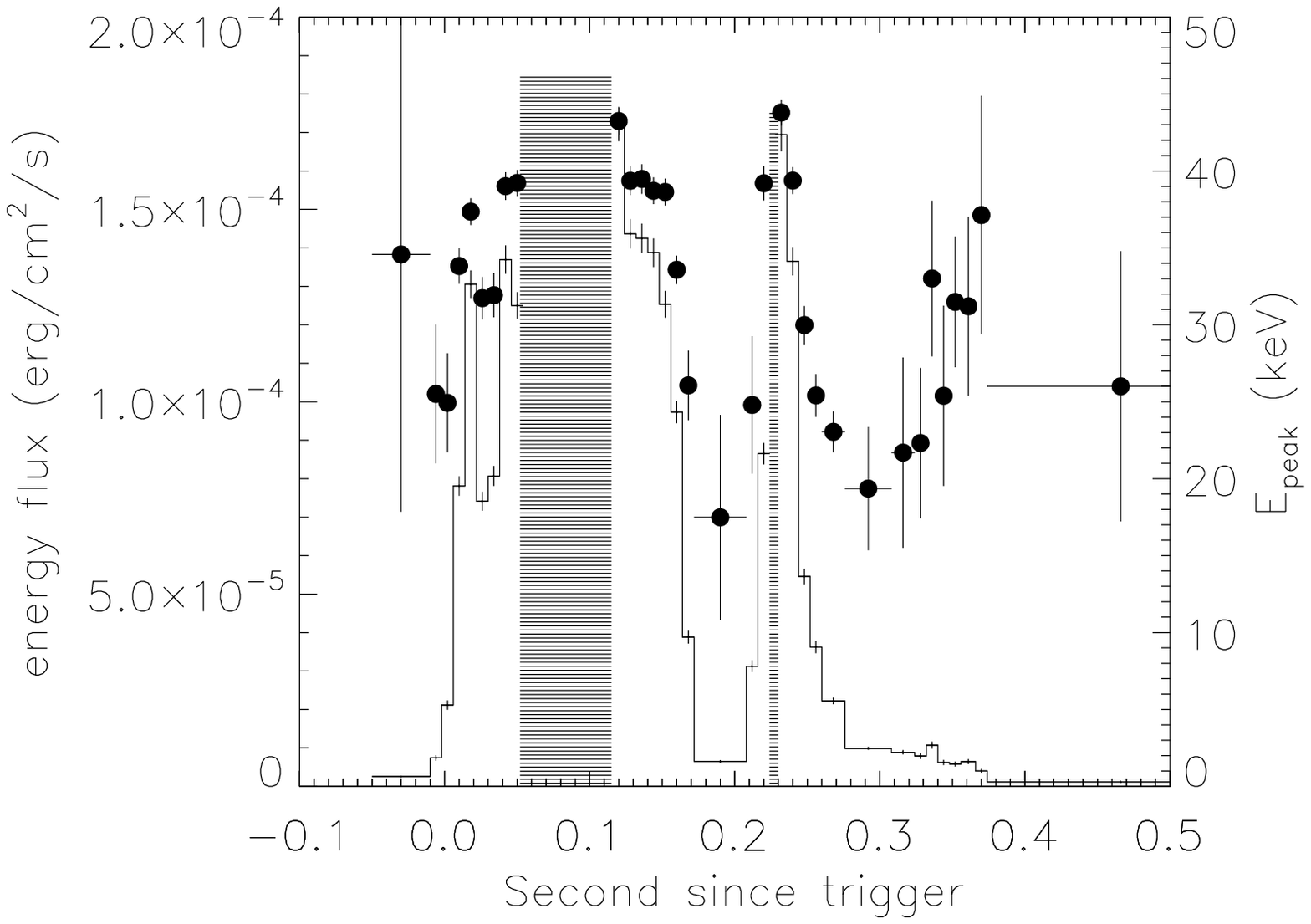}
    \includegraphics*[bb=50 10 450 330, scale=0.35]{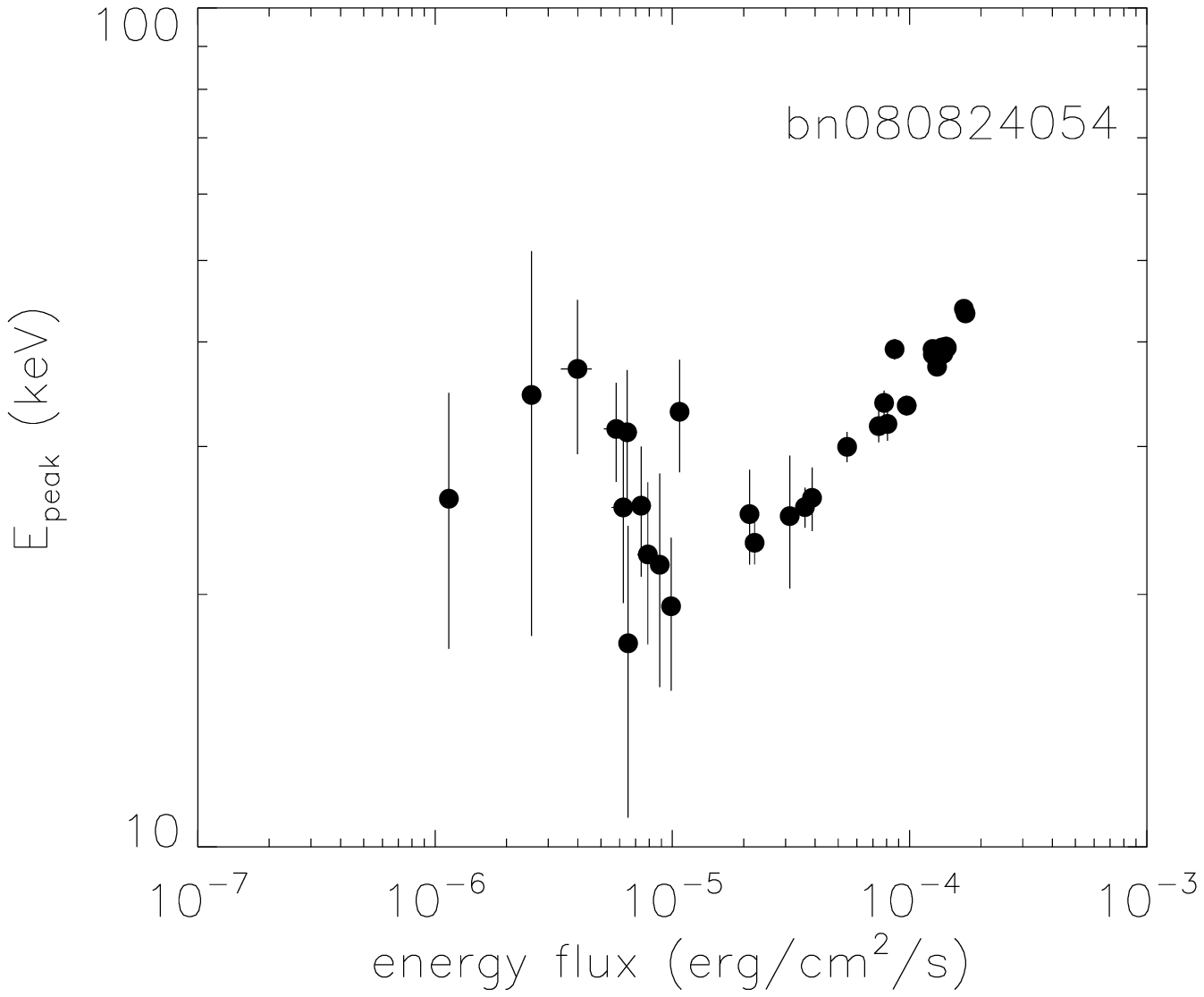}\\
    \includegraphics*[bb=10 10 445 340, scale=0.35]{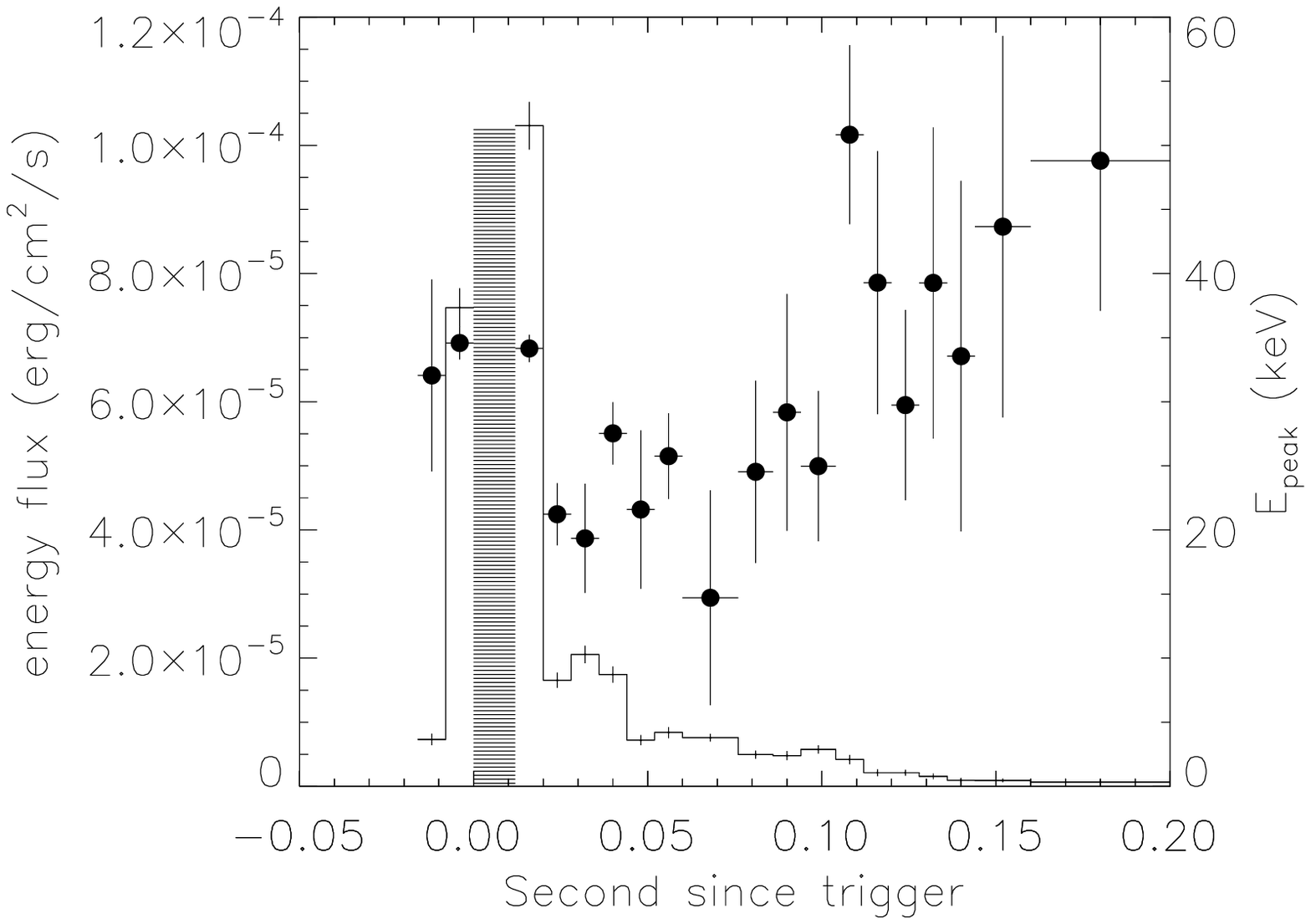}
    \includegraphics*[bb=50 10 450 330, scale=0.35]{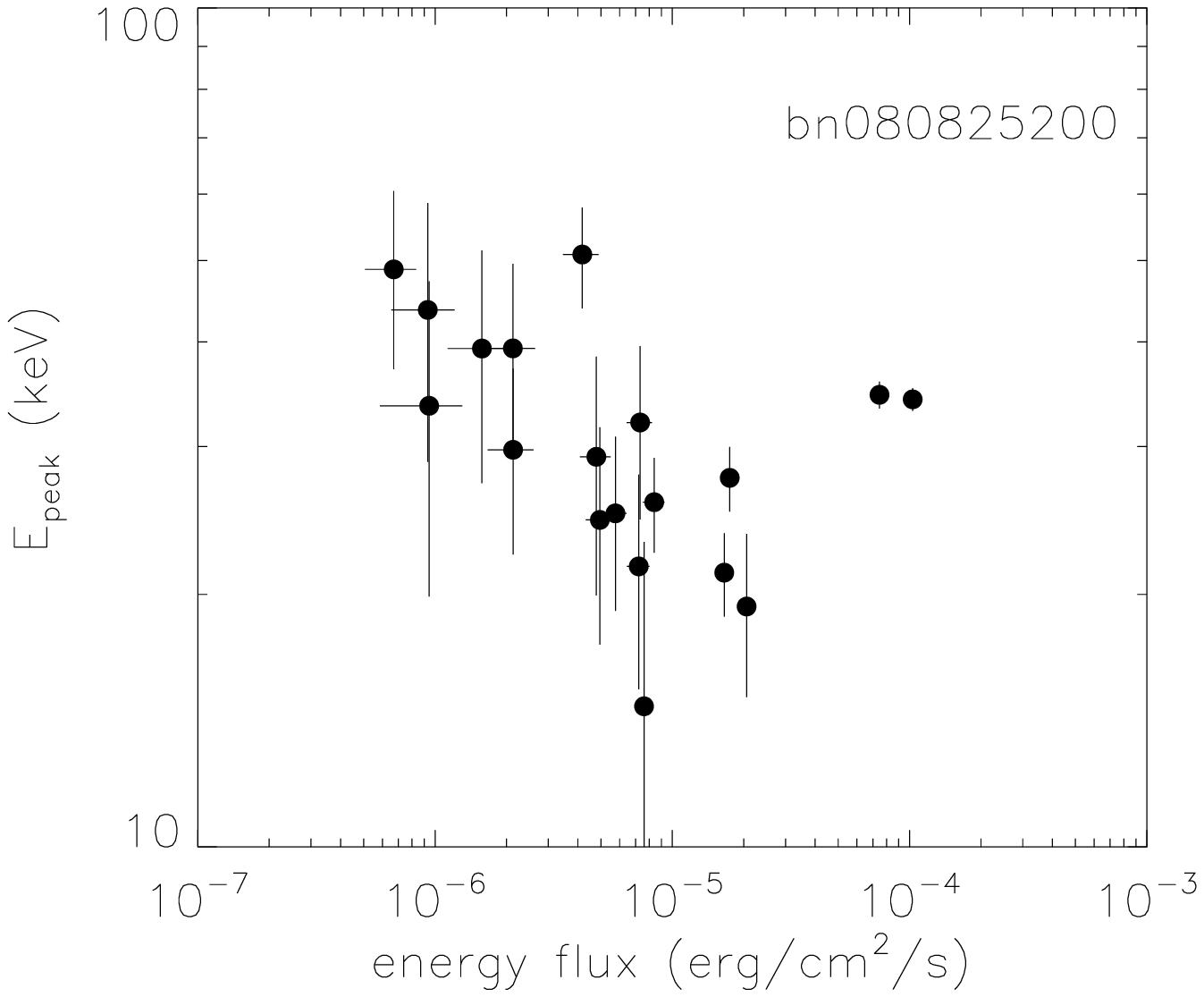}\\
    \includegraphics*[bb=10 10 445 340, scale=0.35]{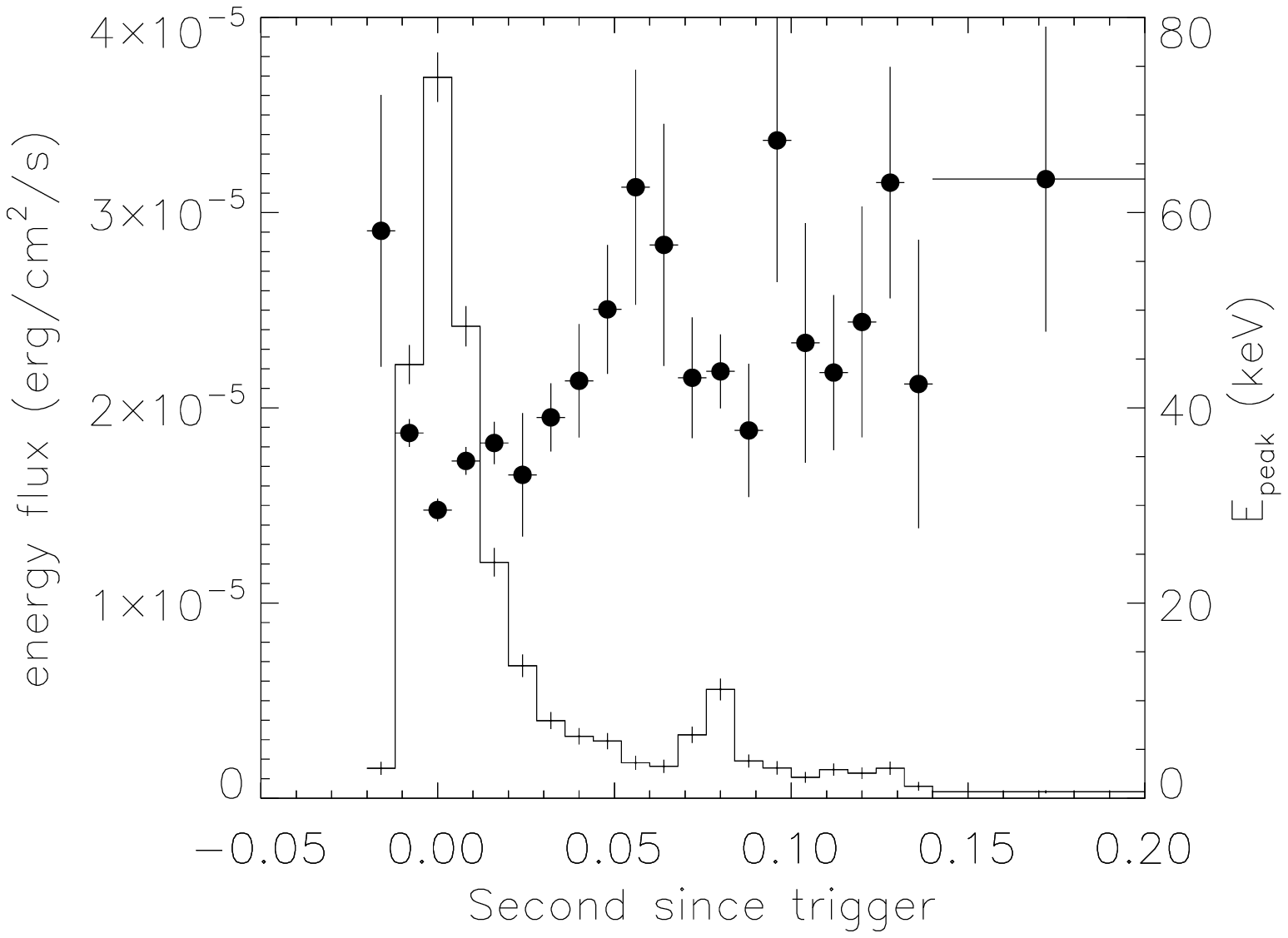}
    \includegraphics*[bb=50 10 450 330, scale=0.35]{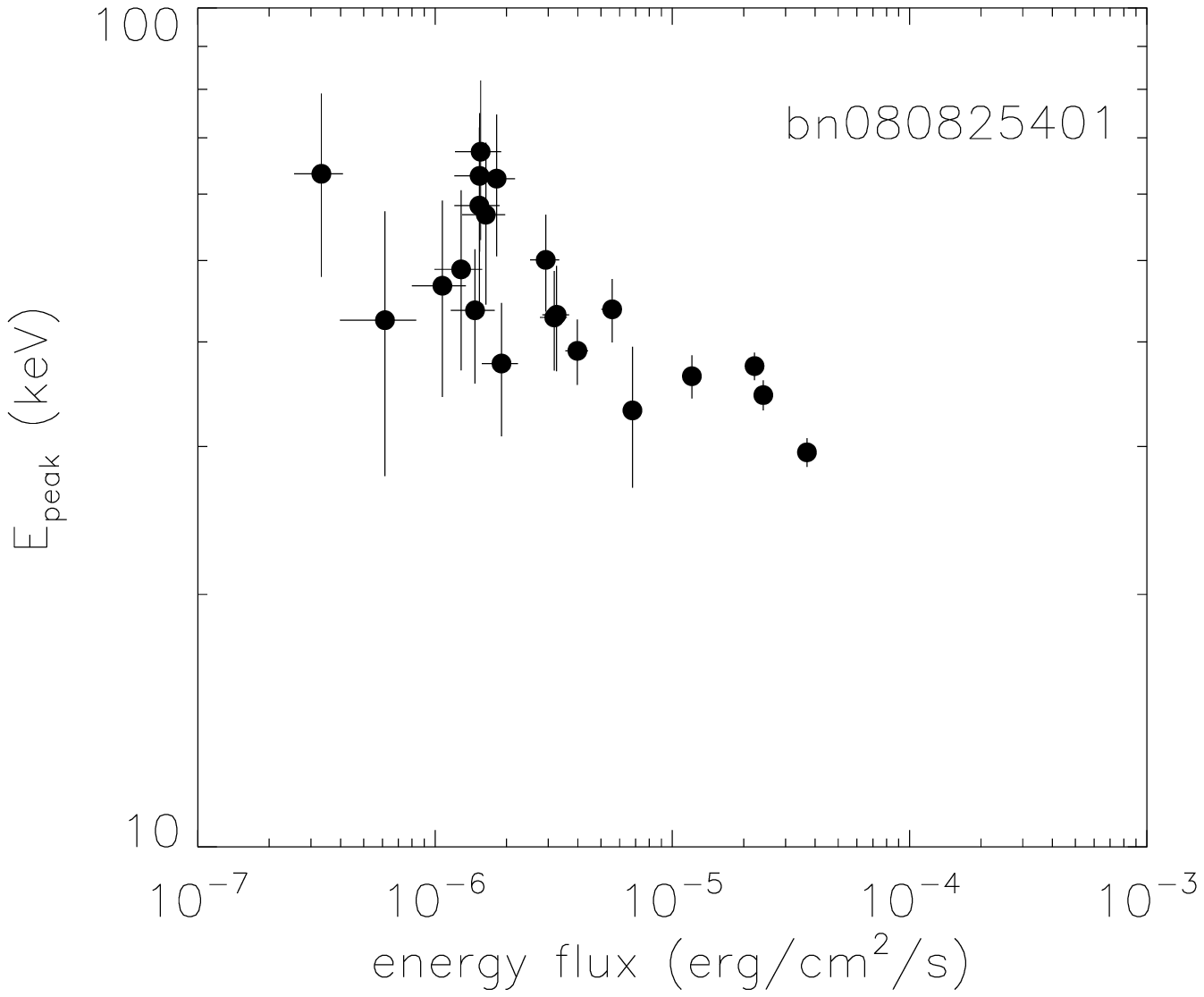}\\
    \includegraphics*[bb=10 10 445 340, scale=0.35]{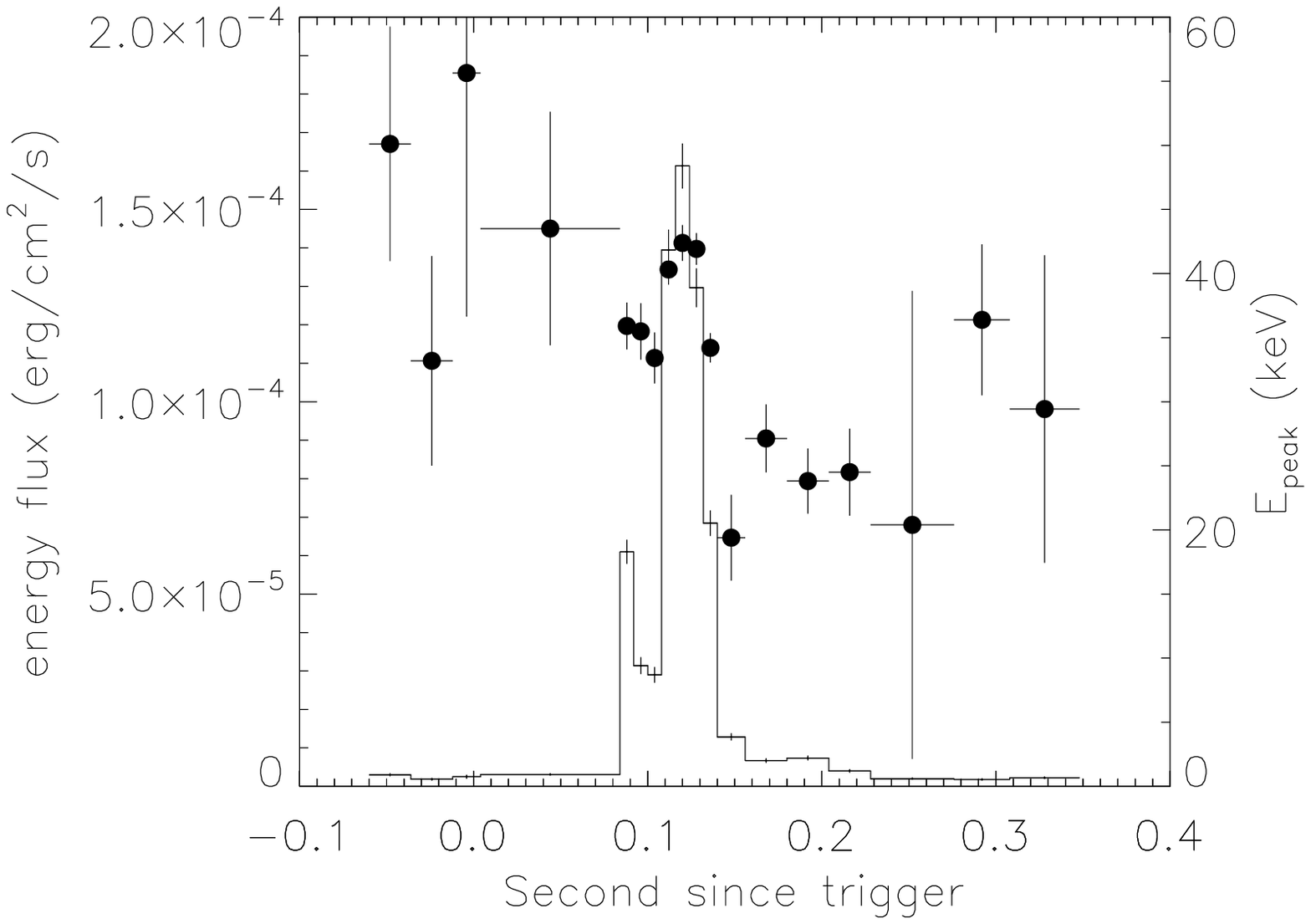}
    \includegraphics*[bb=40 10 440 330, scale=0.35]{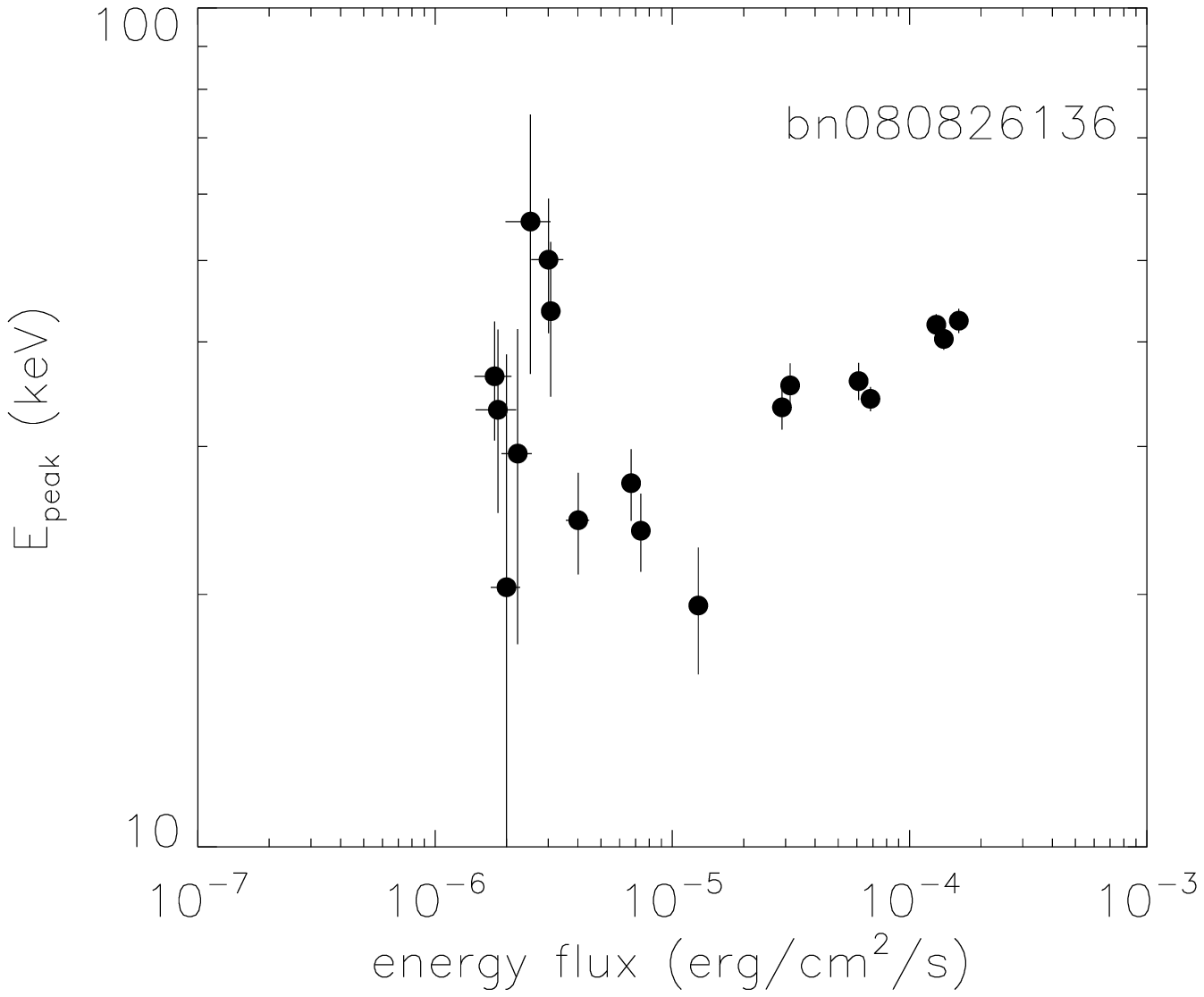}
    \caption{\textit{Left column}: Light curves of the five brightest events from \sgrnos, overplotted with their $E_{\rm{peak}}$ values. The dark areas mark the saturated parts in two of the bursts. \textit{Right column}: Correlation between $E_{\rm{peak}}$ and energy flux in the $8-200$ keV band. \label{tr_spec}}
\end{figure*}

We performed time-resolved spectral analysis with 8\,ms temporal resolution for the five brightest bursts of \sgrnos. We binned the data
requiring a significance of at least $3\sigma$ above background for each bin. Each bin was then fit with the COMPT model, which was the best
model for the time-integrated spectra. Figure \ref{tr_spec} displays the light curves of these five events overplotted with their
$E_{\rm{peak}}$ values (left column) and the correlation between $E_{\rm{peak}}$ and energy flux (right column). We see that the
$E_{\rm{peak}}$ follows the light curve for the brightest part of the burst in four out of five cases (except for the fourth panel from the top), and it rises surprisingly at the beginning and the tail end of each event.

\begin{figure}[p11t]
\includegraphics*[bb=40 10 500 340, scale=0.5]{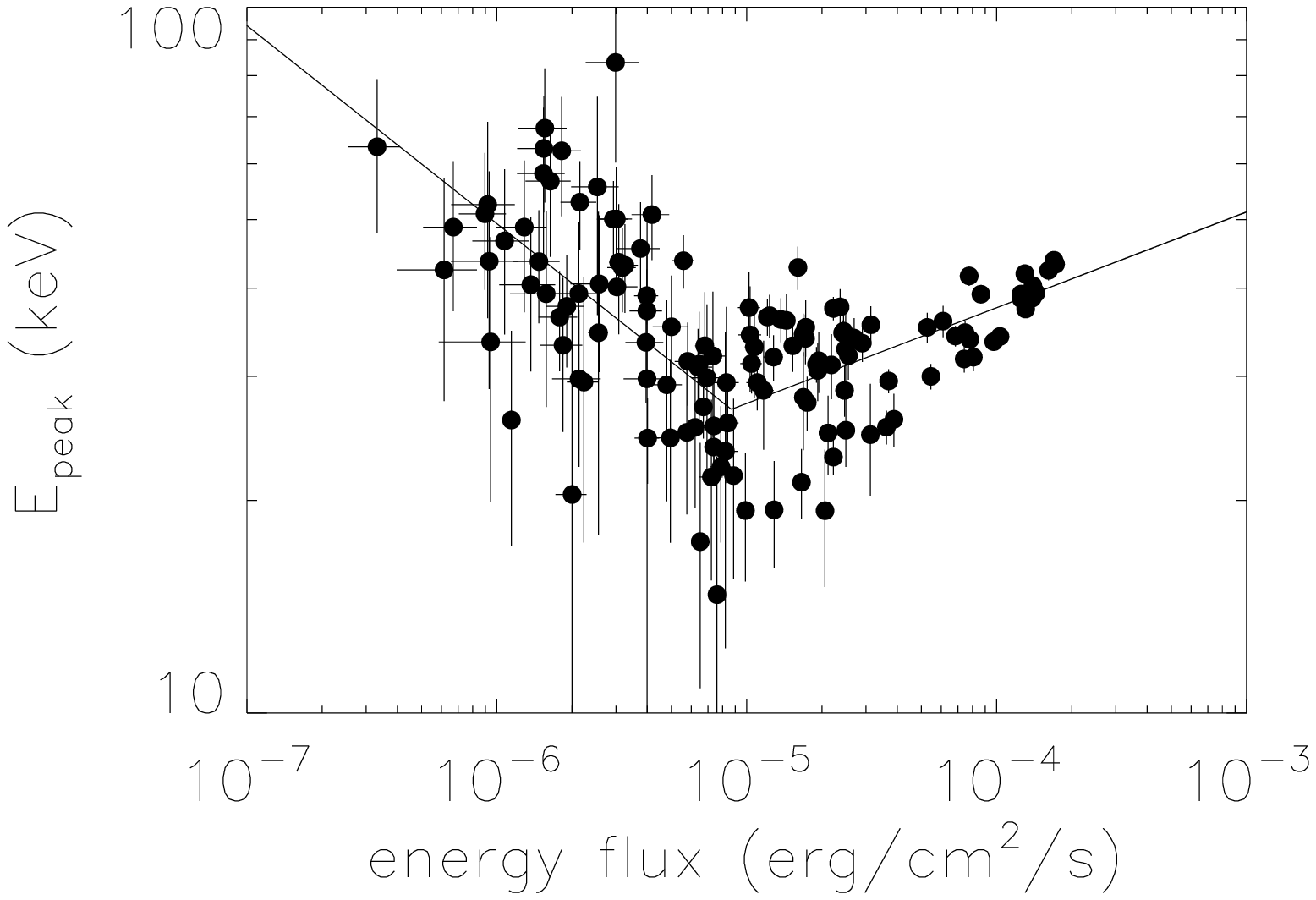}
\includegraphics*[bb=40 10 500 340, scale=0.5]{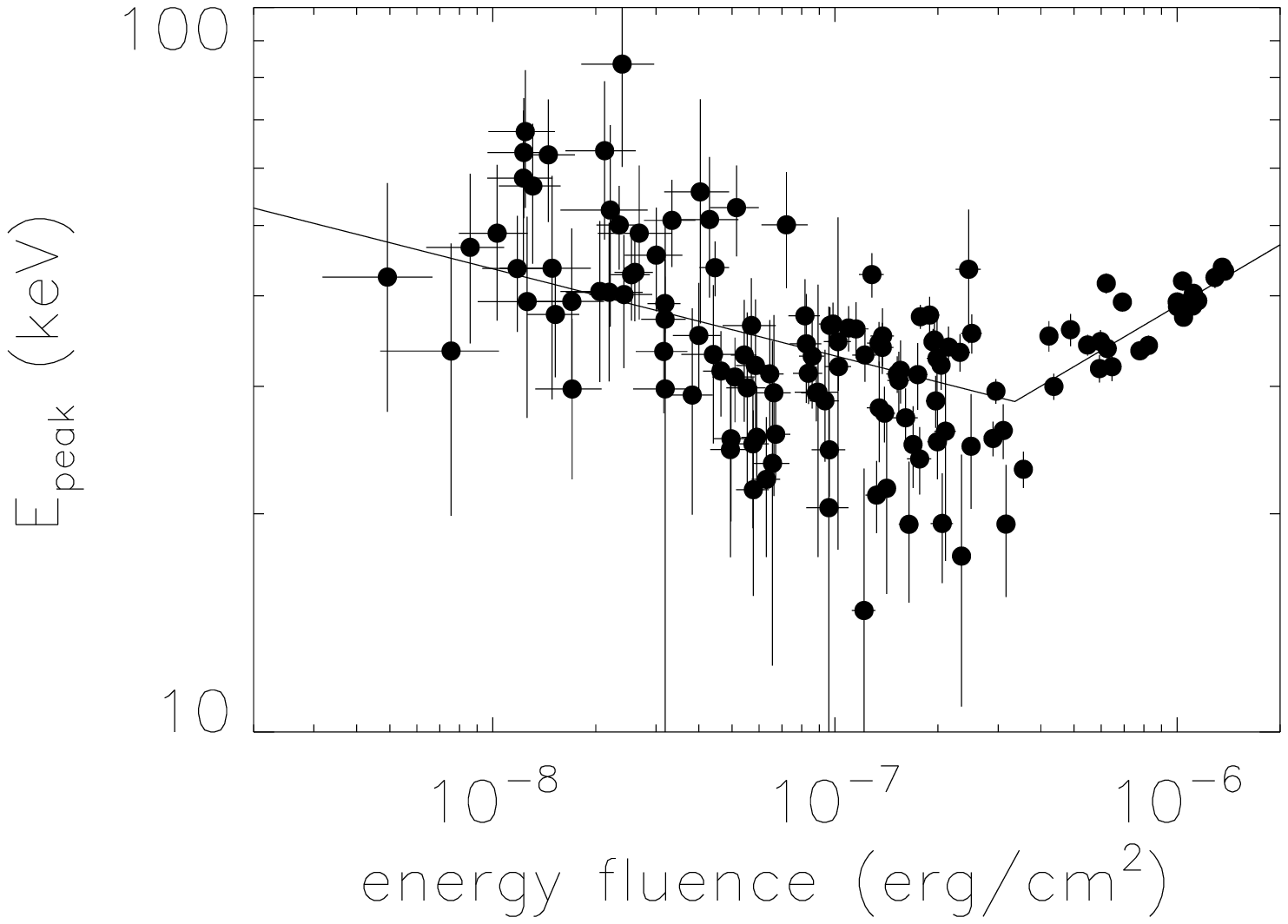}
    \caption{(\textit{top panel}): Correlation between the $E_{\rm{peak}}$ and energy flux in the time-resolved spectra of the five brightest bursts of \sgrnos. (\textit{bottom panel}): Correlation between the $E_{\rm{peak}}$ and energy fluence for the same bursts. The solid lines exhibit the best fit to the data with broken power-law functions. \label{tr_correlate}}
\end{figure}

To investigate this trend, we plotted in Figure \ref{tr_correlate} the combined $E_{\rm{peak}}$ values from all five bursts {\it versus} their fluxes (top panel). We clearly see that $E_{\rm{peak}}$ rises at both high {\it and} low flux values with a minimum determined with a broken power-law fit at $8.7 \pm 0.9 \times 10^{-6}$\,erg $\rm{cm}^{-2}$ $\rm{s}^{-1}$. The same trend is seen in the lower panel between $E_{\rm{peak}}$ and fluence with a minimum at $3.4 \pm 0.2\times 10^{-7}$\,erg $\rm{cm}^{-2}$. The indices below and above this critical flux (fluence) values are $-0.28 \pm 0.03$ and $0.14 \pm 0.01$ ($-0.14 \pm 0.01$ and $0.28 \pm 0.02$), respectively. An F-test comparison of a broken power law and a single power law fit shows that the former provides a better description of the correlation, with a small probability of chance coincidence $1.2\times10^{-7}$ for the flux and $9.8\times10^{-7}$ for the fluence. We also performed a Spearman rank test for both branches of each plot and found a {\it low} flux (fluence) correlation coefficient of $-0.66$ ($-0.59$) with a probability of chance coincidence of $1.7 \times 10^{-9}$ ($2.7 \times 10^{-11}$). The correlation coefficient and probability for the {\it high} flux (fluence) parts are $0.51$ ($0.78$) and $8.7 \times 10^{-6}$ ($3.3 \times 10^{-6}$), respectively. The latter probabilities are less significant than the ones for the low flux/fluence parts, because the Spearman rank tests involve a smaller number of data points. For the full flux (fluence) data sets, the correlations are not very significant, with a coefficient of $-0.23$ (-0.28) and probability of $6.4 \times 10^{-3}$ ($1.3 \times 10^{-3}$), giving further support to changing trends (minima) in the correlations between $E_{\rm{peak}}$ and flux/fluence.

\begin{figure}[h]
    \includegraphics*[bb=50 10 500 340, scale=0.5]{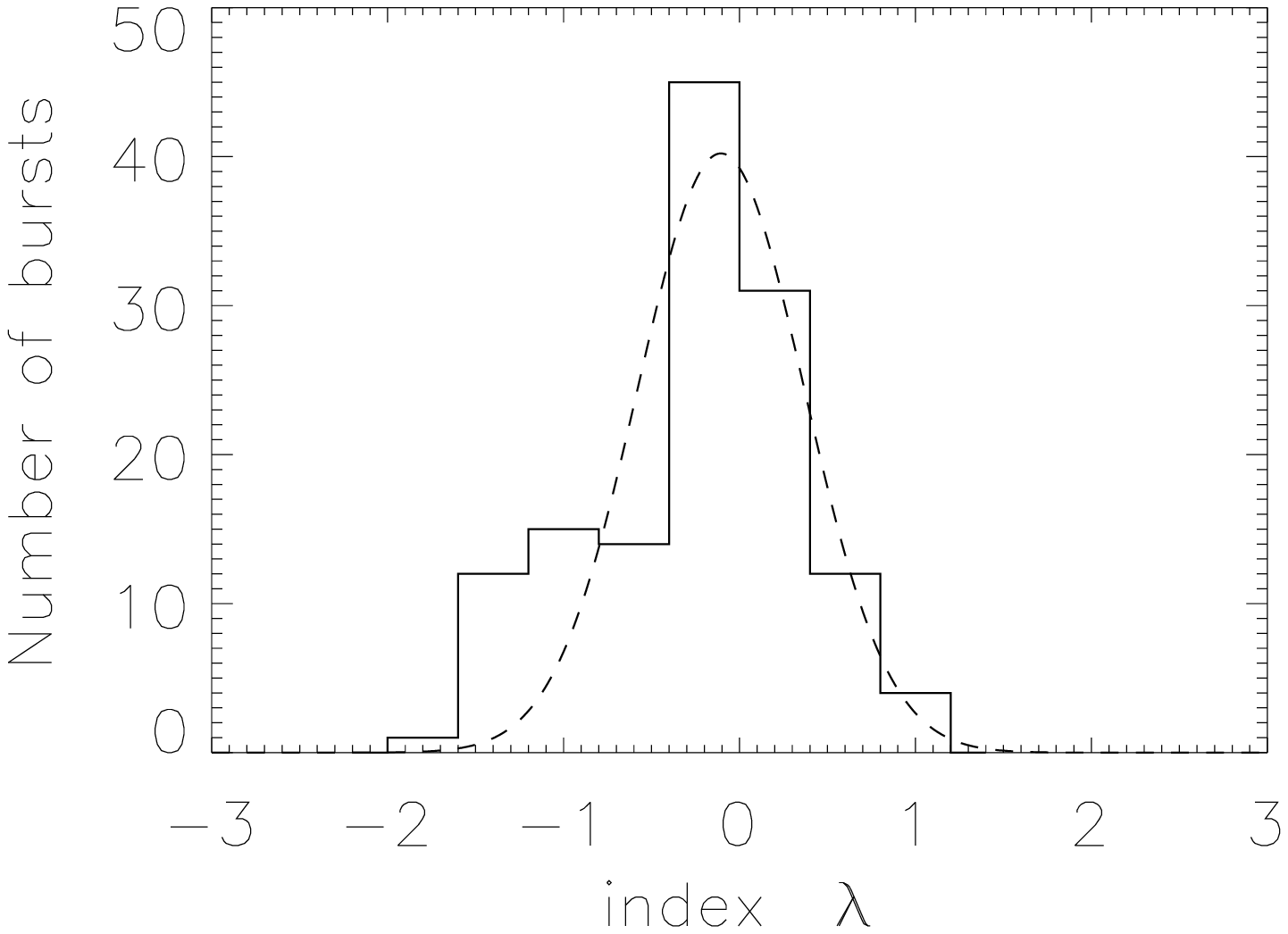}
    \includegraphics*[bb=50 10 500 340, scale=0.5]{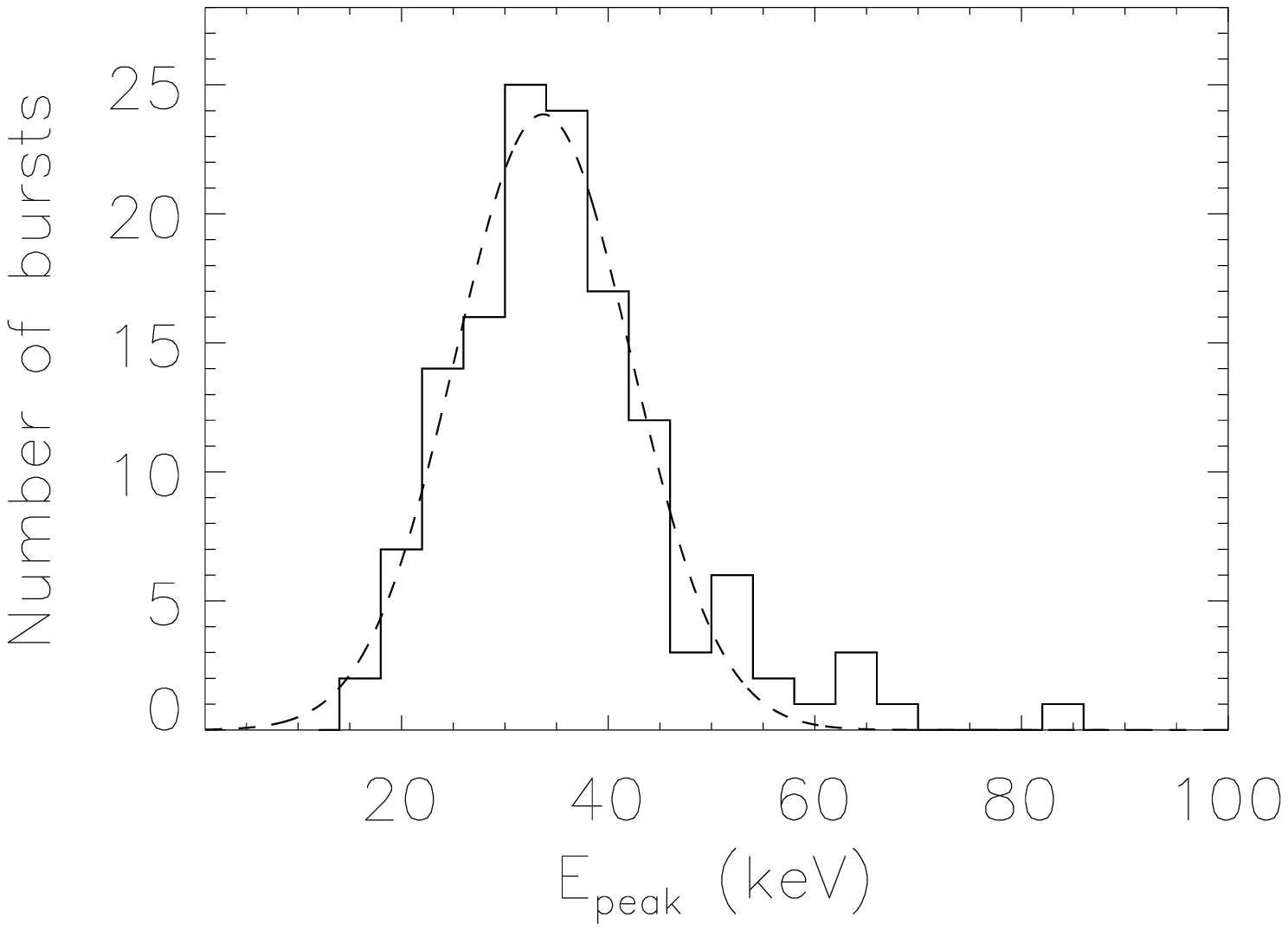}
    \caption{Distribution of index (\textit{top panel}) and $E_{\rm{peak}}$ (\textit{bottom panel}) of the COMPT model for all time-resolved spectra. The dashed lines are the best fits with a normal distribution. \label{tr_compt_dis}}
\end{figure}

Figure \ref{tr_compt_dis} displays the distribution of the COMPT index and $E_{\rm{peak}}$ of all time-resolved spectra. Both distributions were fit with normal functions with $\langle E_{\rm{peak}} \rangle = 33.7 \pm 0.5$ keV ($\sigma = 8.5 \pm 0.5$ keV) and $\langle$index$\rangle = -0.11 \pm 0.09$ ($\sigma = 0.47 \pm 0.09$). These values are very similar to those obtained with the time-integrated fits.

\begin{figure}[h]
    \includegraphics*[bb=5 10 500 340, scale=0.5]{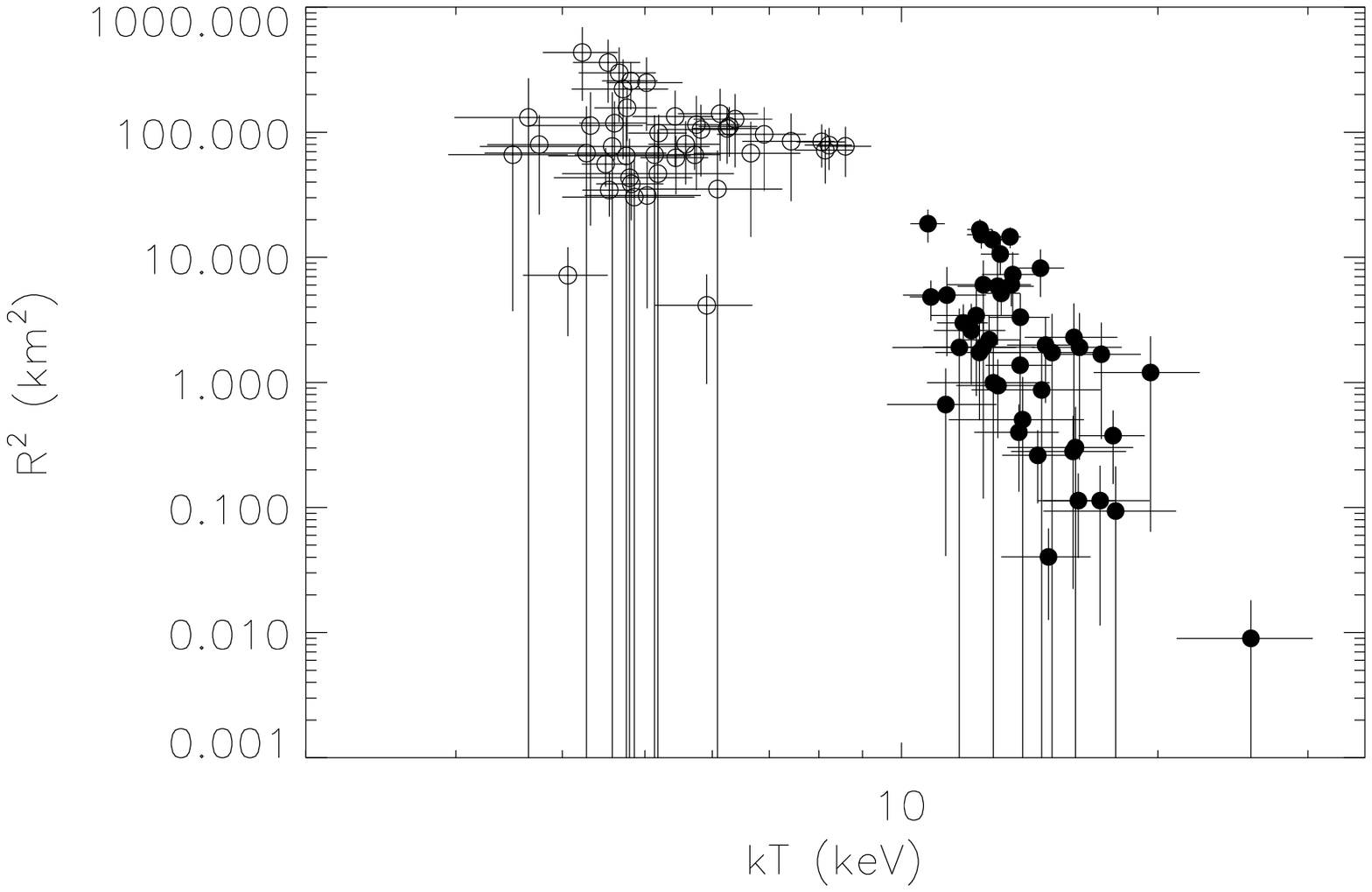}
    \caption{Emission area as a function of black body temperature for time-resolved spectra. The dots mark the black body component with the higher temperature, while the circles represent the lower temperature black body. \label{tr_r2_kt}}
\end{figure}

Finally, we fit the time-resolved spectra with a BB+BB model. Retaining only the fits which constrained the BB+BB parameters, we calculated the
emission areas for each BB component. Similar to the integrated spectra, Figure \ref{tr_r2_kt} shows that the emission areas follow different
behaviors with temperature for the hard and soft BB. Interestingly, a comparison with similar results shown in Figure 5 of \citep{israel2008} for SGR$1900+14$ reveals that although the $kT$ values are very close for the two BBs, the emission areas differ by at least one (and maybe two) orders of magnitude. These results indicate that either the two sources have different burst energetics or that (very unlikely) the distance determination of SGR$1900+14$ is off by a large factor. 

\begin{figure}[h]
    \includegraphics*[bb=10 10 500 340, scale=0.5]{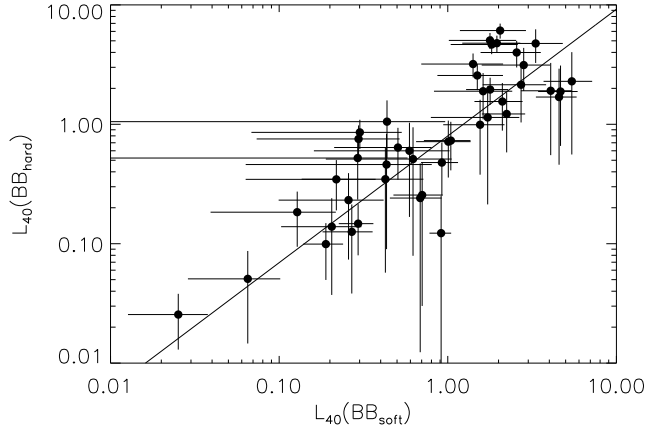}
    \caption{The correlation between the time-resolved luminosities (in $10^{40}$ erg s$^{-1}$) of the soft and hard BB. \label{tr_lbb_hbb}}
\end{figure}

Using the 2 kpc distance for the source, we estimate the isotropic luminosity of the two
BB components shown in Figure \ref{tr_lbb_hbb}. The Spearman rank correlation coefficient is $0.83$ corresponding to a chance coincidence probability of $2.78
\times 10^{-12}$, indicating that these components are well correlated, as also shown with a power-law fit with index $1.1\pm0.1$. \citet{israel2008} have performed a similar analysis for the data of SGR$1900+14$ and found a spectral index of $0.70\pm0.03$.

\section{Discussion \label{discussion}}
\subsection{COMPT {\it versus} BB+BB}
Our spectroscopic fitting clearly indicates that the COMPT and BB+BB
models yield superior fits to the other possibilities. The COMPT model,
with its power-law shape curtailed by an exponential turnover, is
intended to mock-up the classic unsaturated Comptonization spectrum
realized in models of accretion disks such as in Cyg X-1 or in active
galactic nuclei (see Chapter 7 of Rybicki \& Lightman 1979 for a summary
of its development as a solution of the Kompaneets equation).  These
models use hot, thermal electrons in a corona to repeatedly scatter
low-energy photons, heating them gradually up to an energy $E\sim kT_e$
consistent with the electron temperature $T_e$, at which point a
quasi-exponential spectral turnover emerges as further heating becomes
impossible. The power law marks the scale-independence of the Compton
upscattering, and its slope depends only on the mean energy gain per
collision ($\langle \Delta E\rangle =4 kT_e$ for non-relativistic
electrons) and the probability of loss of photons from the scattering
zone, being expressed as
\begin{equation}
    \frac{dN}{dE}\;\propto\; E^{\lambda}
    \quad ,\quad
    -\lambda\; =\; -\frac{1}{2} + \sqrt{\frac{9}{4} + \frac{4}{y}}\quad ,
  \label{Comp_index}
\end{equation}
with the Compton $y$-parameter in the domain $y< 1$. This
parameter is the product of the average fractional energy change per
scattering and the mean number of scatterings, and is $4 kT_e/(m_ec^2)\,
\hbox{max} \{\tau,\, \tau^2\}$ for a non-relativistic situation for
repeated
Compton upscattering.  Here $\tau$ is the scattering optical depth.

In the context of magnetars, such Comptonization can also ensue, but the strong magnetic field now plays an important role.   Ephemeral coronae
of hot electrons could be expected in a dynamic inner magnetosphere. For example, this could be due to intense dissipation of magnetic energy
in the closed field line region via field line twisting, i.e., transient departures from poloidal geometry, as in the considerations of
Thomson, Lyutikov \& Kulkarni (2002), Thompson \& Beloborodov (2005),  Beloborodov \& Thompson (2007), and Nobili, Turolla \& Zane (2008).  Such a pumping of energy into
electrons in low altitude regions would then be subject to irradiation by the intense bath of surface X-ray emission. The electron coronae
would mimic those of their black hole counterparts discussed above, and serve as a Comptonization target for the X-rays. Temporally, the
coronae could be quite variable, resulting in varying or chaotic time profiles such as are observed.  If the deposition of energy in hot
electrons is persistent over many light-crossing timescales, so also can the upscattered hard X-ray emission be.  In this magnetic case, the
spectrum would again naturally be a power law truncated by an exponential tail, whose energy is pinned by $T_e$.  The slope does differ
somewhat in value from the non-magnetic case because the presence of the intense field anisotropizes the environment.  This alters the average
fractional energy change per scattering from the isotropic $4 kT_e$ form in Rybicki \& Lightman (1979). We note that this picture is similar to
that of Lyutikov \& Gavriil (2006; see also Rea et al. 2008), who modeled the steep X-ray tails below 10 keV in quiescent magnetar emission
using a resonant cyclotron scattering picture.

Since the electrons will move along the field lines in the zeroth Landau
level, the exact kinematics that impact the determination of
$\langle\Delta E\rangle$ depend on the colatitude and altitude of the
collisions.  Therefore, the effective {\it magnetic Compton
$y$-parameter}, $y_B$, that could be substituted in
Eq.~(\ref{Comp_index}), would take a somewhat different value from its
non-magnetic cousin, but one anticipates that the range of spectral
realizations would be similar to the $B=0$ case.  Since head-on
collisions generally yield greater heating of photons, and these are
more likely at lower altitudes, it is expected that interaction zones
close to the stellar surface would spawn larger $y_B$ and therefore
flatter photon spectra.  If there were a coronal radius expansion, this
would then translate to an evolutionary steepening of the spectra with
time.  Note that the photon retention probabilities in the coronae will
need to be higher than in the Lyutikov \& Gavriil (2006) scenario to
generate the requisite moderate optical depths $\tau$ to match the flat
outburst spectra discussed in this paper.

The scatterings will take place below the cyclotron resonance unless
their altitude is above around 10 stellar radii (e.g. Lyutikov \&
Gavriil 2006).  Below resonance, the cross section is far inferior to
the Thomson value (e.g. see Herold 1979).  Then, the coronal electron
density must accordingly be much higher than when $B=0$ in order to
establish a sizable optical depth. This provides a possible distinction
between the two classes of magnetar emission: the steady $<10$\,keV signal
may originate at higher altitudes where the resonance is accessed and
the local electron density is low, whereas the bursts reported
here may be triggered closer to the surface in higher density zones that
initially precipitate scattering below the cyclotron resonance. As
photons get energized above 20 keV, the altitude where the resonance is
accessed is lowered, increasing the cross section for scattering and
therefore $y_B$.  Note also, that to a considerable extent, polarization
mode-switching (e.g. Miller 1995) between the two photon polarizations
can help increase the opacity, adding a further nuance to consider when
modeling the scattering environment.

Observationally, it is difficult to discern unambiguously the presence
of the intense field using a Comptonization model scenario in the energy
range of data presented in this paper: details of both the field
strength and the geometry are subsumed in a single parameter $y_B$.
Since the emission in this scenario should be strongly polarized, and
the degree of polarization should depend on the interaction geometry, a
hard X-ray polarimeter would provide insightful probes into the presence
of a strongly anisotropizing super-critical field.

The dual blackbody can also be envisaged as a viable alternative from a
theoretical standpoint.  The moderate Thomson depths required to
generate flat Comptonization spectra could easily be higher, thereby
pushing the electron-photon interactions more towards an equilibration.
The saturation temperature is then controlled by the largely uncertain total
energy dissipated in the inner magnetosphere per hot electron present.
Since any equilibration will be non-uniform over a coronal volume, there
should be a modest temperature gradient throughout, smearing out the
continuum.  The range of temperatures will not be great because the
system is not gravitationally hydrostatic in character.  Moreover,
radiative transfer effects impact the spectral shape and further modify
it.  Accordingly, pure blackbody shapes are not expected.  It is quite
conceivable that a two-component blackbody fit may well approximate the
emergent continuum that is a superposition of distorted blackbodies
spanning a small range of temperatures.  Most probably, due to general
thermodynamic considerations, the base of the coronae (e.g. centered
near the source of magnetic dissipation) should be hotter than the outer
layers (see also Lyubarsky 2002 and Thompson \& Duncan 1995).  Interestingly, the fits here generate a smaller volume
associated with the hotter blackbody contribution, consistent with the
expectation for coronal structure.  Yet, somehow, we must obtain a view
of the hotter zone, therefore indicating a strongly aspherical coronal
geometry.  Modeling this semi-equilibration is a challenging task for
theorists considering the influences of the field, the twisted
magnetospheric geometry, and the inherent anisotropy and
polarization-dependence of the scattering process. At present, it is not
possible to distinguish between this thermal scenario, and a
Comptonization one.

\subsection{Conclusions}
Since the COMPT model fits all events, we used it to derive fluences (time-integrated and time-resolved), and determined -- using the
time-resolved values -- that the hardness-fluence correlation can be described by a broken power law, with a minimum at a fluence of
$1.7\times10^{-7}$ erg cm$^{-2}$. We have used $E_{\rm peak}$ values to characterize the hardness of the events in our sample. Earlier studies
of this correlation used hardness ratios to bypass proper spectral fits due to low quality or insufficient data. Moreover, these studies
\citep{fenimore1994,gogus2001} had a very narrow overlap in fluence space. \citet{fenimore1994} used 95  SGR\,$1806-20$ events detected with
the Interplanetary Sun Earth Explorer-3 (ISEE-3) with fluences ranging between $1.25\times10^{-7} - 8\times10^{-6}$ erg cm$^{-2}$, and found a
slightly positive correlation between the two quantities (hardness increasing with fluence). On the other hand, \citet{gogus2001} analyzed 159
and 385 events from SGRs $1806-20$ and $1900+14$, respectively, observed with {\it RXTE}, with fluences ranging between $1.0\times10^{-9} -
2.0\times10^{-7}$ erg cm$^{-2}$, and found the opposite trend (an anticorrelation between hardness ratios and fluences). Since the GBM data
sample covers both fluence ranges ($2.0\times10^{-8} - 2.0\times10^{-5}$ erg cm$^{-2}$) we were able to establish that both trends are indeed
correct, and to define the turning point in the $E_{\rm peak}$ - fluence diagram.

Only two magnetar candidates were observed with GBM to emit a multitude of bursts, thus allowing us to construct their $E_{\rm peak}$ - flux
diagrams. We find for the time-resolved data of \sgrnos, that $E_{\rm peak}$ reaches a minimum of $\sim30$\,keV at a flux value of
$8.7\times10^{-6}$ erg cm$^{-2}$ s$^{-1}$. The second source (SGR\,J$1550-5418$; van der Horst et al. 2011 in preparation) has flux values of
$4.4\times10^{-6}$ erg cm$^{-2}$ s$^{-1}$ at a similar $E_{\rm peak}$ minimum. In addition we used the hardness ratio - count rate relationship
of SGR\,$1806-20$ found in the {\it Integral} data \citep{gotz2004} to derive an approximate flux value of $1.7\times10^{-6}$ erg cm$^{-2}$
s$^{-1}$ (using their conversion from count rate to flux) at the hardness ratio minimum \citep[Figure 3 in][]{gotz2004}. We then converted
these values into isotropic source luminosity, $L_{\rm iso}$ (see Table \ref{liso}); we note that (with the caveat of small number statistics,
and uncertainties in the distance measurements) these values are comparable [$(0.4-1.5)\times10^{40}$ erg s$^{-1}$], although the $B-$fields
and fluxes at $E_{\rm peak}$ minima vary by approximately a factor of ten between the two GBM sources and SGR\,$1806-20$. Whether these
differences reflect intrinsic source properties or instrumental effects is not yet clear. The $E_{\rm peak}$ trend is, however, clearly
established in at least two different instrument data sets. The physical interpretation of this trend is beyond the scope of this paper.

Part of our sample (18 bursts) was fit with the BB+BB spectral function applied to time-integrated intervals; the remaining 11 events could not be fit due to poor statistics. In addition for five bright events we performed time-resolved spectral analysis. We found that the temperatures, emission areas and fluences (luminosities) of the two thermal components exhibit very similar properties in both cases (time-integrated and -resolved). Our results are consistent with those presented for intermediate and short bursts from SGR $1900+14$ by \citet{olive2004} and \citet{israel2008}; the \sgr burst emission areas and
temperatures fall into the same region with those of the short bursts of SGR $1900+14$ shown in Figure 5 of \citet{israel2008}. We also see a
similar trend in the correlation between the luminosities of the two components as described by \citet{israel2008}, namely that both
luminosities increase in tandem. This behavior indicates that the hot and cool BB components may come from two separate emission regions, as
also pointed out by \citet{israel2008}: a smaller but hotter one from the surface of the magnetar, and a larger, cooler one from the star's
magnetosphere (but see also the discussion in section 5.1). We note here that the time-integrated hot BB emission area of 13 \sgr bursts (i.e.,
not including the 5 brightest events; Figure \ref{r2_kt}) is similar to the emission area ($\sim 0.05$ km$^2$) of the BB component found in the
persistent emission of SGR J1550$-$5418 during one active bursting episode in January 2009, believed to originate from a hot spot on the
neutron star surface \citep{kaneko2010}.

Finally, we performed a detailed temporal analysis of all 29 bursts and estimated their durations ($T_{90}$/$T_{50}$) for the first time in
count {\it and} in photon space. Both estimates agree within statistics and thus validated all earlier (count space) duration estimates of
magnetar candidate bursts. The durations (and for four SGRs also the emission times and duty cycles) of five magnetar candidates follow a
log-normal or normal distribution. We find that \sgr events are very similar in average duration with four more magnetar candidates (three SGRs
and one AXP). However, the $T_{90}$ distribution of the bursts from AXP\,1E$2259+586$ has a factor of two larger dispersion \citep[$\sigma \sim
0.73$,][]{gavriil2004} than those of at least three other SGRs \citep[$\sigma \sim0.34, 0.35$,][ and $\sim0.35$, present work]{gogus2001}. \sgr
bursts have an average duty cycle ($\delta_{90}=0.68$) larger than other SGRs \citep[0.45, 0.46 for SGRs $1900+14$, $1806-20$,
respectively;][]{gogus2001}. The differences in the intrinsic properties of the sources may be due to differences in the size of the active
region responsible for the burst emission. We discuss below the burst energetics and its evolution. Similarly to other magnetars, we also do
not find a correlation between the pulse phase of the source and the burst peak times.

The overall behavior of \sgr during its 13-day active period (2008 August 22 to September 3) is also very interesting.  Although about half of
the bursts (16/29) occurred on one day (2008 August 23), most of the burst energy was emitted from the source during August $24-26$ (see also
Figure \ref{energetic}). The average fluence of the beginning and the later part of the active period was constant and at a lower level. During
these three days, GBM detected seven bursts, five of which are also distinguished in that the emission areas from both BB components are the
largest. Further, one of these five events (bn080824.054) shows a double-peaked structure reminiscent of some bright thermonuclear Type I X-ray
bursts from accreting neutron stars, which has been interpreted as due to photospheric radius expansion (PRE). \citet{watts2010} studied the
effects of PRE in high magnetic fields using this event and find that the predicted flux from PRE theory is consistent with the one observed,
opening the way to determining fundamental parameters of neutron stars, such as their equation of state.

%%%%%%%%%%%%%%%%%%%%%%%%%%%%%%%%%%%%%%%%%%%%%%%%
%%%%%%%%%%%%%%%%%%%%%%%%%%%%%%%%%%%%%%%%%%%%%%%%

\acknowledgments This publication is part of the GBM/Magnetar Key Project (NASA grant NNH07ZDA001-GLAST, PI: C. Kouveliotou). M.G.B. acknowledges support from NASA through grant NNX10AC59A. E.G. and Y.K. acknowledge the support from the Scientific and Technological Research Council of Turkey (T\"UB\.ITAK) through grant 109T755. A.v.K. was supported by the Bundesministeriums f\"ur Wirtschaft und Technologie (BMWi) through DLR grant 50 OG 1101. ALW acknowledges support from a Netherlands Organisation for Scientific Research (NWO) Vidi Grant. RAMJW acknowledges support from the European Research Council via
Advanced Investigator Grant no. 247295.

\clearpage

\begin{deluxetable}{clllccccccc}
\tabletypesize{\scriptsize}
\rotate
\tablecaption{Summary of time-integrated spectral analysis for the 29 GBM bursts from \sgrnos. \label{obs}}
\tablewidth{0pt}
\tablehead{
\colhead{No.} & \colhead{Trigger catalog} & \colhead{Trigger time} & \colhead{Detectors} & \colhead{$T_{90}$} & \colhead{$T_{90}^{\rm{ph}}$} & \colhead{Spectral} & \colhead{$E_{\rm{peak}}$\tablenotemark{e}} & \colhead{C-stat/dof\tablenotemark{e}} & \colhead{Fluence\tablenotemark{e,f}} & \colhead{Peak flux\tablenotemark{e,g}} \\
 & $\#$ & (UT) & & (ms) & (ms) & index\tablenotemark{e} & (keV) & & & \\
}
\startdata
1 & bn080822.529\tablenotemark{a} & 12:41:56.914 & 8, 7, 4 & $86^{+42}_{-24}$ & $80\pm16$ & $1.06 \pm 0.70$ & $40.83 \pm 2.76$ & $209.37/177$ & $7.05 \pm 0.62$ & $1.59 \pm 0.32$ \\
2 & bn080822.647\tablenotemark{\dagger} & 15:36:35.200 & 9, 10 & $216^{+46}_{-20}$ & $226\pm24$ & $-1.32 \pm 0.34$ & $39.83 \pm 5.98$ & $147.70/116$ & $19.3 \pm 1.42$ & $7.04 \pm 0.67$ \\
3 & bn080822.981 & 23:32:57.746 & 2 & $30^{+93}_{-14}$ & $30\pm15$ & $1.48 \pm 1.42$ & $44.86 \pm5.02$ & $47.19/57$ & $4.41 \pm 0.67$ & $2.37 \pm 0.52$ \\
4 & bn080823.020\tablenotemark{a,b} & 00:28:09.904 & 3, 4 & $66^{+52}_{-14}$ & $48\pm7$ & $-1.27 \pm 0.20$ & $36.37 \pm 3.28$ & $149.25/117$ & $25.02 \pm 1.12$ & $6.61 \pm 0.59$ \\
5 & bn080823.091\tablenotemark{a} & 02:11:36.630  & 10, 11 & $676^{+54}_{-98}$ & $554\pm40$ & $-1.17 \pm 0.17$ & $42.09 \pm 2.63$ & $154.94/117$ & $82.84 \pm 3.04$ & $6.49 \pm 0.74$\\
6 & bn080823.174 & 04:10:19.280  & 0, 1 & $447^{+53}_{-99}$ & $330\pm51$ & $-0.51 \pm 0.44$ & $57.53 \pm 7.06$ & $130.57/119$ & $14.13 \pm 1.29$ & $1.63 \pm 0.37$ \\
7 & bn080823.248 & 05:56:31.529  & 2 & $272^{+131}_{-126}$ & $276\pm34$ & $1.03 \pm 0.55$ & $51.94 \pm 3.41$ & $65.11/57$ & $22.18 \pm 1.80$ & $3.42 \pm 0.64$ \\
8 & bn080823.293\tablenotemark{a} & 07:01:09.967  & 3, 0, 1, 5 & $174^{+70}_{-20}$ & $164\pm7$ & $0.52 \pm 0.29$ & $48.13 \pm 1.85$ & $272.41/239$ & $20.10 \pm 0.90$ & $2.89 \pm 0.34$ \\
9 & bn080823.293\tablenotemark{\dagger} & 07:04:22.610  & 3, 0, 1, 5 & $38^{+24}_{-10}$ & $30\pm11$ & $-1.63 \pm 0.26$ & $26.68 \pm 8.25$ & $276.95/240$ & $9.54 \pm 0.56$ & $5.59 \pm 0.36$ \\
10 & bn080823.319 & 07:39:32.257  & 9, 10 & $142^{+76}_{-34}$ & $122\pm25$ & $-0.98 \pm 0.30$ & $36.96 \pm 3.22$ & $14.17/115$ & $19.42 \pm 1.16$ & $4.03 \pm 0.49$ \\
11 & bn080823.330 & 07:55:45.690  & 4, 3, 8, 7 & $192^{+60}_{-36}$ & $162\pm13$ & $-0.79 \pm 0.14$ & $30.10 \pm 1.10$ & $282.62/238$ & $67.05 \pm 1.55$ & $15.24 \pm 0.71$ \\
12 & bn080823.354\tablenotemark{a} & 08:30:01.633  & 11, 8 & $96^{+145}_{-28}$ & $94\pm114$ & $0.09 \pm 0.80$ & $28.10 \pm 2.89$ & $141.75/119$ & $8.62 \pm 0.83$ & $2.97 \pm 0.37$ \\
13 & bn080823.429 & 10:18:13.891  & 0, 1, 3, 5 & $94^{+26}_{-22}$ & $82\pm13$ & $-0.85 \pm 0.24$ & $55.32 \pm 4.73$ & $262.81/238$ & $14.24 \pm 0.76$ & $5.02 \pm 0.40$\\
14 & bn080823.478\tablenotemark{a,c} & 11:27:32.306  & 8, 4 & $264^{+34}_{-18}$ & $246\pm6$ & $-0.12 \pm 0.10$ & $34.50 \pm 0.48$ & $111.42/118$ & $512.6 \pm 6.48$ & $69.62 \pm 2.80$ \\
15 & bn080823.623\tablenotemark{a} & 14:56:23.563  & 10, 11 & $220^{+74}_{-32}$ & $204\pm21$ & $0.88 \pm 0.51$ & $48.70 \pm 2.89$ & $128.45/118$ & $21.12 \pm 1.50$ & $3.02 \pm 0.54$\\
16 & bn080823.714 & 17:08:49.038  & 9, 10 & $406^{+52}_{-28}$ & $398\pm11$ & $1.50 \pm 0.41$ & $46.68 \pm 1.69$ & $133.23/116$ & $33.04 \pm 1.59$ & $3.11 \pm 0.47$\\
17 & bn080823.847 & 20:19:30.659  & 9, 10 & $264^{+96}_{-130}$ & $124\pm11$ & $-1.12 \pm0.14$ & $36.57 \pm 1.77$ & $106.81/116$ & $78.61 \pm 2.26$ & $19.95 \pm 1.04$\\
18 & bn080823.847\tablenotemark{\dagger} & 20:23:42.822  & 9, 10 & $108^{+224}_{-52}$ & $110\pm108$ & $-1.01 \pm 0.23$ & $29.95 \pm 2.22$ & $120.36/116$ & $33.09 \pm 1.31$ & $10.29 \pm 0.71$\\
19 & bn080823.986\tablenotemark{b} & 23:39:24.472  & 9, 11, 7, 6 & $60^{+36}_{-18}$ & $30\pm10$ & $-0.37 \pm 0.60$ & $46.62 \pm 5.84$ & $251.91/241$ & $4.37 \pm 0.49$ & $1.43 \pm 0.28$\\
20 & bn080824.054\tablenotemark{*} & 01:17:55.394  & 2, 5 & $260^{+6}_{-8}$ & $250\pm3$ & $-0.25 \pm 0.04$ & $36.01 \pm 0.24$ & $216.46/119$ & $1537 \pm 9.88$ & $185.90 \pm 6.49$\\
21 & bn080824.346 & 08:18:24.418  & 3, 4 & $34^{+68}_{-12}$ & $28\pm11$ & $-0.22 \pm 0.48$ & $57.33 \pm 7.0$ & $107.47/118$ & $5.70 \pm 0.61$ & $3.29 \pm 0.50$\\
22 & bn080824.828 & 19:52:51.264  & 2, 10 & $82^{+80}_{-20}$ & $62\pm16$ & $0.5 \pm 0.84$ & $43.22 \pm 3.98$ & $127.93/116$ & $6.39 \pm 0.72$ & $1.64 \pm 0.36$\\
23 & bn080825.200\tablenotemark{c,*} & 04:48:27.405  & 4 & $110^{+12}_{-10}$ & $102\pm8$ & $-0.36 \pm 0.12$ & $30.32 \pm 0.7$ & $73.1/58$ & $213.6 \pm 4.10$ & $103.07 \pm 3.70$\\
24 & bn080825.401 & 09:37:42.158  & 4, 3, 8 & $128^{+18}_{-14}$ & $114\pm4$ & $-0.03 \pm 0.13$ & $37.76 \pm 0.66$ & $211.69/176$ & $104.8 \pm 2.01$ & $36.99 \pm 1.28$\\
25 & bn080826.136\tablenotemark{c,d} & 03:16:14.773  & 8 & $160^{+74}_{-26}$ & $146\pm7$ & $-0.09 \pm 0.1$ & $36.51 \pm 0.56$ & $74.49/58$ & $507.3 \pm 7.78$ & $161.30 \pm 5.85$\\
26 & bn080826.236 & 05:40:19.425  & 9, 10 & $88^{+68}_{-36}$ & $100\pm16$ & $0.03 \pm 0.36$ & $51.88 \pm 3.41$ & $90.78/116$ & $17.08 \pm 1.06$ & $4.74 \pm 0.57$\\
27 & bn080828.875 & 20:59:39.966  & 1, 0, 5, 3 & $72^{+44}_{-24}$ & $44\pm22$ & $-0.85 \pm 0.47$ & $43.41 \pm 5.69$ & $245.95/239$ & $5.28 \pm 0.51$ & $1.48 \pm 0.24$\\
28 & bn080903.421 & 10:06:35.329  & 4, 5 & $50^{+68}_{-22}$ & $60\pm83$ & $-1.07 \pm 0.48$ & $47.35 \pm 7.63$ & $134.69/119$ & $10.96 \pm 1.09$ & $4.06 \pm 0.67$\\
29 & bn080903.787 & 18:53:48.775 & 2, 10 & $100^{+74}_{-32}$ & $80\pm6$ & $-0.95 \pm 0.39$ & $33.68 \pm 3.75$ & $128.86/117$ & $13.88 \pm 1.01$ & $6.68 \pm 0.65$\\
\enddata
\tablecomments{$^*$ saturation burst; $^\dagger$ untriggered burst\\
$^{a}$ Simultaneously detected with \textit{Swift}/BAT \\
$^{b}$ Simultaneously detected with \textit{RXTE}/PCA \\
$^c$Simultaneously detected with \textit{KONUS}/Wind \\
$^d$Simultaneously detected with \textit{Suzaku} \\
$^e$Calculated with the COMPT model \\
$^f$In $10^{-8}$erg cm$^{-2}$ between $8-200$\,keV \\
$^g$In $10^{-6}$erg cm$^{-2}$ s$^{-1}$ for 8\,ms between $8-200$\,keV
}
\end{deluxetable}

\clearpage

\begin{deluxetable}{lcccccccc}
\tabletypesize{\scriptsize} \tablecaption{Parameters of duration distributions and weighted mean durations for 29 burst from \sgrnos.
\label{durations}} \tablewidth{0pt} \tablehead{
 & \colhead{$T_{90}$} &  \colhead{$T_{50}$} & \colhead{$T_{90}^{\rm ph}$} & \colhead{$T_{50}^{\rm ph}$} & \colhead{$\tau_{90}$} & \colhead{$\tau_{50}$} & \colhead{$\delta_{90}$} & \colhead{$\delta_{50}$} \\
}
\startdata
Mean\tablenotemark{a} & $122.6_{-7.5}^{+7.9}$ & $31.6_{-2.3}^{+2.5}$ & $124.2_{-15.2}^{+17.3}$ & $27.6_{-1.7}^{+1.8}$ & $70.3_{-6.5}^{+7.2}$ & $20.9_{-2.3}^{+2.5}$ & $0.68 \pm 0.03$ & $0.68 \pm 0.02$ \\
$\sigma$\tablenotemark{b} & $0.35 \pm 0.03$ & $0.30 \pm 0.03$ & $0.38 \pm 0.06$ & $0.21 \pm 0.03$ & $0.39 \pm 0.04$ & $0.30 \pm 0.05$ & $0.14 \pm 0.03$ & $0.19 \pm 0.02$ \\
Weighted mean\tablenotemark{c} & $138.3_{-20.5}^{+1.07}$ & $32.4_{-0.8}^{+0.9}$ & $161.2 \pm 1.6$ & $49.2 \pm 0.8$ & & & & \\
\enddata
\tablecomments{$^{a}$ ms in columns $1-6$, dimensionless in columns 7 \& 8; $^{b}$ in the log-frame except for $\delta_{90}$ and $\delta_{50}$;
$^{c}$ in ms.}
\end{deluxetable}

\begin{deluxetable}{lcccc}
%\tabletypesize{\scriptsize}
\tablecaption{$L_{\rm iso}$ at the hardness turnover for three magnetar candidates. \label{liso}}
\tablewidth{0pt}
\tablehead{
 & \colhead{Distance} & \colhead{Flux} & \colhead{$L_{\rm iso}$} & \colhead{$B_{\rm surface}$} \\
 & (kpc) & ($10^{-6}$ erg cm$^{-2}$ s$^{-1}$) & ($10^{40}$ erg s$^{-1}$) & ($10^{14}$ G) \\
}
\startdata
\sgr & 2\tablenotemark{a} & 8.7 & 0.41 & 2.0\tablenotemark{b} \\
SGR\,J$1550-5418$ & 3.9\tablenotemark{c}  & 4.4 & 0.82 & 2.2\tablenotemark{d}  \\
SGR\,$1806-20$ & 8.7\tablenotemark{e} & 1.7 & 1.5 & 20.6\tablenotemark{f} \\
\enddata
\tablecomments{$^{a}$ \citet{xu2006}; $^{b}$ \citet{gogus2010}; $^c$ \citet{tiengo2010}; \\
$^d$ \citet{camilo2007}; $^e$ \citet{bibby2008}; \\
$^f$ The mean surface dipole field $B_{\rm surface} = 3.2\times10^{19}\sqrt{P\dot{P}}$ G, $P$ and $\dot{P}$ from \citet{mereghetti2005}.}
\end{deluxetable}

\end{document}